\documentclass[pdftex, letter, 11pt ]{article}
\usepackage[left=.9in, right=.9in, top=1in, bottom=1in]{geometry}

\usepackage[numbers,sort&compress]{natbib}
\usepackage{etoolbox}
\AtBeginEnvironment{thebibliography}{\linespread{1.0}\selectfont}

\usepackage[tracking]{microtype} 

\usepackage[T1]{fontenc}
\usepackage{amsmath, amssymb, bm, gensymb}
\usepackage{palatino}
\usepackage[sc,osf]{mathpazo}  
\usepackage[euler-digits,small]{eulervm}
\usepackage{textcomp}

\usepackage{graphicx, geometry, float, epstopdf, wrapfig, caption, subcaption} 
\usepackage{xcolor} 
\usepackage{enumitem, lscape, multirow}
\usepackage[colorlinks,
			linkcolor={red!100!black},
			citecolor={blue!100!black},
			urlcolor={blue!80!black}]{hyperref}
\usepackage{calc, comment}
\graphicspath{ {figures/}}
\newsavebox{\largestimage}

\usepackage{titlesec}
\titleformat{\section}
{\normalfont\fontsize{14}{15}\bfseries}{\thesection}{1em}{}

\titleformat{\subsection}
{\normalfont\fontsize{12}{15}\itshape}{\thesubsection}{1em}{}

\titleformat{\subsubsection}
{\normalfont\fontsize{12}{15}\itshape}{\thesubsubsection}{1em}{}

\titleformat{\paragraph}
{\normalfont\normalsize\bfseries}{\theparagraph}{1em}{}
\titlespacing*{\paragraph}
{0pt}{3.25ex plus 1ex minus .2ex}{1.5ex plus .2ex}

\titlespacing{\section}{0pt}{5pt}{*0}
\titlespacing{\subsection}{0pt}{5pt}{2pt}
\titlespacing{\subsubsection}{0pt}{5pt}{2pt}
\titlespacing{\paragraph}{0pt}{*0}{*0}

\usepackage{fancyhdr}
\setlist{noitemsep}
\usepackage{setspace, blindtext}

\makeatletter
\renewcommand{\maketitle}{\bgroup\setlength{\parindent}{0pt}
	\begin{flushleft}
		\textbf{\@title}
		
		\@author
	\end{flushleft}\egroup
}
\makeatother

\title{\Large Lattice Boltzmann method for simulation of diffusion magnetic resonance imaging physics in multiphase tissue models }

\author{\vspace{8pt} Noel M. Naughton$^{1}$, Caroline G. Tennyson$^2$, and John G. Georgiadis$^{1,3}$
	\footnotesize
	\\ \singlespacing \textit{$^1$Department of Mechanical Science and Engineering, University of Illinois at Urbana-Champaign, Urbana, IL 61801, USA; \\[2pt]
	$^2$McKinsey \& Company, 609 Main St, Houston, TX 77002; \\[2pt]
	$^3$Department of Biomedical Engineering, Illinois Institute of Technology, Chicago, IL  60616,  USA.}}

\newcommand\stacklines[2]{\genfrac{}{}{0pt}{}{#1}{#2}}

\begin{document} 
	
\onecolumn
\noindent

{\let\newpage\relax\maketitle}
\thispagestyle{empty}
\begin{abstract}
	\singlespacing \vspace{-12pt}
	We report the first implementation of the lattice Boltzmann method (LBM) to integrate the Bloch-Torrey equation, which describes the evolution of the transverse magnetization vector and the fate of the signal of diffusion magnetic resonance imaging (dMRI). Motivated by the need to interpret dMRI experiments in biological tissues, and to offset the small time step limitation of classical LBM, a hybrid LBM scheme is introduced and implemented to solve the Bloch-Torrey equation. A membrane boundary condition is presented which is able to accurately represent the effects of thin curvilinear membranes typically found in biological tissues.  As implemented, the hybrid LBM scheme accommodates piece-wise uniform transport, dMRI parameters, periodic and mirroring outer boundary conditions, and finite membrane permeabilities on non-boundary-conforming inner boundaries. By comparing with analytical solutions of limiting cases, we demonstrate that the hybrid LBM scheme is more accurate than the classical LBM scheme. The proposed explicit LBM scheme maintains second-order spatial accuracy, stability, and first-order temporal accuracy for a wide range of parameters. The parallel implementation of the hybrid LBM code in a multi-CPU computer system, as well as on GPUs, is straightforward and efficient. Along with offering certain advantages over finite element or Monte Carlo schemes, the proposed hybrid LBM constitutes a flexible scheme that can by easily adapted to model more complex interfacial conditions and physics in heterogeneous multiphase tissue models and to accommodate sophisticated dMRI sequences. 
\end{abstract}

\section{Introduction}

Random molecular motion in the presence of tailored magnetic field gradients imparts a phase dispersion in the nuclear spin transverse magnetization. The resulting signal loss has been employed to quantify the statistics of that motion and probe microscopic diffusion barriers in heterogeneous media \cite{paul1993principles}. Both diffusion-weighted nuclear magnetic resonance (NMR) and diffusion-weighted magnetic resonance imaging (dMRI) exploit this phenomenon to non-invasively probe the microscopic structure of porous media such as sedimentary rocks \cite{behroozmand2015review,vogt2002self} and biological tissues \cite{brownstein1979importance,le2015diffusion}. In biological tissues, dMRI has successfully been used to sensitize the measured NMR signal to microstructural restrictions to the free diffusion of water within the tissue. This measurement has successfully been made in a variety of different tissues such as the brain \cite{le1986mr,pierpaoli1996diffusion,moseley1990early}, skeletal muscle \cite{cleveland1976nuclear,tanner1979self}, cardiac muscle \cite{garrido1994anisotropy}, breast tissue \cite{englander1997diffusion,sinha2002vivo}, liver \cite{kim1999diffusion,taouli2010diffusion}, and cancerous tumors \cite{chenevert2002diffusion,lyng2000measurement}. The present work is motivated by the need to interpret the signal measured during water diffusion through heterogeneous biological tissues in terms of the underlying microstructure.

The spin ensemble physics of dMRI are mathematically described by the Bloch-Torrey partial differential equation, which is a semi-classical model describing the evolution of the bulk magnetization of the spin ensemble in space and time \cite{torrey1956bloch}. The Bloch-Torrey equation is a linear diffusion-reaction equation, with a reaction term that is a function of space and time. This equation can fully accommodate dMRI physics by modeling the effect of externally applied magnetic field gradients (whose timing defines the dMRI sequence) and the diffusion and bulk flow of spins. Spin ensemble physics undergoing diffusion can also be described though the use of a diffusion propagator subjected to magnetic gradients \cite{stallmach2007spin}. Many reduced dMRI models have been employed based on limiting cases \cite{norris2001effects}, a-priori Brownian motion statistics \cite{grebenkov2007nmr}, or effective medium models \cite{novikov2010effective,novikov2011random,sigmund2014time,novikov2014revealing}. Reviews of the different approaches to microstructure modeling in dMRI may be found in \cite{norris2001effects,kiselev2017fundamentals,novikov2018modeling,novikov2019quantifying,alexander2019imaging,jelescu2017design} with an emphasis on neural microstructure. 

Notable progress has been made in developing analytical models that describe the evolution of the signal \cite{grebenkov2007nmr}; however, our focus here is on numerical models of dMRI based on realistic representations of individual cell geometry and tissue microstructure, which are contained in tissue-based representative elementary volumes (REV). Such numerical models are useful in at least two ways: (i) simulating the signal of a dMRI experiment within a given microscopic reconstruction of the tissue microstructure (a tissue model), or (ii) developing and validating more accurate reduced-order dMRI models. In particular, numerical models can be used to simulate the evolution of the dMRI signal in complex tissue geometries, such as domains based on histological images of tissue \cite{berry2018relationships, rose2019novel}, for which no such analytical models exist. 

Previous numerical schemes for the solution of the Bloch-Torrey equation include Monte Carlo \cite{szafer1995theoretical,balls2009simulation,fieremans2010monte,baxter2013computational,yeh2013diffusion,berry2018relationships,bates2017monte,rose2019novel,hall2009convergence}, finite difference \cite{chin2002biexponential, hwang2003image, xu2007numerical, russell2012finite}, and finite element \cite{van2014finite,nguyen2018partition,beltrachini2015parametric} methods, with Monte Carlo methods being the most widely employed. The majority of schemes employ a forward-Euler temporal discretization \cite{szafer1995theoretical} of the Bloch-Torrey reaction term, which is first order in time. Higher order temporal discretization schemes using finite elements have been introduced recently, such as an explicit Runge-Kutta \cite{van2014finite,li2013numerical} and a second-order implicit scheme based on Crank-Nicolson \cite{beltrachini2015parametric,nguyen2018partition}. Based on boundary-conforming finite elements, these schemes have definite advantages over finite differences in terms of describing complex boundaries. This advantage is shared with Monte Carlo methods, which are very simple to code but require careful optimization in order to run efficiently \cite{hall2009convergence,chubynsky2012optimizing}.

The objective of the present work is to revisit and develop a novel simulation method for the numerical integration of the Bloch-Torrey equation in a specified tissue-based continuum REV with an applied linear gradient based on the lattice Boltzmann method (LBM). As a mesoscopic method based on the discrete Boltzmann equation, LBM is particularly efficient for simulating transport processes in complex heterogeneous biological tissue, whereby each lattice node can be assigned unique physics or transport properties. LBM is competitive relative to other computational methods because it involves uncomplicated algorithms, handles complex boundary conditions efficiently and accurately \cite{noble1995consistent,gallivan1997evaluation,li2013boundary,zhang2018consistent}, and is naturally amenable to parallelization \cite{zhao2008lattice,clausen2010parallel,campos2016lattice}. Some of the challenges of LBM are the constraint it imposes on the time step (typical of explicit schemes) and the requirement to derive special boundary conditions for the probability distribution functions in order to preserve the consistency and accuracy of the numerical scheme \cite{noble1995consistent,hiorth2009lattice}.

This work introduces a hybrid implementation of the LBM to integrate the Bloch-Torrey equation in heterogeneous tissue models, which carries the following advantages:
\begin{itemize}
\item  Obviates the problem of the classical LBM implementations, which require small temporal steps when applied to reaction-diffusion problems with dominant reaction terms.

\item  Retains the second-order spatial accuracy in multi-compartmental domains containing complex permeable interfaces.
 
\item  The numerical algorithm can be easily parallelized and executed efficiently with high parallel efficiency in multi-core computer systems. Based on spatial domain decomposition, the expectation is that the kinetic nature of the LBM and the locality of the operations involved result in execution times that scale linearly with the number of cores.
\end{itemize}
The overarching aim of this work is to support the claim that the proposed scheme is accurate, fast, and can accommodate complex geometries of relevance to more general tissue models.

\section{Methods}
\subsection{Diffusion-weighted imaging}

The governing equation describing hydrogen proton (\textsuperscript{1}H) spin dynamics in the presence of diffusion during an MRI experiment is the Bloch-Torrey equation \cite{torrey1956bloch}.  Neglecting coherent (advective) fluid transport, the Bloch-Torrey differential equation can be formulated in a coordinate frame rotating at a fixed Larmor frequency (determined by the MR scanner permanent magnetic field) as follows: 
\begin{equation}
\label{LBM-B-Teq}
\frac{\partial \bm{M}}{\partial t} = 
-i \gamma [ \bm{x} \cdot \bm{G}(t)] \bm{M} - 
\frac{\bm{M}}{T_2 ( \bm{x} )} + 
\nabla \cdot (D ( \bm{x} ) \nabla \bm{M} ) {;} 
\ \ \ 
\bm{M} ( \bm{x} ,t)= 
\Re(  \bm{M} ( \bm{x} ,t) ) + 
i \Im (  \bm{M} ( \bm{x} ,t) ) {,}
\end{equation}			 
where $\bm{M}(\bm{x},t)$ is a complex variable representing the bulk (transverse) magnetization of the spins, $i$ is the imaginary unit, $\gamma$ is the gyromagnetic ratio for \textsuperscript{1}H, $\bm{x}$ is the spin position vector, $\bm{G}(t)$ is the time-varying magnetic field gradient vector used to encode diffusion, $T_2$ is the spin-spin relaxation time, and $D$ is the diffusion coefficient. Except for $\gamma$, all variables listed above are local, in that they represent the ensemble average of spin behavior at a given spatial location. The problem described by Eq. (\ref{LBM-B-Teq}) is supplemented with an initial condition $\bm{M}(\bm{x},0)$ and appropriate boundary conditions, which will be discussed in section \ref{BC_section}. Unless explicitly stated, the initial condition throughout is $\bm{M}(\bm{x},0) = 1+i0$. Decomposing the transverse magnetization $\bm{M}(\bm{x},t)$, as shown in Eq. (\ref{LBM-B-Teq}), the Bloch-Torrey equation yields two coupled reaction-diffusion equations for $\Re(\bm{M}(\bm{x},t))$ and $\Im(\bm{M}(\bm{x},t))$, respectively. The coupling occurs through the first term of the right-hand side of Eq. (\ref{LBM-B-Teq}), which depends on the specific dMRI sequence. 

\begin{figure}[t]
	\includegraphics[height=2in]{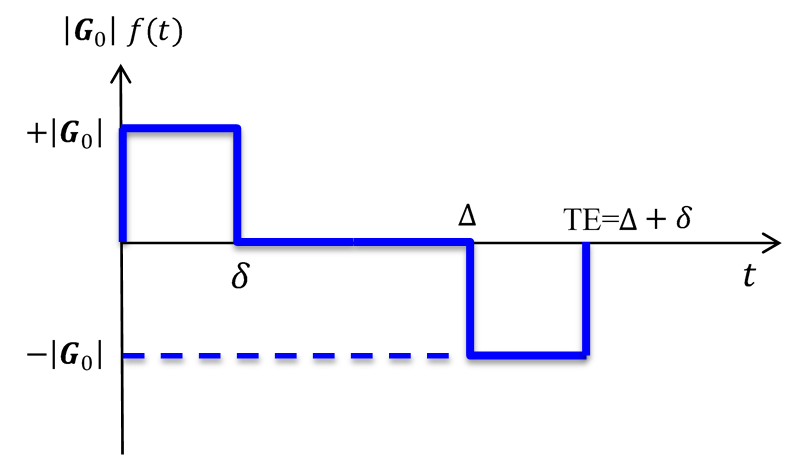}
	\caption{Two-pulse Stejskal-Tanner PGSE sequence with rectangular bipolar gradient waveforms for diffusion MRI.}
	\label{fig2}
\end{figure}

A typical dMRI sequence is the Stejskal-Tanner Pulse-Gradient-Spin-Echo (PGSE) sequence \cite{stejskal1965spin}. Although more sophisticated sequences are in use, we will employ PGSE here since it is adequate to represent MR physics and study local diffusion without image formation. This sequence involves a bipolar magnetic gradient pulse (diffusion gradients), with the gradient vector $\bm{G}(t)$ controlled by the operator, cf. Figure \ref{fig2}. The resulting evolution of $\bm{M}(\bm{x},t)$ generates a time-varying magnetic flux which constitutes the dMRI signal and is acquired upon the appearance of a “spin echo” at time $t$ = TE. Assuming spatially uniform spin density, the dMRI signal $S$ is obtained by integrating $|\bm{M}|$ over the domain, here defined as a representative elementary volume (REV). The gradient magnitude is typically constant in space but varies in time, so it is convenient to express it in separable form, $\bm{G}(t)=\bm{G}_0 \ f(t)$. By judiciously choosing a set of vectors $\bm{G}_0$, each oriented along a specific, non-collinear direction, the signal can be sensitized to probe the dynamics of diffusion along these directions. In addition to the gradient orientation, three parameters describe the PGSE sequence: the diffusion gradient amplitude $|\bm{G}_0|$, the gradient pulse duration $\delta$, and the delay $\Delta$ between the gradient pulses, cf. Figure \ref{fig2}. These parameters can be grouped to define a diffusion decay factor, $b=\gamma^2\ |\bm{G}_0|^2\delta^2(\Delta - \delta/3)$. Additionally, in dMRI it can be useful to define a parameter $\bm{q}$ as $\bm{q}=(\gamma \bm{G}_0 \delta /2\pi)^2$. By taking measurements with multiple q-vectors, the average diffusion propagator can be reconstructed \cite{kiselev2017fundamentals}.

\begin{figure}[t]
	\centering
	\begin{subfigure}[t]{0.49\textwidth}
		\centering
		\includegraphics[height=2.3in]{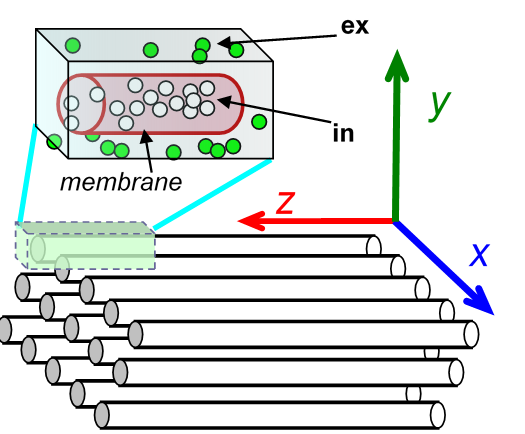}
		\caption{}
		\label{fig2_a}
	\end{subfigure}\hfill
	\begin{subfigure}[t]{0.49\textwidth}	
		\centering
		\includegraphics[height=2.3in]{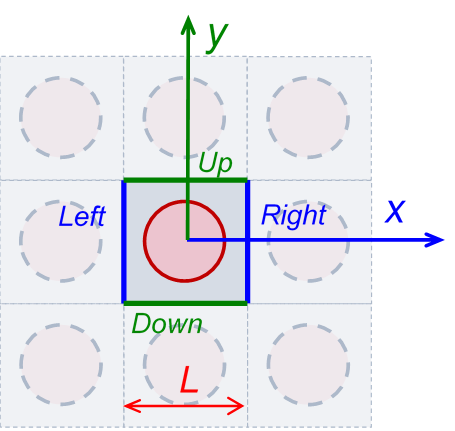}
		\caption{}
		\label{fig:square_domain}
		\label{fig2_b}
	\end{subfigure}
	
	\caption{(a) Periodic parallel fiber model of heterogeneous tissue, with intracellular (in) and extracellular (ex) compartments and membrane of infinitesimal thickness. (b) Two-dimensional periodic computational domain and representative elementary volume (REV) of size. }
	\label{fig1}
\end{figure}

Here heterogeneous tissues are considered to be fibrous inclusions encased in an extracellular matrix. The fibers are surrounded by thin permeable membranes cf. Figure \ref{fig2_a}. Such domains are commonly found in biological tissues and much effort has been devoted to understanding their influence on the dMRI signal in order to use dMRI as a probe of the tissue structure. While such work is important to the larger goal of relating the dMRI signal to the underlying tissue microstructure, the focus of this paper is firmly on the forward problem of solving the Bloch-Torrey equation in tissues with thin permeable membranes. Here the Bloch-Torrey equation is solved in a representative elementary volume (REV) containing intracellular (in) and extracellular (ex) subdomains, cf. Figure \ref{fig2_b}. We will consider problems with isotropic diffusion and piece-wise uniform $T_2$ and $D$. Referring to the two subdomains in Figure \ref{fig2_b}, for example, there are two diffusion coefficients, $D_{in}$ and $D_{ex}$, for intracellular and extracellular compartments, respectively. This property notation will be suppressed in the following, until it is explicitly reinstated. 

\subsection{The lattice Boltzmann method}

Historically, numerical methods of analyzing dMRI physics have been split between particle-tracking based Monte Carlo methods and continuum-based finite difference and finite element methods. Lattice Boltzmann methods (LBM) are mesoscale methods that exist between microscopic Monte Carlo methods, which consider the dynamics of individual particles and macroscopic methods like finite elements, which directly discretize the continuum-based Bloch-Torrey equation. In contrast, the LBM is based on simplified kinetic models that incorporate the necessary microscopic physics to allow the averaged properties to obey desired macroscopic equations. Setting aside the Bloch-Torrey equation for a moment, LBM solves a discretized version of the Boltzmann distribution through consideration of a discrete-velocity distribution function ($g_i$) that describes the distribution of particle velocities in the different lattice directions \cite{kruger2017lattice}. This discretization of the Boltzmann equation leads to the lattice Boltzmann equation, 
\begin{equation}
\label{lattice_eq}
g_i(\bm{x} + \bm{e_i}\cdot \delta t, t+\delta t) = {g}_i(\bm{x},t) + \Omega_i(\bm{x},t) {,}
\end{equation}
which describes how the particles $g_i(\bm{x},t)$ move to the neighboring node $g_i(\bm{x} + \bm{e_i}\cdot \delta t, t+\delta t)$ with velocity $\bm{e_i}$ after being influenced by the collision operator $\Omega_i(\bm{x},t)$, which models the collision and subsequent redistribution of fictitious particles at each node. Through proper selection of $\Omega_i(\bm{x},t)$, often by adopting the Bhatnagar-Gross-Krook (BGK) form of the collision operator \cite{bhatnagar1954model}, it is possible to recover the Navier-Stokes equations of fluid mechanics \cite{qian1992lattice}, leading to the LBM's success in modeling a variety of different fluid mechanical domains \cite{chen1998lattice}. In the case of the Bloch-Torrey equation, there is no bulk fluid velocity, and the Bloch-Torrey equation can be viewed as a reaction-diffusion equation, for which LBM schemes have previously been presented \cite{ayodele2011lattice} (see Appendix A). Reviews of the LBM for fluid flow may be found in \cite{chen1998lattice, aidun2010lattice} while reviews focused on LBM solutions of heat and mass transfer problems, which have a similar formulation to the Bloch-Torrey equation, are available in \cite{perumal2015review,he2019lattice}.

We note that that although traditional methods of modeling dMRI often consider the evolution of a diffusion propagator, the connection between such a propagator and the discrete-velocity distribution function discussed here is a superficial one. LBM considers a fictitious particle distribution, which is a numerical technique that enables solutions to both reaction-diffusion equations as well as problems that include advection, in which case the lack of a diffusion propagator analogue is clear. There is no direct connection to the actual diffusing spin packets, though such a connection may be an interesting future avenue of inquiry.

\subsection{Hybrid lattice Boltzmann method}\label{hybrid-LBM-section}

An order of magnitude analysis of the terms of the right hand side of Eq. (\ref{LBM-B-Teq}) is presented in Appendix A. The phase of the magnetization vector $\bm{M}(\bm{x},t)\ $ exhibits fast oscillations when the term $\bm{x}\cdot \bm{G}(t)$ becomes large. As the truncation error analysis in Appendix A shows, the implementation of the classical LBM method to solve the reaction-diffusion equation (\ref{LBM-B-Teq}) introduces a truncation error term that grows with the square of the REV length size. To remove this dependence on domain size, a hybrid lattice Boltzmann method involving the factorization of the operator in terms of a reaction (slow) and a diffusion (fast) operator is introduced here where, for each timestep,
\begin{equation}
\label{hybrid_formulation}
\bm{M}(\bm{x},t) = \exp\left( -i \gamma [ \bm{x} \cdot \bm{G}(t)] \delta t^\prime - \frac{1}{T_2} \delta t^\prime \right) \bm{M}^\prime (\bm{x},t) {,}
\end{equation}
with $\bm{M}^\prime(\bm{x},t) $ as an intermediate function. The exponent in Eq. (\ref{hybrid_formulation}), with $ \delta t^\prime$ denoting the reaction time step (employed in the discretization of the reaction term), has been reported first in \cite{szafer1995theoretical} and has since been used in most schemes to integrate the Bloch-Torrey equation \cite{balls2009simulation,fieremans2010monte,baxter2013computational,yeh2013diffusion,berry2018relationships,chin2002biexponential,hwang2003image,xu2007numerical}. As shown in Appendix A, this functional form of the exponent is appropriate for $\bm{G}(t)$ piece-wise constant in time, like in the case of the PGSE gradient pulse sequence, cf. Figure \ref{fig2}. For sequences with gradient pulses of different time-dependence, treatment can be generalized. Eq. (\ref{LBM-B-Teq}) is recovered from Eq. (\ref{hybrid_formulation}), accurate to first order in $\delta t^\prime$ (Appendix A), if $\bm{M}^\prime$ obeys the following diffusion equation
\begin{equation}
\label{diff_eq}
\frac{\partial\bm{M}^\prime}{\partial t}= \nabla \cdot (D \nabla \bm{M}^\prime) {.}
\end{equation}

\begin{figure}[t]
	\begin{subfigure}[t]{0.22\textwidth}
		\centering
		\raisebox{.1\height}{%
			\includegraphics[height=1.2in]{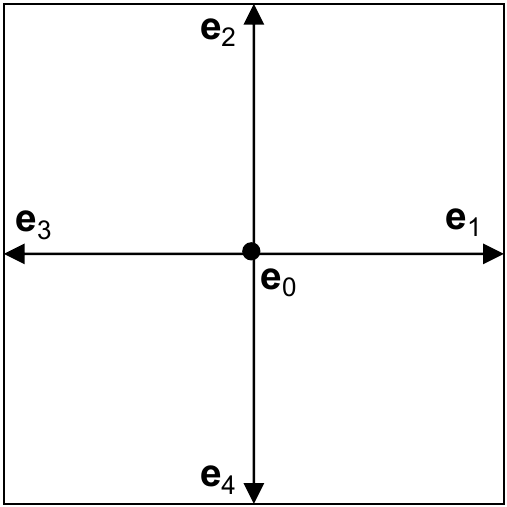}}
		\caption{}
	\end{subfigure}
	\begin{subfigure}[t]{0.22\textwidth}
		\centering
		\includegraphics[height=1.4in]{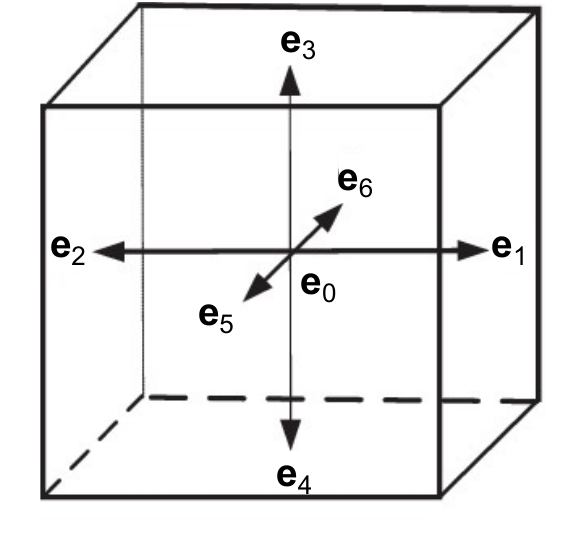}
		\caption{}
	\end{subfigure}
	\caption{ Stencils for Lattice Boltzmann scheme. (a) Two-dimensional, five-speed (D2Q5) and (b) three-dimensional, seven-speed (D3Q7).  }
	\label{fig3}
\end{figure}

Eq. (\ref{diff_eq}) is then integrated with the classical LBM algorithm over a diffusion time step $\delta t$, as shown below. The proposed hybrid LBM scheme for the integration of Eq. (\ref{LBM-B-Teq}) is essentially a time splitting scheme: 
\begin{equation}
\label{LBM_process_list}
\left. \begin{array}{c}
\text{Diffusion split during }[t,t+\delta t]: \ \bm{M}(\bm{x},t)\rightarrow \overline{\bm{M}}(\bm{x},t+\delta t) \text{ from Eq. \ref{diff_eq},} \\
\text{then, initialize } \bm{M}^\prime (\bm{x},t) = \overline{\bm{M}}(\bm{x},t+\delta t) \text{, and} \\
\text{Reaction split during }[t,t+\delta t^\prime]: \  \bm{M}^\prime (\bm{x},t) \rightarrow  \bm{M}(\bm{x},t+\delta t^\prime) \text{ from Eq. \ref{hybrid_formulation}}
\end{array} \right\}
\end{equation}
As written, the practical implementation of the time splitting scheme (\ref{LBM_process_list}) requires that the diffusion and reaction time steps are identical ($\delta t=\delta t^\prime$), but this is not necessary. In fact, one of the advantages of separating diffusion and reaction steps is that different time steps or time-splitting schemes can be used depending on the stiffness of the Eqs. (\ref{hybrid_formulation}) - (\ref{diff_eq}) \cite{alemani2005lbgk}. For example, using $\delta t=k\ \delta t^\prime,\ $ with an integer $k > 1$, solve Eq. (\ref{diff_eq}) for 1 step, and then solve Eq. (\ref{hybrid_formulation}) for $k$ steps so that the timing is consistent. As the analysis in Appendix A indicates, $\delta t=\delta t^\prime$ is a choice that is consistent with the range of physical and numerical parameters pertinent to the present work. In the following, when we refer to the time step, $\delta t=\delta t^\prime$ is assumed. 

The integration of Eq. (\ref{diff_eq}) during the diffusion split is performed with the classical LBM algorithm presented in Eq. \ref{lattice_eq}. The most common version of this algorithm is based on a single relaxation parameter (Bhatnagar–Gross–Krook model) and can be expressed as: 
\begin{equation}
\label{lattice_BGK_eq}
\bm{g}_i(\bm{x} + \bm{e}_i\cdot \delta t, t+\delta t) - \bm{g}_i(\bm{x},t) = -\frac{1}{\tau}[\bm{g}_i(\bm{x},t)-\bm{g}_i^{eq}(\bm{x},t)]
\end{equation}
where $\bm{g}_i$\ is the particle probability distribution function defined on a discrete lattice, $i$ denotes the lattice direction, $\bm{x}$ is the space coordinate on the lattice, $\delta x$ is the lattice spacing (grid size), $\bm{e}_{i}$ is the lattice (speed) vector, $\delta t$ is the diffusion time step, $\bm{g}_i^{eq}$ is the discretization of the fully-mixed equilibrium state for $\bm{g}_i$, and $\tau$ is the dimensionless relaxation time. In this case, like the magnetization $\bm{M}$, the function $\bm{g}_i$ is a complex variable ($\bm{g}_i = \Re(  \bm{g}_i  + i \Im (  \bm{g}_i )$). Since the zero-th moment of $\bm{g}_i$ is equal to $\bm{M}$, the magnetization vector components are recovered by taking the sum of these functions over the lattice directions. Following the Chapman-Enskog analysis of Eq. (\ref{lattice_BGK_eq}), the diffusion equation (\ref{diff_eq}) can be recovered, accurate to $O(\delta t,\ {\delta x}^2)$ if the relaxation time parameter $\tau$ is defined as 
\begin{equation}
\label{tau_def}
\tau = \frac{1}{2} + \frac{\delta t}{\varepsilon_D(\delta x)^2}D {,}
\end{equation} 
where $\varepsilon_D\ $ is a positive constant related to the weighting factors $\omega_i$. Because advection is neglected, the equilibrium distribution function for the LBM scheme (irrespective of whether the reaction term is included) is given by
\begin{equation}
\label{g_eq}
\bm{g}_i^{eq}(\bm{x},t)=\omega_i\ \bm{M}(\bm{x},t) {.}
\end{equation}
Here, $\omega_i$ are weighting factors chosen to allow Eq. (\ref{diff_eq}) to be recovered \cite{yoshida2010multiple}. To be consistent with the LBM lattice topology, $\omega_i$ must satisfy the isotropy and symmetry conditions
\begin{equation}
\sum_i \omega_i \bm{e}_i = 0; \quad \sum_i \omega_i \bm{e}_{ia} \bm{e}_{ib}= \varepsilon_D \delta_{ab}; \quad \text{and} \quad \sum_i \omega_i = 1 ,
\end{equation}
where $\bm{e}_{ia}$ and $\bm{e}_{ib}$ are the spatial components of $\bm{e}_i$ \cite{huber2010lattice}.

For clarity, the theoretical development reported in this section is confined to 2-D isotropic diffusion, thus a 2-D square lattice, 5-speed model (D2Q5), as shown in Figure \ref{fig3}, has sufficient symmetries for a consistent spatial discretization of Eq. (\ref{diff_eq}). In fact, an analysis of the 2D advection-diffusion equation \cite{li2017lattice} indicates that the D2Q5 stencil produces more accurate and robust results than D2Q9, which is the 9-speed stencil. The extension of this scheme to 3D is straight-forward, as is demonstrated in Section \ref{results}, so is not considered in detail here.  
 
The D2Q5 lattice speed vectors in Eq. (\ref{lattice_BGK_eq}) are given by
\begin{equation}
\bm{e}_i =
\left\{
\begin{array}{cl}
(0,0) & (i=0)\\
(\pm 1,0),(0,\pm 1) & (i = 1,2,3,4)
\end{array}
\right. {.}
\end{equation}
For a D2Q5 lattice, $\varepsilon_D=\frac{1}{3}$, and the weighting factors for the equilibrium distribution are
\begin{equation}
\label{omega_i}
\omega_i =
\left\{
\begin{array}{cl}
\varepsilon_D & (i=0)\\
\frac{\varepsilon_D}{2} & (i = 1,2,3,4)
\end{array}
\right. {.}
\end{equation}
Under the classical LBM scheme, the evolution equation, Eq. (\ref{lattice_BGK_eq}), is integrated in two steps over $\delta t$, a collision step ($\bm{g}_i \rightarrow \hat{\bm{g}}_i$) followed by a streaming step ($\hat{\bm{g}}_i \rightarrow \bar{\bm{g}}_i$). The collision step is 
\begin{equation}
\label{collision_LBM}
\hat{\bm{g}}_i(\bm{x},t) = \bm{g}_i(\bm{x},t)-\frac{1}{\tau}\left[\bm{g}_i(\bm{x},t)-\bm{g}_i^{eq}(\bm{x},t)\right] {,}
\end{equation}
where $\bm{g}_i$\ is the initial particle distribution at the beginning of the time step, $\bm{g}_i^{eq}$ is the equilibrium particle distribution Eq. (\ref{g_eq}), and {$\hat{\bm{g}}_i$ is the particle distribution function following the collision step, which is the input to the streaming step. The streaming step is  
\begin{equation} 
\label{streaming_LBM}
\bar{\bm{g}}_i (\bm{x}+\delta x \bm{e}_i\ ,t+\delta t) = \hat{\bm{g}}_i(\bm{x},t) {.}
\end{equation}
From here the reaction step is initialized as $\bm{g}_i^\prime(\bm{x}_n,t^k) = {\bar{\bm{g}}}_i\ \left(\bm{x}_n,t^{k+1}\right),$ where $t^k=k\ \delta t^\prime.$ The magnetization vector $\bm{M}^\prime(\bm{x},t)$ is computed by the zero-th moment of the particle probability distribution function
\begin{equation} 
\label{M'summation}
\bm{M}^\prime\left(\bm{x},t\right)=\ \sum_{i}{\bm{g}_i^\prime(\bm{x},t)} {.}
\end{equation} 
As proven in \cite{huber2010lattice}, the above LBM scheme is unconditionally stable for $\tau>1/2$, which is always satisfied given that the diffusion coefficient D in Eq. (\ref{tau_def}) is positive.	

Turning to the reaction step, the integration of Eq. (\ref{hybrid_formulation}) during the reaction split accounts for the effects of the diffusion gradient pulse and $T_2$ relaxation on the magnetization. We start with a discretized version of Eq. (\ref{hybrid_formulation})
\begin{equation}
\label{descrete_hybrid}
\bm{M}\left(\bm{x}_{n}\ ,t^k+\delta t^\prime\right)=\exp{\left(-i\ \gamma\ \left[\bm{x}_n \cdot \bm{G}\left(t^k\right)\right]\delta t^\prime\right)}\exp{\left(\frac{-\delta t^\prime}{T_2}\right)}\bm{M}^\prime\left(\bm{x}_{n}\ ,t^k\right)\ 
\end{equation}
where $\bm{x}_n$ denotes the coordinate location, and  $\bm{G}\left(t^k\right)=\bm{G}_0\ f(t^k)$ is the gradient vector at time $t^k$. By substituting Eq. (\ref{M'summation}) into Eq. (\ref{descrete_hybrid}), the distribution function after the completion of the reaction step at ${t^{k+1}=t}^k+\delta t^\prime$ becomes	
\begin{equation}
\label{reaction_LBM}
\bm{g}_i\left(\bm{x}_n ,t^{k+1}\right)=\exp{\left(-i\ \gamma\left[\bm{x}_n\cdot\bm{G}_0f\left(t^k\right)\right]\delta t^\prime\right)\exp{\left(\frac{-\delta t^\prime}{T_2}\right)}}\bm{g}_i^\prime\left(\bm{x}_n,t^k\right) {.}
\end{equation}
The hybrid LBM scheme is summarized below in terms of the sequence of the particle distribution functions ($\bm{g}_i\rightarrow{\hat{\bm{g}}}_i\rightarrow{\bar{\bm{g}}}_i\rightarrow\bm{g}_i^\prime$) computed at each step
\begin{equation}
\label{LBM_process_list_2}
\left. \begin{array}{lcl}

\text{Collision at $t^k$:} &
\bm{g}_i\left(\bm{x}_n,t^k\right) \rightarrow{\hat{\bm{g}}}_i\left(\bm{x}_n,t^k\right) 
&\text{Eq. \ref{collision_LBM};} \\

\text{Streaming at $t^k$:} &  {\hat{\bm{g}}}_i\left(\bm{x}_n,t^k\right)\rightarrow{\bar{\bm{g}}}_i\ \left(\bm{x}_n+\delta x\ \bm{e}_{i}\ ,t^{k+1}\right) 
&\text{Eq. \ref{streaming_LBM};} \\

\multicolumn{2}{c}{\text{then, initialize }
\bm{g}_i^\prime\left(\bm{x}_n,t^k\right)={\bar{\bm{g}}}_i\ \left(\bm{x}_n,t^{k+1}\right).}\\

\text{Reaction at $t^k$:} &
\bm{g}_i^\prime\left(\bm{x}_n,t^k\right) \rightarrow\bm{g}_i\left(\bm{x}_n\ \ ,t^{k+1}\right) 
&\text{Eq. \ref{reaction_LBM}}
\end{array} \right\} {.}
\end{equation}

The phase in the first term on the right hand side of Eq. (\ref{reaction_LBM}), which is a function of space and time, couples the computation of the distribution functions. Assume, for example, that a diffusion gradient $\bm{G}_0$ is applied in the (x,y) plane, and there are $[N+1]\times[N+1]$ lattice nodes in the 2D REV shown in Figure \ref{fig2_b}. The phase in Eq. (\ref{reaction_LBM}) at time $t^k$ and location $\bm{x}_n$ becomes
\begin{equation}
\label{delta_phi}
-\Delta \varphi^k_n = \gamma [ \bm{x}_n \cdot \bm{G}_0]f(t^k)\delta t^\prime \text{, \ \ \ with } \ n\in [0,1,2,\ldots N] \times [0,1,2,\ldots N] {.}
\end{equation}
In this implementation of the LBM code, the distribution function $\bm{g}_i$\ is then updated in terms of its real $\Re(\bm{g}_i$) and imaginary $\Im(\bm{g}_i$) components, as follows
\begin{equation}
\label{Re_Im_reaction}
\begin{aligned}
\Re \left( \bm{g}_i(\bm{x}_n,t^{k+1})\right) = \left[ \Re\left( \bm{g}_i(\bm{x}_n,t^{k})\right) \cos{(\Delta \varphi^k_n)} - \Im\left( \bm{g}_i(\bm{x}_n,t^{k})\right) \sin{(\Delta \varphi^k_n)} \right] \exp(-\delta t^\prime/T_2) 
\\
\Im \left( \bm{g}_i(\bm{x}_n,t^{k+1})\right) = \left[ \Im\left( \bm{g}_i(\bm{x}_n,t^{k})\right) \cos{(\Delta \varphi^k_n)} + \Re\left( \bm{g}_i(\bm{x}_n,t^{k})\right) \sin{(\Delta \varphi^k_n)} \right] \exp(-\delta t^\prime/T_2) 
\end{aligned}
\end{equation}
For simplicity, the bold font notation for $\bm{g}_i$ and the other sequence members is henceforth suppressed. 

\subsection{Boundary Conditions}\label{BC_section}

In the following, we show how the boundary conditions for the presented LBM scheme are expressed in terms of the distribution functions for external boundaries, which are periodic, and internal boundaries consisting of the mathematical membranes separating the “in” and “ex” subdomains.

\subsubsection{Membrane boundary condition}

\begin{figure}[t]
	\includegraphics[width=3.2in]{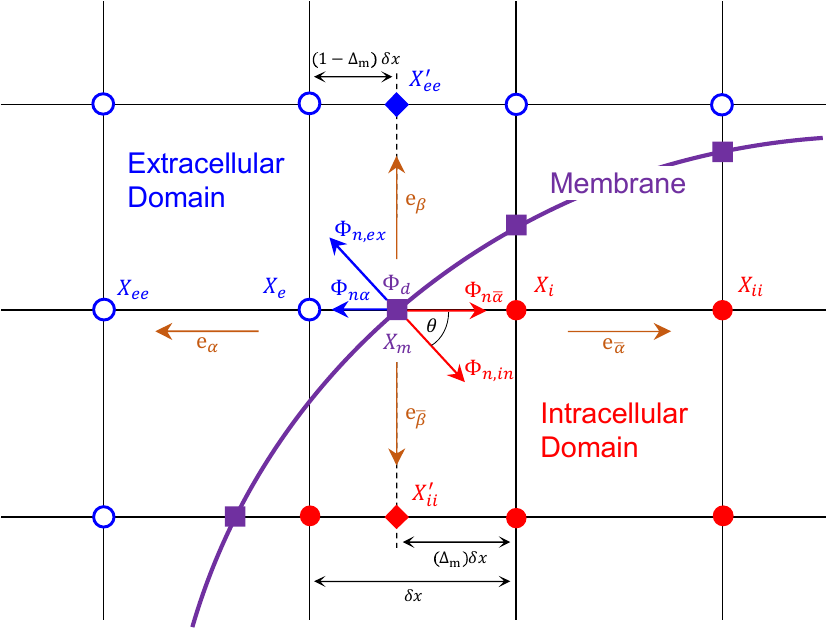}
	\caption{The particle probability distribution functions near the membrane interface and relationship to the local lattice. Adapted from Li et al. \cite{li2014conjugate}.}
	\label{figB1}
\end{figure}

Biological cells are delineated by thin semi-permeable membranes that are weakly diamagnetic, which means that they do not disturb the magnetization significantly. The boundary condition at a thin permeable membrane involves the conservation of mass flux of water spins carrying the scalar $\bm{M}$, without any loss in the membrane. Assuming equal spin density on both sides of the membrane implies conservation of magnetization flux across the membrane. Letting $\bm{n}$ denote the unit vector normal to the membrane and pointing towards the extra-cellular space, and introducing the membrane permeability $\kappa$, this conservation principle imposes the following boundary condition  
\begin{equation}
\label{permeability_flux}
D_{ex}\bm{n}\cdot\nabla \bm{M}_{ex}{=D}_{in}\bm{n}\cdot\nabla \bm{M}_{in}=\ \kappa\left(\bm{M}_{ex}-\bm{M}_{in}\right) {,}
\end{equation}
where $\bm{M}_{in}$ and $\bm{M}_{ex}$ denote the values on the intra- and extra-cellular sides of the membrane, respectively.  This interfacial condition, which is a mixed boundary condition, needs to be reformulated in terms of the particle distribution functions in order to be integrated in the LBM scheme. 

The second-order Dirichlet and Neumann boundary conditions presented by Li et al. \cite{li2013boundary} are used in this work. These boundary conditions are based on the idea of `bounce back' from the membrane. Additionally, their use of spatial interpolation allows preservation of the membrane geometry and their application to curved geometries. These boundary conditions are presented in detail in \cite{li2013boundary,li2014conjugate,guo2015lattice}. Li et al. \cite{li2013boundary} derived second-order accurate Dirichlet and Neumann boundary conditions for general curved boundaries, and later Li et al. \cite{li2014conjugate} extended these boundary conditions to develop an interfacial treatment for conjugate heat and mass transfer. Guo et al. \cite{guo2015lattice} further developed boundary conditions for jumps in concentration or flux at the interface. The extension presented here is the case of the membrane boundary condition, Eq. (\ref{permeability_flux}), used in place of the continuity equation ($\bm{M}_{in}=\bm{M}_{ex}$). 

The variable distance from the lattice point in the intracellular region to the point at which the membrane cuts the lattice link is denoted by $\Delta_m\ \delta x$. So, $\Delta_m$ expresses the dimensionless distance between the internal node closest to the membrane and the membrane ($0<\Delta_m<1$). Here we keep the LBM collision $\rightarrow$ streaming nomenclature with the distributions denoted by $g_i\rightarrow{\hat{g}}_i\rightarrow g_i^\prime$. Each membrane boundary condition is enforced at the end of the LBM collision step, so extra subscripts are necessary to distinguish particle distributions based on the direction the particles move. Figure \ref{figB1} illustrates the nomenclature used in this section. Four lattice velocities are defined ($\bm{e}_\alpha$, $\bm{e}_{\bar{\alpha}}$, $\bm{e}_\beta$, and $\bm{e}_{\bar{\beta}}$), with $\alpha$ and $\beta$ denoting lattice velocities moving in the direction of the intracellular to extracellular domain while $\bar{\alpha}$ and $\bar{\beta}$ denote lattice velocities in the opposite direction.

A detailed derivation of this reformulation is presented in Appendix B. The final equations of the membrane boundary condition are
\begin{equation}
\label{B2a_text}
\begin{split}
{g^\prime}_{\bar{\alpha}}\left(\bm{x}_i,t\right)= \
& 
A_1^i{\hat{g}}_\alpha\left(\bm{x}_i,t\right)+
A_2^i{\hat{g}}_\alpha\left(\bm{x}_{ii},t\right)+
A_3^i{\hat{g}}_{\bar{\alpha}}\left(\bm{x}_i,t\right)+\\ 
&
B_1^i{\hat{g}}_{\bar{\alpha}}\left(\bm{x}_e,t\right)+\ 
B_2^i{\hat{g}}_{\bar{\alpha}}\left(\bm{x}_{ee},t\right)+\ 
B_3^i{\hat{g}}_\alpha\left(\bm{x}_e,t\right)+ \\
&
C_1^i{\hat{g}}_\beta\left({\bm{x}^\prime}_i,t\right)+\ 
C_2^i{\hat{g}}_\beta\left({\bm{x}^\prime}_{ii},t\right)+ 
C_3^i{\hat{g}}_{\bar{\beta}}\left({\bm{x}^\prime}_i,t\right)+\\ 
&
D_1^i{\hat{g}}_{\bar{\beta}}\left({\bm{x}^\prime}_e,t\right)+\ 
D_2^i{\hat{g}}_{\bar{\beta}}\left({\bm{x}^\prime}_{ee},t\right)+\ 
D_3^i{\hat{g}}_\beta\left({\bm{x}^\prime}_e,t\right)
\end{split}
\end{equation}
and
\begin{equation}
\label{B2b_text}
\begin{split}
{g^\prime}_\alpha\left(\bm{x}_e,t\right)= \
&
A_1^e{\hat{g}}_{\bar{\alpha}}\left(\bm{x}_e,t\right)+\
A_2^e{\hat{g}}_{\bar{\alpha}}\left(\bm{x}_{ee},t\right)+\
A_3^e{\hat{g}}_\alpha\left(\bm{x}_e,t\right)+\\
&
B_1^e{\hat{g}}_\alpha\left(\bm{x}_i,t\right)+\ 
B_2^e{\hat{g}}_\alpha\left(\bm{x}_{ii},t\right)+\
B_3^e{\hat{g}}_{\bar{\alpha}}\left(\bm{x}_i,t\right)+\\
&
C_1^e{\hat{g}}_{\bar{\beta}}\left({\bm{x}^\prime}_e,t\right)+\ 
C_2^e{\hat{g}}_{\bar{\beta}}\left({\bm{x}^\prime}_{ee},t\right)+\ 
C_3^e{\hat{g}}_\beta\left({\bm{x}^\prime}_e,t\right)+\\
&
D_1^e{\hat{g}}_\beta\left({\bm{x}^\prime}_i,t\right)+\ 
D_2^e{\hat{g}}_\beta\left({\bm{x}^\prime}_{ii},t\right)+\ 
D_3^e{\hat{g}}_{\bar{\beta}}\left({\bm{x}^\prime}_i,t\right)  {.}
\end{split}	
\end{equation}
where the coefficients are given in Eqs. (\ref{B3a}) and (\ref{B3b}). It is noted that in the case of infinite permeability, these boundary conditions match those presented in Li et al. \cite{li2014conjugate} for conjugate heat and mass transfer. If the membrane is impermeable, the boundary conditions reduce to those originally presented by Li et al. \cite{li2013boundary}.

\subsubsection{Half-lattice link membrane boundary condition}
While the above boundary condition is valid for general membrane geometries, in the case of a straight membrane that is perpendicular to the lattice direction ($\theta = 0$), then $\Phi_{n,\bar{\alpha}} =\ \Phi_{n,in}$ and $\Phi_{n,\alpha} =\ \Phi_{n,ex}$, yielding a simplified version of the coefficients 
\begin{equation}
\label{B4a}
\begin{split}
&A_j^i=\ \Big(c_{d4}c_{d4}^\ast c_{nj}\epsilon_D\delta x+(c_{d4}^\ast c_{n4} c_{dj}+c_{n4}^\ast c_{d4} c_{nj})\kappa\ \delta t\Big)\Big/F \\
&B_j^i=\ \Big(c_{d4}c_{n4}\ \left({c_{nj}^\ast-c}_{dj}^\ast \right) \kappa \ \delta t \Big)\Big/F \\
&C_j^i = \ D_j^i = \ 0
\end{split}
\end{equation}
and
\begin{equation}
\label{B4b}
\begin{split}
&A_j^e=\ \Big(c_{d4}c_{d4}^\ast c_{nj}^\ast\epsilon_D\delta x+(c_{d4}c_{n4}^\ast c_{dj}^\ast+c_{n4} c_{d4}^\ast c_{nj}^\ast )\kappa\ \delta t\Big)\Big/F \\
&B_j^e=\ \Big(c_{d4}^\ast c_{n4}^\ast\left({c_{nj} -c}_{dj} \right) \kappa \ \delta t\Big)\Big/F \\
&C_j^e = \ D_j^e = \ 0
\end{split}
\end{equation}
with 
\begin{equation}
\label{B4c}
F\ =\ c_{d4} c_{d4}^\ast \epsilon_D\delta x+\Big(c_{d4} c_{n4}^\ast+c_{n4}c_{d4}^\ast\Big)\kappa\ \delta t {.}
\end{equation}
In such a simplified case, it is worthwhile to place the membrane at the half-way point between nodes, setting $\Delta=0.5$. In this case, and for the coefficients chosen in Eq. (\ref{c_n}) and Eq. (\ref{c_d}), Eqs. (\ref{B4a}) - (\ref{B4c}) further reduce to  
\begin{equation}
\label{B5}
A_1^i = A_1^e = \frac{P}{1+P} \quad \text{and} \quad 
B_1^i = B_1^e = \frac{1}{1+P} 
\end{equation}
with 
\begin{equation}
\label{P_def}
P=\frac{\epsilon_D}{2\kappa}\frac{\delta x}{\delta t}
\end{equation}
and all other coefficients equal to zero, allowing the boundary condition to be expressed as
\begin{equation}
\label{permeability_BC}
\begin{aligned}
&g_{\bar{\alpha}}^\prime\left(\bm{x}_i,t\right)=\ \frac{1}{1+P} \ \hat{g}_{\bar{\alpha}}\left(\bm{x}_e,t\right)+\frac{P}{1+P} \ \hat{g}_\alpha\left(\bm{x}_i,t\right)
\\
&g_{\alpha}^\prime\left(\bm{x}_e,t\right)=\ \frac{P}{1+P} \ \hat{g}_\alpha\left(\bm{x}_e,t\right)+\frac{1}{1+P} \ \hat{g}_{\bar{\alpha}}\left(\bm{x}_i,t\right)
\end{aligned} {.}
\end{equation}
The boundary condition is applied after the collision step of the LBM algorithm and in place of the streaming step. It is expressed only in terms of distribution functions at nodes $\bm{x}_{i}$ and $\bm{x}_{e}$, reducing the complexity and computational cost of the boundary condition as only the nearest neighboring nodes are necessary and the coefficients ($P/1+P$ and $1/1+P$) are constants that can be precomputed to increase efficiency. We conclude this section with a physical interpretation of the factors in Eq. (\ref{permeability_BC}) by considering the particle distribution functions involved. The presence of the membrane splits the population of the particles moving towards the membrane (from either side) into a portion $\frac{1}{1+P}$ that cross and a portion $\frac{P}{1+P}$ that is reflected back. In the limit of infinite permeability ($\kappa\rightarrow\infty$  and $P\rightarrow0$), Eq. (\ref{permeability_BC}) reduces to Eq. (\ref{streaming_LBM}). Conversely, for the limit of impermeability ($\kappa\rightarrow0$  and $P\rightarrow\infty$), Eq. (\ref{permeability_BC}) reduces to the standard bounce-back condition for a homogeneous Neumann boundary \cite{li2013boundary}. 
	
\subsubsection{Modified periodic boundary condition}
	
	The typical method to terminate the prescribed external boundary conditions for the REV in 2D is to consider an infinite periodic solution domain exhibiting a spatial translation symmetry along x and y, as shown in Figure \ref{fig2_b}. Given the the signal phase's linear spatial dependence, owing to the dMRI gradient term $\left[\bm{x}\cdot\bm{G}_0\ f(t)\right] $ in Eq. (\ref{LBM-B-Teq}), conventional periodic conditions do not apply. Let us consider a spatial period $L$ (length of the REV), and the two (vertical) boundaries marked “Left” and “Right” in Figure \ref{fig2_b}. As demonstrated in \cite{xu2007numerical}, the magnetization on these boundaries obeys the following constraint
	\begin{equation}
	\label{modified_periodic_BC}
	\begin{split} &
	\bm{M}\left(\bm{x}_{Left},t\right)=\exp{\left[i\ \varphi(L,t)\right]}\bm{M}\left(\bm{x}_{Right},t\right);\ \ \\ & \varphi(L,t)=\gamma\ \left[(\bm{x}_{Right}-\bm{x}_{Left})\cdot \bm{G}_0\right]\int_{0}^{t}{f\left(t^\prime\right)dt^\prime } {.}
	\end{split}
	\end{equation}
	This means that the periodic boundary condition results in a phase difference that is proportional to the component of $\bm{G}_0 $ that is perpendicular to the boundaries, the spatial period $L$, and an integral factor that varies with time. If $\bm{G}_0 $ is applied only along x, then the periodic boundary condition along the two (horizontal) boundaries marked “Up” and “Down” in Figure \ref{fig2_b} reduces to the conventional form 
	\begin{equation}
	\label{periodic_BC}
	\bm{M}\left(\bm{x}_{Down},t\right)=\bm{M}\left(\bm{x}_{Up},t\right) {.}
	\end{equation}
	To further simplify the presentation, we set $t^k=k\ \delta t,\ $ and keep only the x-coordinate dependence below, with $\bm{x}_n=n\ \delta x,\ $ where\ $n\in\left[0,1,2,\ \ldots N\right]$. Since the spatial period is $L=N\ \delta x$, the phase in Eq. (\ref{modified_periodic_BC}) can be discretized by approximating the integral by a sum (low order approximation is consistent with Eq. (\ref{A9}))
	\begin{equation}
	\label{periodic_phi}
	\varphi(L,t^k)=\gamma L \left|\bm{G}_0\right|\sum_{m=0}^{k-1}{f(t^m)\ \delta t} {.}
	\end{equation}
	
	As with the membrane boundary condition, all external boundary conditions are applied between the collision and streaming step. The modified periodic boundary condition has been adapted to the LBM scheme as follows. Two external “buffer” lattice rows are introduced at $n\in\left[-1,N+1\right]\ $ in order to complete the streaming step at n=0 and n=N. The following assignments are applied to these rows after the collision step, 
	\begin{equation}
	\label{collision_BC}
	\begin{aligned}
	&{\hat{g}}_i\left(\bm{x}_{-1},t^k\right)=\exp{\left[i\ \varphi(L,t^k)\right]}\ {\hat{g}}_i\left(\bm{x}_{N-1},t^k\right) 
	\\
	&{\hat{g}}_i\left(\bm{x}_{N+1},t^k\right)=\exp{\left[-i\ \varphi(L,t^k)\right]}\ {\hat{g}}_i\left(\bm{x}_1,t^k\right) {.}
	\end{aligned}
	\end{equation}
	After the streaming, Eq. (\ref{streaming_LBM}), and reaction initialization step, the correct phase difference, Eq. (\ref{periodic_phi}), is maintained for the distributions at the nodes on the “Left” and “Right” boundaries at n=0 and n=N, respectively,
	\begin{equation}
	\label{post_BC}
	g_i^\prime\left(\bm{x}_0,t^k\right)=\exp{\left[i\ \varphi(L,t^k)\right]}\ g_i^\prime\left(\bm{x}_N,t^k\right) {.}
	\end{equation}
	During the reaction step, each distribution in Eq. (\ref{post_BC}) gains phase according to Eqs. (\ref{reaction_LBM}) and (\ref{delta_phi}), 
	\begin{equation}
	\label{reaction_at_boundaries}
	\begin{aligned}
	&g_i\left(\bm{x}_{0},t^{k+1}\right)=\exp{\left(i\ \Delta\varphi^k_0\right)} \exp(-\delta t^\prime / T_2) \ g_i^\prime \left(\bm{x}_N,t^k\right) 
	\\
	&g_i\left(\bm{x}_{N},t^{k+1}\right)=\exp{\left(i\ \Delta\varphi^k_N\right)} \exp(-\delta t^\prime / T_2) \ g_i^\prime \left(\bm{x}_0,t^k\right) {.}
	\end{aligned}
	\end{equation}
	Considering Eqs. (\ref{delta_phi}) and (\ref{periodic_phi}), it is straightforward to show that
	\begin{equation}
	\label{phase_combo}
	\exp{\left(-i\ \Delta\varphi^k_0\right)}
	\exp{\left[i\ \varphi(L,t^k)\right]}
	\exp{\left(i\ \Delta\varphi^k_N\right)} = 
	\exp{\left[i\ \varphi(L,t^{k+1})\right]} {.}
	\end{equation}
	By incorporating Eq. (\ref{phase_combo}), Eqs. (\ref{post_BC}) and (\ref{reaction_at_boundaries}) yield 
	\begin{equation} 
	\label{final_BC_modified}
	g_i\left(\bm{x}_0,t^{k+1}\right)=\exp{\left[i\ \varphi(L,t^{k+1})\right]}\ g_i\left(\bm{x}_N,t^{k+1}\right) {,}
	\end{equation}
	which is consistent with the modified periodic boundary condition Eq. (\ref{modified_periodic_BC}). This implies that, by making the assignments from Eq. (\ref{collision_BC}) to the nodes on buffer rows after the collision step, the correct phase difference, Eq. (\ref{periodic_phi}), is maintained for the distributions at the appropriate boundary nodes and at the completion of each time step. For the “Up” and “Down” boundaries, the conventional periodic condition given by Eq. (\ref{periodic_BC}) is satisfied if we make the following assignment after the collision step: 
	\begin{equation} 
	\label{final_BC_periodic}
	{\hat{g}}_i\left(\bm{x}_{Down},t^k\right)=\ {\hat{g}}_i(\bm{x}_{Up},t^k) {.}
	\end{equation}
	
	\subsubsection{Mirroring boundary condition}
	The modified periodic boundary condition allows implementation of the periodic boundary condition, however, it can also be adapted to implement a mirroring boundary condition \cite{fieremans2018physical}. A mirroring boundary condition reflects the domain across a boundary, effectively doubling the analyzed REV. Such a boundary condition is particularly useful when considering non-uniform geometries that have cells cross the REV's boundaries, such as when considering domains derived from tissue micrographs. The mirroring boundary condition avoids the possibility of geometrical discontinuities without having to manually edit the image to make both sides of the domain agree, as is necessary if a periodic boundary condition is imposed \cite{sharafi2010micromechanical}. In the mirroring boundary condition, the buffer node ($\bm{x}_m$) geometry is equal to the geometry of the node on the boundary of the domain ($\bm{x}_n$). The mirroring boundary condition is similar to the modified boundary condition described in Eq. (\ref{modified_periodic_BC}), however, there are two notable differences. The first relates to the exchange of the lattice velocities. Because the node is mirrored instead of translated, the lattice directions are also mirrored. Recalling the notation used for the membrane boundary condition, this means that the lattice velocities at the buffer node in direction $\bm{e}_i$ are computed using the lattice velocities in direction $\bm{e}_{\bar{\imath}}$ from the source nodes. The second notable aspect of the mirroring boundary condition is related to this swapping of the lattice velocities. The lattice velocity distribution at each node is influenced by the gradient direction. Under the mirroring boundary condition, the mirrored node is effectively subject to a gradient that is also mirrored, and thus in a different direction than the gradient direction at all other nodes in the domain. To account for this, Eq. (\ref{modified_periodic_BC}) is modified to `unwind' the magnetization at the boundary back to the origin ($\bm{x}_o$), defined as the location where $\bm{G}_0 \cdot \bm{x}=0$,
	\begin{equation}
	\label{mirror_periodic_BC}
	g_i\left(\bm{x}_{o},t\right)=
	g_{\bar{\imath}}\left(\bm{x}_{n},t\right)
	\exp{\left(i \gamma (\bm{x}_{n}-\bm{x}_{o})\cdot \bm{G}_0 \int_{0}^{t}{f\left(t^\prime\right)dt^\prime }\right)} {.}
	\end{equation}
	Here, the $0^{th}$ order moment of the lattice velocity distribution $\bm{M}$ corresponds to the expected value if no diffusion-sensitizing gradient were gradient. However, the lattice velocity distribution will still exhibit the effect of the gradient direction. If the gradient direction is perpendicular to the boundary's edge, the direction of the gradient for the mirrored boundary node is a reflection of the original gradient direction. In this case, the complex conjugate ($g_{\bar{\imath}}^\ast$) describes the lattice velocity distribution in the gradient direction necessary to describe the mirrored node. Thus, the magnetization can be `rewound' to the buffer node, 
	\begin{equation}
	\label{mirror_periodic_BC2}
	g_i\left(\bm{x}_{m},t\right)=
	g^\ast_{\bar{\imath}}\left(\bm{x}_{o},t\right)
	\exp{\left(i \gamma (\bm{x}_{o}-\bm{x}_{m})\cdot \bm{G}_0 \int_{0}^{t}{f\left(t^\prime\right)dt^\prime }\right)}{.}
	\end{equation}
	Eqs. (\ref{mirror_periodic_BC}) and (\ref{mirror_periodic_BC2}) can be combined to describe the mirroring boundary condition when applied after the collision step as
	\begin{equation}
	\begin{split}
	\Re\big(\hat{g}_i\left(\bm{x}_{m},t\right)\big) =& \ \Re\left[ \hat{g}_{\bar{\imath}}\left(\bm{x}_{n},t\right) \exp\left( i \gamma \left( \bm{x}_n+\bm{x}_m-2\bm{x}_{o} \right) \cdot \bm{G}_0 \int_{0}^{t} \right) \right] \\
	\Im\big(\hat{g}_i\left(\bm{x}_{m},t\right)\big) =& \ \Im\left[ \hat{g}^\ast_{\bar{\imath}}\left(\bm{x}_{n},t\right) \exp\left( i \gamma \left( \bm{x}_n+\bm{x}_m-2\bm{x}_{o} \right) \cdot \bm{G}_0 \int_{0}^{t} \right) \right] {.}
	\end{split}
	\end{equation}
	In the case of the gradient applied parallel to the boundary edge, no reflection of the gradient direction is necessary, leading to the boundary condition being described by
	\begin{equation}
	\hat{g}_i\left(\bm{x}_{m},t\right) = \  \hat{g}_{\bar{\imath}}\left(\bm{x}_{n},t\right) \exp\left( i \gamma \left( \bm{x}_n-\bm{x}_m \right) \cdot \bm{G}_0 \int_{0}^{t} \right) {.}
	\end{equation}
	For the more general case of the gradient not aligning with one of the boundary edges, a more complicated rotation of $g_i\left(\bm{x}_{o},t\right)$ is necessary, which is not considered here. 
	
\subsection{Parallelization of LBM hybrid scheme}
Considering the length scales probed in a typical dMRI experiment ($\sim$$ 1-100 \ \mu m$) relative to the typical size of a dMRI voxel  ($\sim$$ 1 \ mm^3$), the simulation of the entire magnetization field in a voxel would result in a very large computational problem. Although the quasi-periodic structure of many tissues can be exploited to reduce the size of this problem somewhat, it is still necessary to develop an efficient code adapted to LBM to solve this problem directly. To address this, a parallel implementation of the hybrid LBM scheme based on domain decomposition \cite{wang1991parallel} and a Fortran code with message passage interface (MPI) is presented. We employ a 3D version of the hybrid LBM code (D3Q7 stencil) on an REV that is a thin rectangular prism with grid size $N \times N \times 1$. The REV is decomposed into multiple non-overlapping domains that are assigned to separate MPI processes. In this case, the REV is partitioned into strips, and each strip is handled by one MPI process, which in turn is assigned to one core. For every time step, each core executes the hybrid LBM algorithm over the assigned domain and exchanges boundary information with adjacent strips while the two strips at the edges of the REV exchange boundary information with each other subject to the applied external boundary conditions. Two quantities must be minimized to maximize parallelization efficiency: the number of neighboring domains each individual MPI process communicates with and the amount of information passed between each domain. Here we focused on minimizing the number of neighboring domains, while holding the length of the boundary between strips fixed. The performance of the parallel implementation is quantified in terms of the following ratios (p denotes the number of cores)
\begin{equation}
\label{speedup}
\text{Speedup} = \frac{\text{Exceution time with 2 cores}}{\text{Exceution time with p cores}}; 
\ \ \ \ \
\text{Parallel efficiency} = \frac{\text{Speedup}}{\text{p}} {.}
\end{equation}
Due to the particulars of the MPI implementation, it is not possible to run the code with only one core, hence the definition of speedup ratios in Eq. (\ref{speedup}).

\begin{figure}[t!]
	\centering
	\begin{subfigure}[t]{0.48\textwidth}
		\centering		
		\includegraphics[height=2.3in]{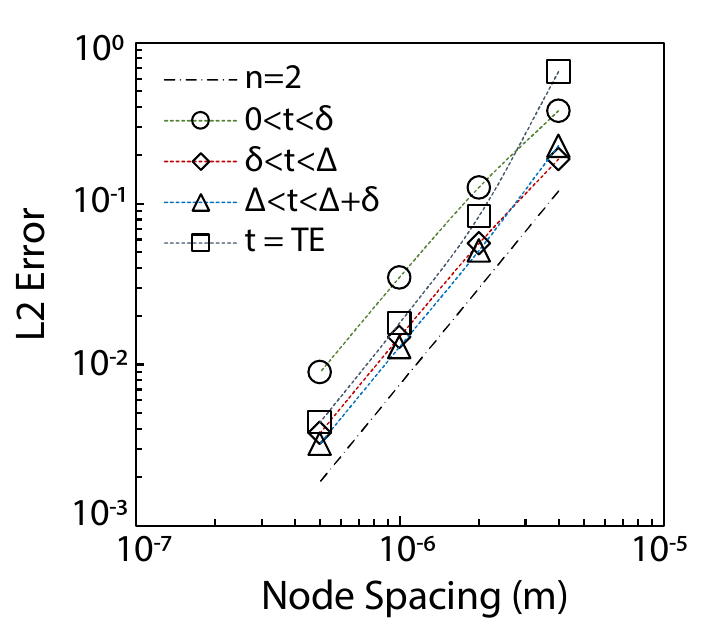}
		\caption{}
	\end{subfigure}\hfill
	\begin{subfigure}[t]{0.48\textwidth}
		\centering
		\includegraphics[height=2.3in]{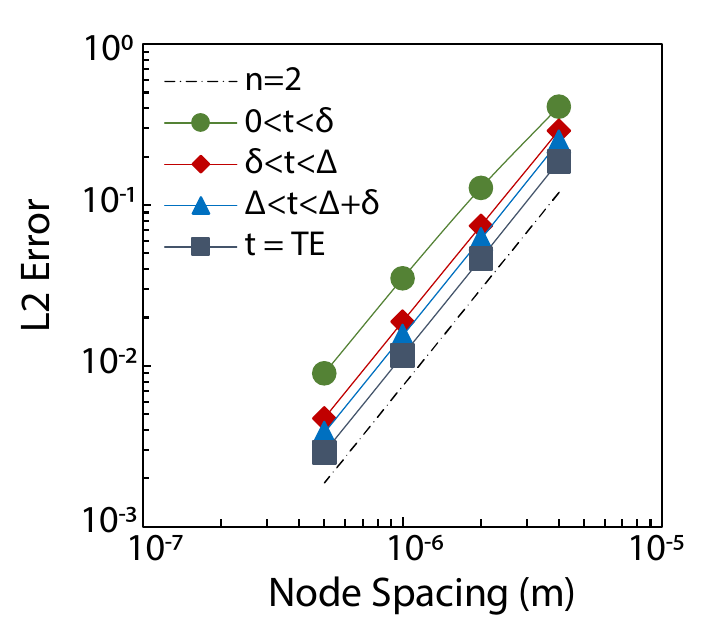}
		\caption{}
	\end{subfigure}
	\caption{Grid convergence rate and spatial truncation error of (a) classical and (b) hybrid LBM schemes, for the impulse response test case, for a 2D square domain (L = 200  $\mu m$)  with $\tau$ = 0.625, b = 1000 s/mm\textsuperscript{2}, $\delta$ = 3 ms, $\Delta$ = 6 ms, and TE = 12 ms. The trend line marked with n=2 corresponds to quadratic convergence.}
	\label{fig4}	
\end{figure}

\section{Results}\label{results}
In this section, we first analyze the accuracy and convergence of the proposed hybrid LBM scheme. This is done by first comparing the hybrid LBM scheme, summarized by Eq. (\ref{LBM_process_list_2}), to that of the classical LBM scheme, which is described by replacing Eq. (\ref{tau_def}) with Eq. (\ref{A12}). All computations involve the numerical integration of the Bloch-Torrey equation (Eq. \ref{LBM-B-Teq}) to simulate the evolution of dMRI signal under the PGSE sequence without imaging gradients (Figure \ref{fig2}). These results are then compared with analytical solutions of the Bloch-Torrey equation. We then assess the accuracy of the hybrid LBM scheme applied to solve the Bloch-Torrey equation as a function of spatial resolution and conclude by examining the accuracy of the proposed membrane boundary conditions. 

Following this analysis, we demonstrate the ability of the hybrid LBM scheme to match solutions of the Bloch-Torrey equation in a number of limiting cases for which analytical solutions exist. We also demonstrate the ability of the hybrid LBM scheme to efficiently scale in a parallel implementation of the computer code as well as the straightforward extension of the scheme to 3D. We conclude this section with an analysis of the error introduced by making various assumption about the orientation and location of the membrane and provide a demonstration of the utility of the hybrid LBM scheme in analyzing complex multiphase domains, such as those typical in biological domains as represented by micrographs. Based on the discussion following Eq. (\ref{LBM_process_list}), we set the time step for the reaction operator identical to that used for the diffusion operator, $\delta t^\prime= \delta t$. 

\begin{figure}[t]
	\centering
\begin{subfigure}[t]{0.48\textwidth}
	\centering		
	\includegraphics[height=2.3in]{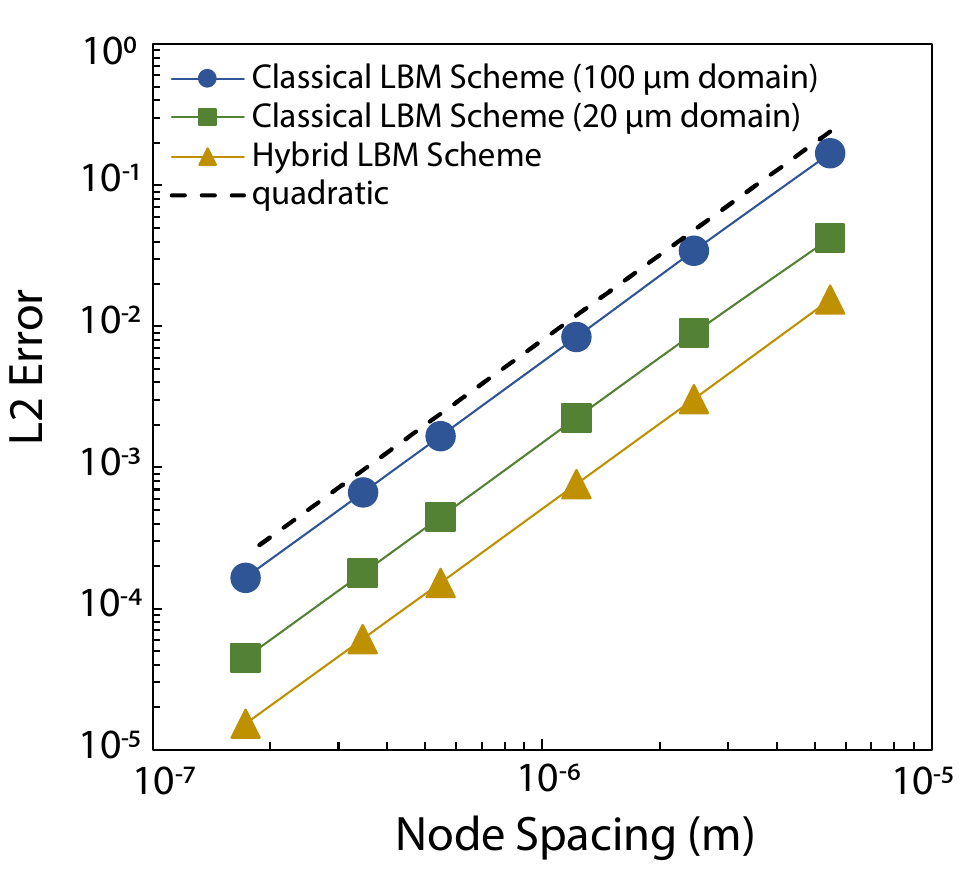}
	\caption{}
\end{subfigure}\hfill
\begin{subfigure}[t]{0.48\textwidth}
	\centering
	\includegraphics[height=2.3in]{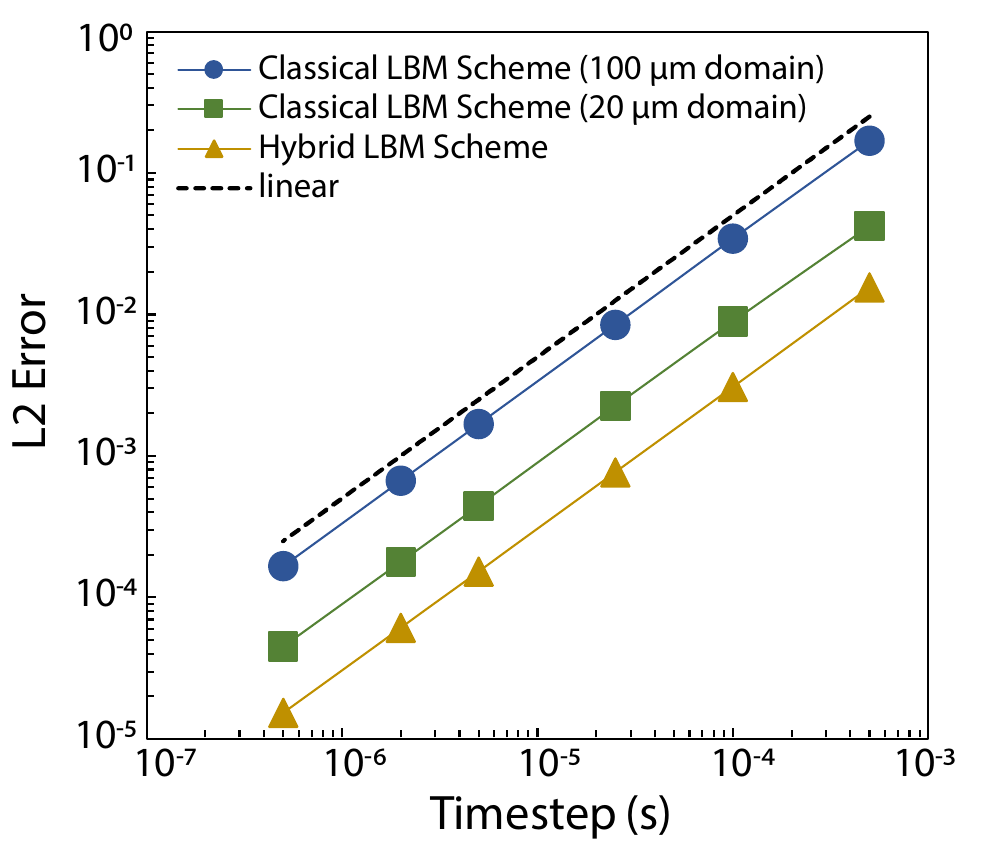}
	\caption{}
\end{subfigure}
	\caption{Comparison of spatial (a) and temporal (b) convergence rates for hybrid and classical LBM schemes for the uniform gradient test case at t = 20 ms, with G = 23.16 mT/m and $\tau$ = 0.60.}
	\label{fig5}
\end{figure}

\subsection[Comparison of truncation error]{Comparison of truncation error for classical and hybrid LBM schemes}

Here, we study the truncation error relative to analytical solutions of Eq. (\ref{LBM-B-Teq}) obtained first with an impulse initial condition and second with a uniform initial condition. The predictions of the classical and hybrid LBM schemes were compared to the exact solution of the Bloch-Torrey equation in a periodic domain with uniform diffusion coefficient and subject to an initial Dirac delta distribution $\bm{M}\left(\bm{x},t=0\right)=\bm{\delta}\left(\bm{x}\right)$ \cite{van2014finite,kenkre1997simple}. A square REV with L=200 $\mu m$ was used, and the remaining physical parameters were b = 1000 s/mm\textsuperscript{2}, $\delta$ = 3 ms, $\Delta$ = 6 ms and TE = 12 ms. The LBM simulations were performed on a $N\times N$ grid for N = 50, 100, 200, and 400 lattice points in each direction, and $\tau$ = 0.625 was kept fixed. The normalized L2 norms of the errors are plotted in Figure \ref{fig4} for four time points: $t_1 = \delta/2$ = 1.5 ms, $t_2 = (\Delta + \delta)/2$ = 4.5 ms, $t_3 = \Delta + \delta/2$ = 7.5 ms, and $t_4$ = TE = 12 ms. A trend line corresponding to L2 error $\sim\delta x^n$ for $n = 2$ is also included for reference. The results indicate that the spatial convergence of both schemes is second order in space, with the exception of the classical scheme at $t_4$. In that case, the scheme reaches the asymptotic convergence regime only for grid sizes smaller than $\delta x = 2$ $\mu m$.
 
In order to highlight the difference between the classical and hybrid LBM scheme as the size $L$ of the domain increases, the numerical solution of the above homogeneous problem was repeated with a uniform initial condition $\bm{M}\left(\bm{x},0\right)=1+i0$. For this case, the solution is trivial: $\bm{M}\left(\bm{x},t\right)=\exp\left(-bD\right) \ \exp{\left(-t/T_2\right)}$. The L2 norms of the errors are plotted in Figure \ref{fig5} as a function of lattice spacing and time step. For the classical method, the error is given for two domain sizes (L=20 $\mu m$, and L=100 $\mu m$) to illustrate how the size of the domain affects accuracy. This effect is not manifested for the hybrid scheme. Although both schemes are second order in space, the accuracy of the hybrid scheme is independent of $L$ and higher than that of the classical scheme. Figure \ref{fig5} indicates that both schemes are first-order accurate in time, as per the discussion in Appendix A. 

To analyze the accuracy of the membrane boundary condition, a periodic domain with a permeable membrane at an angle $\varphi$ was constructed as illustrated in Figure \ref{fig:membrane_BC_domain}. The domain was defined with a distance between the membranes being approximately 40 $\mu m$. In order to allow a direct analysis of the convergence of the scheme, this distance was slightly adjusted such that the number of nodes in each direction of the REV was an integer, thus allowing consistent refinement of the domain with no other changes to the REV geometry. The hybrid LBM scheme results were compared with the analytical short pulse approximation solutions for periodic, permeable membranes \cite{sukstanskii2004effects,tanner1978transient}. To allow comparison with the short pulse approximation, a gradient duration of 1 $\mu s$ was used, or, if such a gradient duration was less than the prescribed time step, $\delta = \delta t$ was used instead. Along with the straight, angled membrane, a packed disk domain similar to Figure \ref{fig2_b} with a diameter of 40 $\mu m$ and intracellular volume fraction of 0.50 was also simulated to examine the convergence of the scheme for curved boundaries. There is currently no analytical solution for such a domain so the scheme was compared with a highly refined solution. Results for both the angled membrane and the packed disked are shown in Figure \ref{fig:membrane_BC}. For both domains, simulation parameters were D = 2.3 $\mu m^2/ms$, $\kappa$ = 50 $\mu m/s$, $\Delta$ = 20 ms, TE = 25 ms and b-value = 1000 $s/mm^2$. The results for the angled membrane domain are second order for all angles while the packed disk is slightly less than second order.

\begin{figure}[t]
	\begin{subfigure}[t]{0.45\textwidth}
		\centering
		\raisebox{0.02\height}{%
			\includegraphics[height=2.3in]{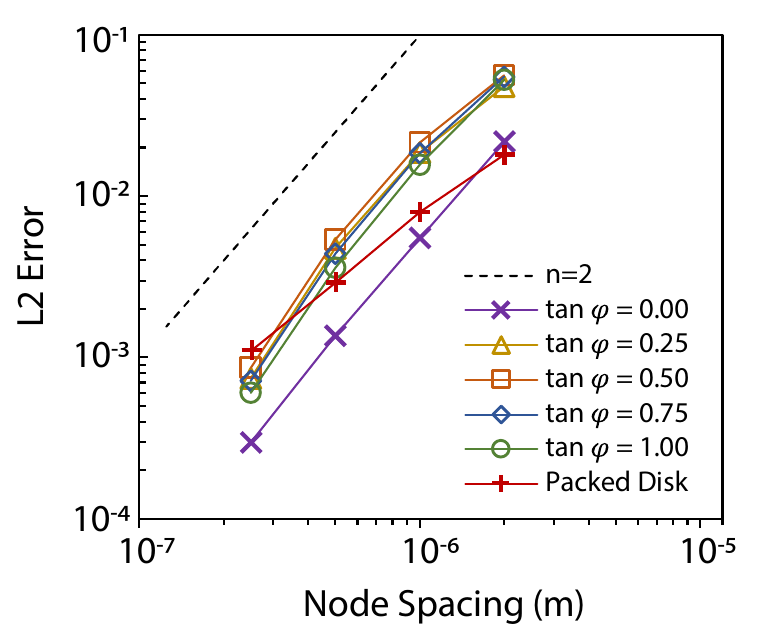}}
		\caption{}
		\label{fig:membrane_BC}
	\end{subfigure}
	\hfill
	\begin{subfigure}[t]{0.5\textwidth}
		\centering
		\raisebox{.2\height}{%
			\includegraphics[height=1.8in]{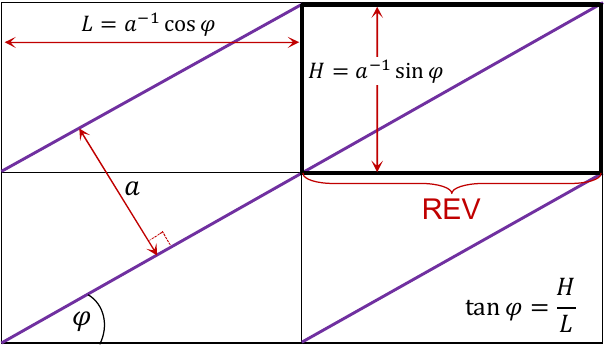}}
		\caption{}
		\label{fig:membrane_BC_domain}
	\end{subfigure}
	\caption{(a) Convergence of packed disk geometry and angled membrane geometry for different angles of $\varphi$ and b) schematic of angled membrane geometry.}
	\label{fig:angled_membrane}
	\label{fig6}
\end{figure}

Given that the first term in the factor  $\exp{\left(-i\ \gamma\left[\bm{x}\cdot\bm{G}\left(t\right)\right]\ \delta t^\prime-\left(T_2\right)^{-1}\delta t^\prime\right)}$ in Eq. (\ref{hybrid_formulation}) changes only the phase of $\bm{M}$, while the second term decays in time, the numerical stability of the hybrid LBM scheme is controlled by the stability of the classical LBM scheme for Eq. (\ref{diff_eq}). LBM is unconditionally stable for the diffusion equation \cite{huber2010lattice}, however, this conclusion does not cover the effect of the membrane boundary condition on the stability of the numerical scheme. In their stability analysis of the Dirichlet and Neumann boundary conditions used to develop the membrane boundary condition, Li et al. \cite{li2013boundary} show that if $-1<c_{d1}<1$, the boundary conditions are stable. Further, they show that the lower bound of stability can be extended below $-1$ depending on the chosen relaxation coefficient $\tau$ and $\Delta_m$. In the implementation presented here $c_{d1} = 2(\Delta_m-1)$, so the lowest possible value of $c_{d1}$ occurs for $\Delta_m=0$ when $c_{d1}=-2$. Due to the coupling between domains, this value of $c_{d1}$ must be in the stability regimes for both $\Delta_m=0$ and $\Delta^\ast_m=1$. In this case, the boundary conditions will be stable if $\tau \gtrsim  0.6$. In the opposite direction, the largest possible value of $c_{d1}$ is $c_{d1}=1$ at $\Delta_m=1$, which is within the original stability regime and so always stable. For the simplified case of $\Delta_m$ fixed at $\Delta_m=0.5$ then $c_{d1}=2\left(\Delta-1\right)=-1$, so the LBM scheme with membrane boundary conditions located halfway between boundaries is always stable, however, for membrane boundaries not located halfway between lattice nodes, there is a restriction on $\tau$ in order to maintain numerical stability.

\subsection{Three-dimensional version of hybrid LBM scheme}

Previous results thus far have been confined to 2D using a D2Q5 stencil, however, the extension of the hybrid LBM scheme to 3D is straightforward. To demonstrate, a D3Q7 stencil (Figure \ref{fig3}) was used to simulate the dMRI signal by solving a 3D periodic array of permeable cylinders with circular cross section whose axis is aligned with the z-coordinate (Figure \ref{fig1}a). The cylinder diameter is 55 $\mu m$ and the packing fraction is 0.65, while the physical parameters are $D_{in}$ = 1.5 $\mu m^2$/ms, $D_{ex}$ = 2 $\mu m^2$/ms, $T_{2,in}$ = 30 ms, $T_{2,ex}$ = 10 ms and $\kappa$ = 10 $\mu m$/s. The REV is a rectangular prism with grid size $55 \times 55 \times 5$. The temporal and spatial steps are $\delta t$ = 0.025 ms, and $\delta x$ = 1.0 $\mu m$. Modified periodic boundary conditions were implemented on all external boundaries, and a PGSE sequence with b = 1000 s/mm\textsuperscript{2}, TE = 24 ms, $\Delta$ = 20 ms, and $\delta$ = 4 ms was used. The gradient $\bm{G}_0$ was applied along an oblique direction with directional cosines $\left(\frac{1}{\sqrt3},\frac{1}{\sqrt3},\frac{1}{\sqrt3}\right)$. The field map of the magnetization at t = TE is shown in Figure \ref{fig13}. Because the cylinder axis is aligned with the z-direction and the boundary conditions are periodic, there is no z-dependence in the signal, even though the gradient $\bm{G}_0$ has a nonzero z-component. This spatial symmetry of the result is a consequence of the symmetry of the inner boundaries and outer boundary conditions and does not constitute a limitation of the general 3D implementation of the LBM scheme.

\begin{figure}[t]
	\includegraphics[height=2.5in]{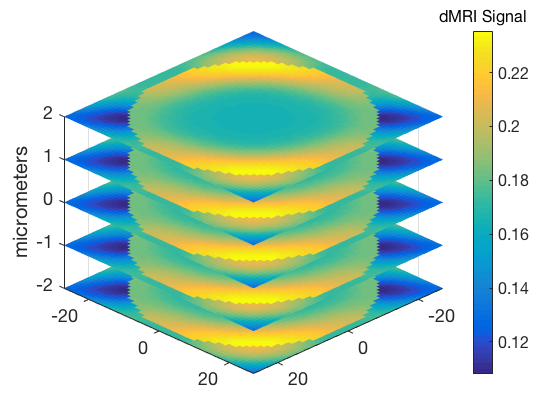}
	\caption{Field map of 3D magnetization on a cross-section at each grid point in the z-direction (exploded view). Due to the cylinder alignment and periodic boundary conditions there is no z-dependence in the signal, even though the gradient was applied obliquely.}
	\label{fig13}
\end{figure}

\subsection[Comparison with analytical solutions]{Comparison with analytical solutions of the Bloch-Torrey equation}
\begin{figure}[t]
	\begin{subfigure}[t]{0.48\textwidth}
		\centering
		\includegraphics[height=2.3in]{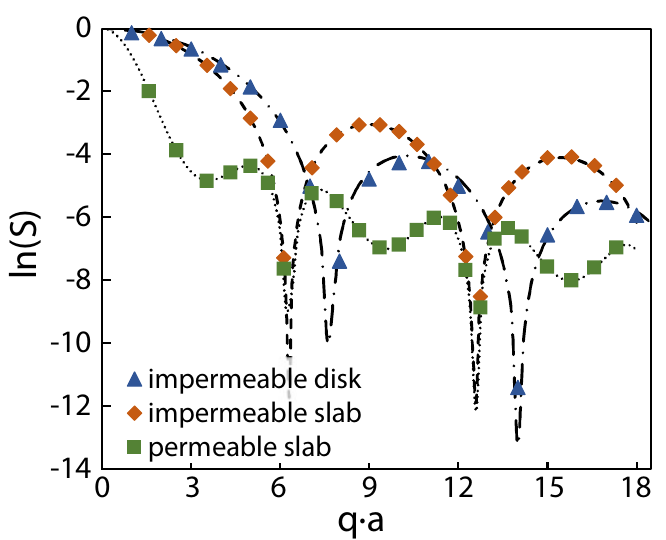}
		\caption{}
		\label{fig:analytical_1}
	\end{subfigure}\hfill 
	\begin{subfigure}[t]{0.48\textwidth}
		\centering
		\includegraphics[height=2.3in]{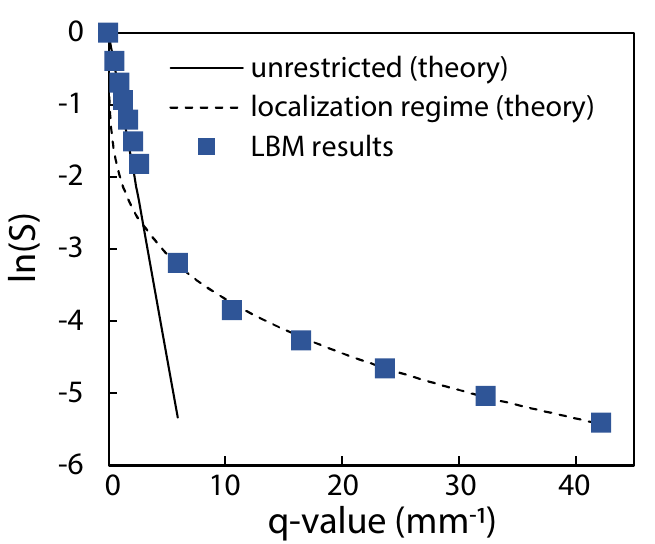}
		\caption{}
		\label{fig:strong_grad}
	\end{subfigure}
	\caption{a) Comparison of LBM scheme with analytical solutions (dotted lines) for impermeable and permeable slabs and an impermeable disk for increasing q-values. b) Comparison of the LBM scheme with the analytical solution of a periodic membrane from \cite{grebenkov2007nmr} for increasing gradient strength as the signal enters the localization regime wherein the signal transitions to a $\sim$$\exp(-q^{1/3})$ dependence. }
\end{figure}

\begin{figure}[t]
\centering
	\begin{subfigure}[t]{0.48\textwidth}
	\centering
		\includegraphics[height=2.3in]{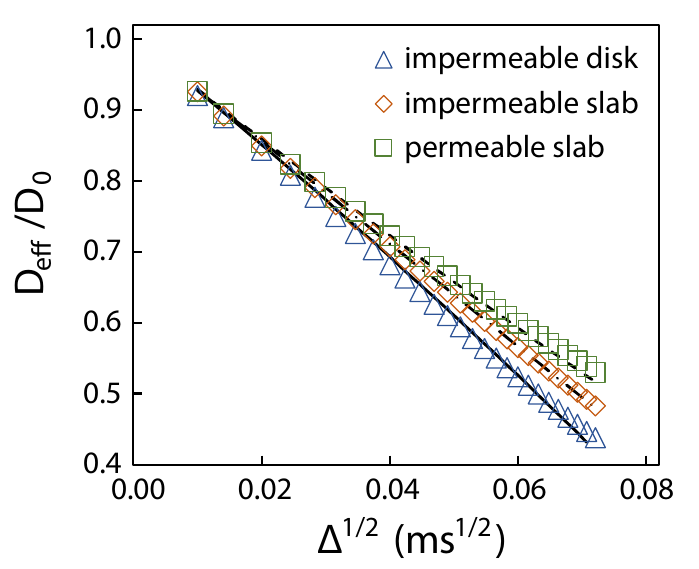}
		\caption{}
		\label{fog:short-time}
	\end{subfigure}\hfill
	\begin{subfigure}[t]{0.48\textwidth}
	\centering
		\includegraphics[height=2.3in]{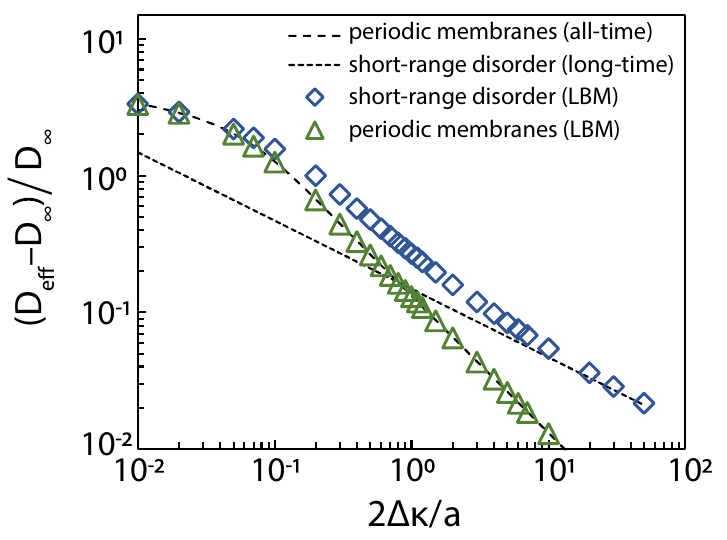}
		\caption{}
		\label{fig:long-time}
	\end{subfigure}
\caption{a) Demonstration of the LBM scheme's ability to match the analytical short time solution (solid and dashed lines) for permeable and impermeable slabs as well as an impermeable disk when the signal exhibits a dependence on the surface-to-volume ratio. D = 2.3 $\mu m^2/ms$, q = 50 $mm^{-1}$, $\delta$ = 50 $\mu s$, TE = $\Delta$ + 1.0 ms, $\delta x$ = 0.5 $\mu m$ and $\delta t$ = 10 $\mu s$. b) LBM scheme in the long time limit for a periodic geometry and a geometry with short-range disorder. In the long time limit, the LBM scheme matches the analytical solution presented in \cite{novikov2014revealing} for the short-range disorder as well as the periodic membrane solution given by \cite{sukstanskii2004effects}, which is valid for all times. D = 2.3 $\mu m^2/ms$, $\kappa$=50 $\mu m/s$, b = 100 $s/mm^2$, $\delta$ = 50 $\mu s$, TE = $\Delta + \delta$, $\delta x$ = 0.5 $\mu m$ and $\delta t$ = 10 $\mu s$.}
\end{figure}

There are a number of analytical solutions to the Bloch-Torrey equation that involve permeable membranes which the hybrid LBM scheme can be compared with. Here we present four benchmarks where the hybrid LBM scheme is compared with known analytical solutions, demonstrating the LBM scheme's ability to accurately match such analytical solutions in a variety of different cases. The hybrid scheme is first compared with the analytical solution for a periodic slab geometry with both permeable and impermeable membranes \cite{sukstanskii2004effects, tanner1978transient} as well as for an impermeable cylinder \cite{soderman1995restricted}. All three cases consider the effect of increasing gradient strength and assume the short gradient pulse approximation ($\delta \rightarrow 0$). Simulations were performed for diameters of 5.0 $\mu m$ with D = 2.3 $\mu m^2/ms$, $\Delta$ = 100 ms, and TE = 110 ms. For the permeable slab case, $\kappa$ = 50 $\mu m/s$. For the disk, $\delta x$ = 0.1 $\mu m$, $\delta t$ = 0.333 $\mu s$, and $\delta$ = 1.0 $\mu s$. For the slabs, $\delta x$ = 0.1 $\mu m$, $\delta t$ = 0.714 $\mu s$, and $\delta$ = 5.0 $\mu s$. Figure \ref{fig:analytical_1} shows that the hybrid scheme is able to successfully match the phase cancellations of the signal, leading to the observed diffraction patterns for all three cases \cite{paul1993principles}. 

For strong gradients, the signal enters a so-called localization region where the signal departs from the Gaussian behavior of the signal and instead illustrates a $\sim$$\exp(-q^{1/3})$ dependence on the gradient strength \cite{stoller1991transverse, grebenkov2007nmr}. To verify that the LBM scheme is able to recreate this behavior, simulations for an impermeable slab with a diameter of 160 $\mu m$ were performed with D = 2.3 $\mu m^2/ms$, $\Delta$ = 60 ms, $\delta$ = 60 ms, TE = 120 ms, $\delta x$ = 0.5 $\mu m$, and $\delta t$ = 12.2 $\mu s$. Figure \ref{fig:strong_grad} shows the hybrid LBM scheme correctly replicates the transition to the localization regime as the gradient increases. 

Solutions of Bloch-Torrey equation demonstrates time-dependent behavior in both the long- and short-time limits. In the short-time limit ($\Delta << D/2a$), the signal demonstrates a dependence on the surface-to-volume ratio that the hybrid LBM scheme is able to match for both a impermeable and permeable ($\kappa$=50 $\mu m/s$) slab with a diameter of 10 $\mu m$ \cite{sukstanskii2004effects} as well as an impermeable 10 $\mu m$ diameter disk \cite{mitra1993short}. The signal behavior was examined for $\Delta$ between 0.1 and 5.2 ms. The effective diffusion coefficient was computed using the low b-value representation of the diffusion coefficient: $D_{eff} = -\ln(S)/b$ \cite{kiselev2017fundamentals}.

In the long time limit, dMRI signal exhibits a power law dependence $t^{-\gamma}$, with the exponent related to the organization of the membranes \cite{novikov2014revealing}. For periodic membranes the effective diffusion coefficient exhibits a $t^{-1}$ dependence while for short-range disorder the signal exhibits a $t^{-1/2}$ dependence. Slabs with both periodic membranes and short-range disorder were simulated with an average diameter of 10 $\mu m$ allowing comparison of the LBM scheme with the 1D results from \cite{novikov2014revealing}. Figure \ref{fig:long-time} shows that the hybrid LBM scheme is able to accurately exhibit the expected long-time behavior for both domains.

\subsection{Parallelization of hybrid LBM scheme}

\begin{figure}[t]
	\centering
	\begin{subfigure}[t]{0.48\textwidth}
		\centering		
		\includegraphics[height=2.3in]{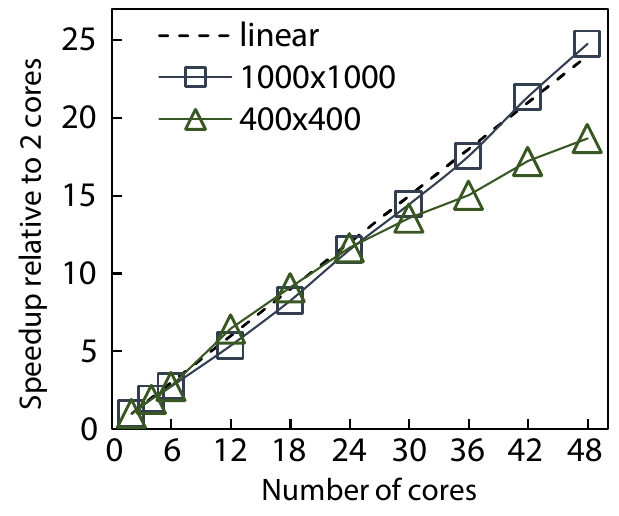}
		\vspace{2pt}
		\caption{}
		\label{fig_parallel_a}
	\end{subfigure}\hfill
	\begin{subfigure}[t]{0.48\textwidth}
		\centering
		\includegraphics[height=2.3in]{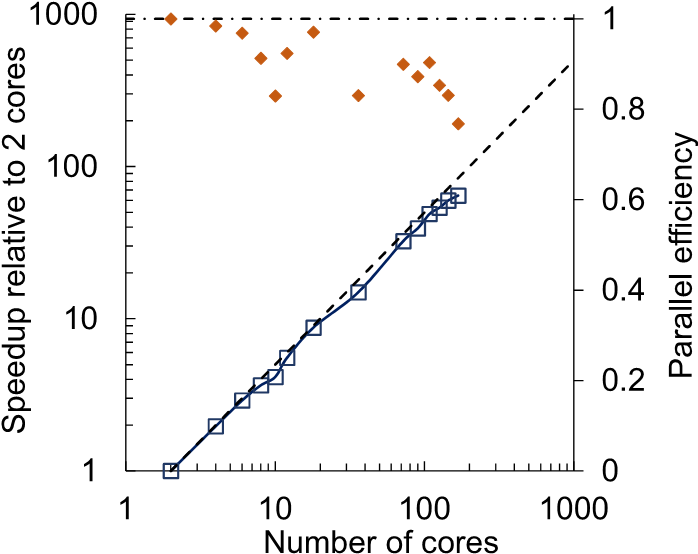}
		\vspace{2pt}
		\caption{}
		\label{fig_parallel_b}
	\end{subfigure}
	\caption{Performance of hybrid LBM code after domain decomposition and parallelization with one-to-one mapping between domains, MPI processes and computer cores. (a) Speedup for N=400 and N=1000 with 2 to 48 cores. (b) Speedup (open symbols) and parallel efficiency (solid symbols) for N=1000 with 2 to 168 cores. Domain and timing parameters were D = 2.0 $\mu m^2$/ms, $T_2$ = 100 ms, q = 40 mm\textsuperscript{-1}, TE = 24 ms, $\Delta$ = 20 ms, and $\delta$ = 4 ms, $\tau$ = 0.58, $\delta t$ = 0.010 ms, and $\delta x$ = 1.0 $\mu m$.}
	\label{fig14}
\end{figure}

To investigate speedup due to parallelization, a parallel code was implemented using Fortran 90 with Intel’s IFORT v14.0.2 compiler and MVAPICH2 v2.1. Simulations were run on SDSC's Comet cluster \cite{towns2014xsede}, which consists of 1944 nodes with 2 x 12 core CPU processors (Intel Xeon E5-2680 v3 2.5 Ghz), and 128 GB DDR4 DRAM running CentOS 6.7. Two homogeneous REVs with square cross-sections and side lengths of 0.4 mm and 1.0 mm were simulated. These domains correspond to $N \times N \times 1$ grids with sizes $N = 400$ and $N = 1000$, respectively. LBM algorithms require high memory throughput and, as such, are often limited by memory bandwidth \cite{kruger2017lattice}. To examine the scaling of the algorithm beyond this known limitation, simulations were preformed using up to 48 cores and adjusting the number of cores per node so the entire problem could be held in cache when possible. All simulations were repeated 5 times and the average execution time was used to determine scaling performance. The results are shown in Figure \ref{fig_parallel_a} and indicate that for $N=1000$ the speedup is linear over the considered range. For $N=400$, there is linear speedup up to 24 cores before the performance begins to degrade. For the current partitioning of the REV, the cost of message passing between cores scales as $\sim$$ N$ while the number of operations per core (p) scales as $\sim$$ N^2/p$. Thus, the cost of message passing relative to operation count per core scales as $\sim \text{p}/N$, implying that performance should degrade at a lower number of cores for coarser grids, explaining the performance degradation for $N=400$. Simulations were also performed for $N=1000$ using up to 168 cores. The results, plotted in Figure \ref{fig_parallel_b}, indicate that the code exhibits ideal (linear) speedup, with a minimum parallel efficiency of 77\% as defined in Eq. (\ref{speedup}).

\begin{figure}[t]
	\centering
	\begin{subfigure}[t]{0.48\textwidth}
		\includegraphics[height=2.3in]{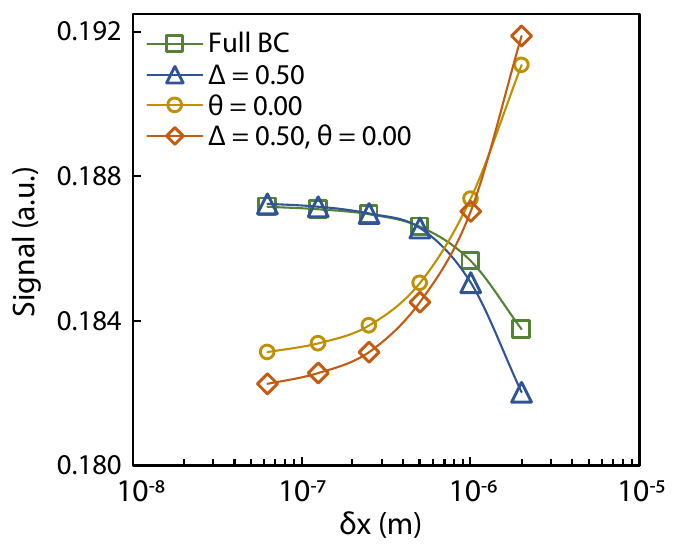}
		\caption{}
		\label{fig:membrane_bc_error_disk}
	\end{subfigure}
	\begin{subfigure}[t]{0.48\textwidth}
		\includegraphics[height=2.3in]{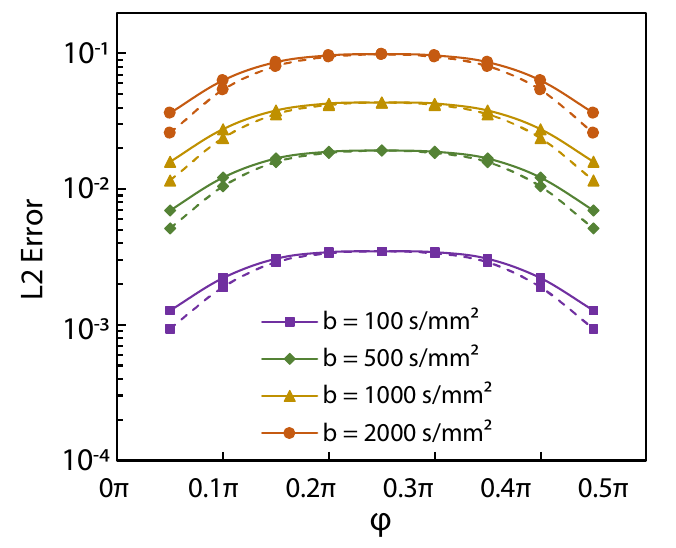}
		\caption{}
		\label{fig:angled_error}
	\end{subfigure}\hfill
	\caption{a) Comparison of the convergence behavior for four versions of the membrane boundary condition. The full boundary condition, the boundary condition with $\Delta_m=0.5$, with $\theta=0$ and with both $\Delta_m=0.5$ and $\theta=0$. b) L2 error between the membrane boundary conditions for $\theta=0$ (dashed lines) and both $\Delta_m=0.5$ and $\theta=0$ (solid lines) for different angles and increasing b-values.}
	\label{fig:membrane_bc_error}
\end{figure}

\subsection[Effect of simplified membrane boundary condition]{Effect of simplified membrane boundary condition assumptions}

While Figure \ref{fig:membrane_BC} demonstrates the second order accuracy of the proposed membrane boundary condition, it is worthwhile to consider the accuracy of simplified versions of the membrane boundary condition, such as when $\Delta_m=0.5$ or $\theta =0$. These boundary conditions are both easier to implement as well as less computationally expensive. If these simplified versions provide sufficient accuracy, it may be preferable to use them, particularly for complex domains when computing $\Delta_m$ and $\theta$ is not straightforward. To begin, the convergence of a 40 $\mu m$ diameter packed disk with an intracellular volume fraction of 0.50 is examined for four different versions of the membrane boundary condition. They are 1) the full membrane boundary condition, 2) the boundary condition with $\Delta_m=0.5$, 3) with $\theta=0$, and 4) with both $\Delta_m = 0.5$ and $\theta=0$. A convergence study was performed with D = 2.3 $\mu m^2/ms$, $\kappa$ = 50 $\mu m/s$, $\Delta$ = 20 ms, $\delta$ = 5 ms, TE = 30 ms and b-value = 1000 $s/mm^2$. Results are shown in Figure \ref{fig:membrane_bc_error_disk}. The membrane boundary condition for $\Delta_m=0.5$ converges with the full boundary condition as the grid is refined while the two membrane boundary conditions that assume $\theta=0$ demonstrate zeroth-order accuracy as they converge to a different value than the full boundary condition, though the result is within 2\% of the result from the full membrane boundary condition. These results indicate that the assumption of $\theta = 0$ is the more limiting of the two assumptions. 

To better understand the error introduced by the assumption of $\theta=0$, the angled domain of Figure \ref{fig:membrane_BC_domain} was reexamined using simplified membrane boundary conditions that assume either $\theta=0$ or both $\theta=0$ and $\Delta_m=0.5$. Multiple angles $\varphi$ were examined for increasing b-values. The angle $\varphi$ was varied between 0 and $\pi/2$ for b-values between 100 and 2000 $s/mm^2$. Simulation parameters were the same as for the packed disk as well as $\delta x$ = 0.5 $\mu m$ and $\delta t$ = 12.25 $\mu s$. These results were  compared with the full membrane boundary condition to quantify the L2 error introduced by these assumptions. Figure \ref{fig:angled_error} shows that the error increases with gradient strength (b-value). It should be noted that the magnitude of the dMRI signal decreases with b-value, and the error for all b-values was always less than 1\% of the original signal value at t=0. The error is greatest for angles of $\varphi=\pi/4$, which, due to symmetries of the geometry, is the greatest deviation possible from $\theta=0$. 

The combined results of Figure \ref{fig:membrane_bc_error} suggest that the assumption of $\theta=0$ should be avoided when possible. However, avoiding such an assumption is not always possible. Particularly when dealing with complex domains, such as those patterned off of realistic biological tissues, it may not be straightforward to compute the local angle $\theta$. While Figure \ref{fig:angled_error} demonstrates that for high b-values this assumption will lead to errors in the simulated dMRI signal, for b-values <1000 $s/mm^2$ the error introduced by this assumption is on the order of <4\% and <1\% for b-values <500 $s/mm^2$. Considering typical SNR values of dMRI measurements are often in the range of 20-50, this suggests that, for b-values <1000 $s/mm^2$, the error introduced by the assumption of $\theta=0$ will be less than the noise in the dMRI signal. Further, any comparisons of simulations with realistic tissue structures will require some approximations of the tissue shape that will also introduce errors. Here, the assumption of $\theta=0$ greatly simplifies the analysis as computing $\theta$ based on images in non-trivial. Further work is necessary to better understand how the error introduced from the assumption of $\theta=0$ is influenced by changes in the b-value or small variations in the tissue structure, however, for low to moderate b-values, these errors do not necessarily preclude its use in analyzing complex tissue geometries.

\subsection{Simulations of histology-informed domains}

\begin{figure*}[ht]
	\centering
	\begin{subfigure}[t]{0.44\textwidth}
		\includegraphics[height=2.1in]{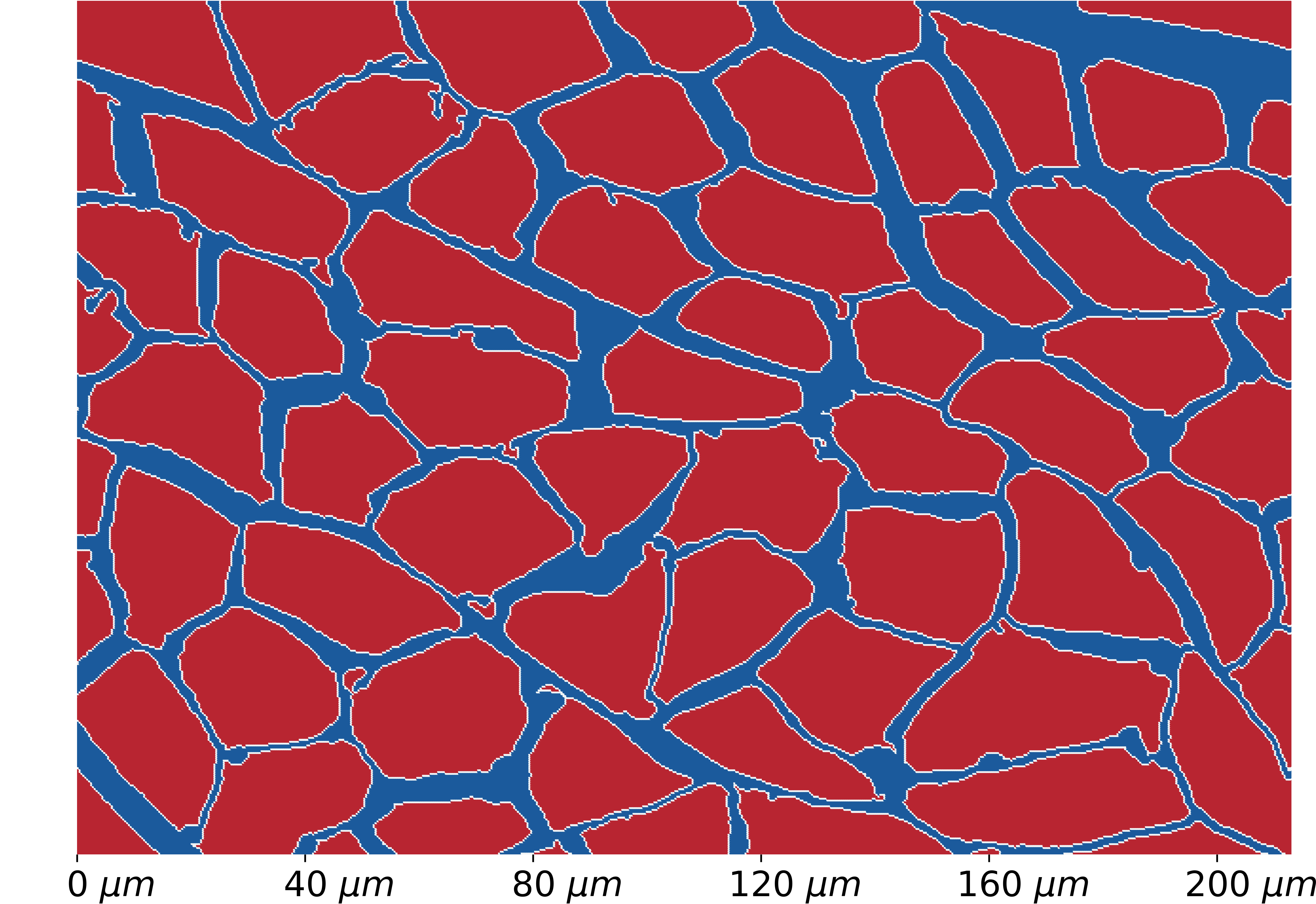}
		\caption{}
		\label{fig:biphasic_Geo}
	\end{subfigure}\hfill
	\begin{subfigure}[t]{0.525\textwidth}
		\includegraphics[height=2.1in]{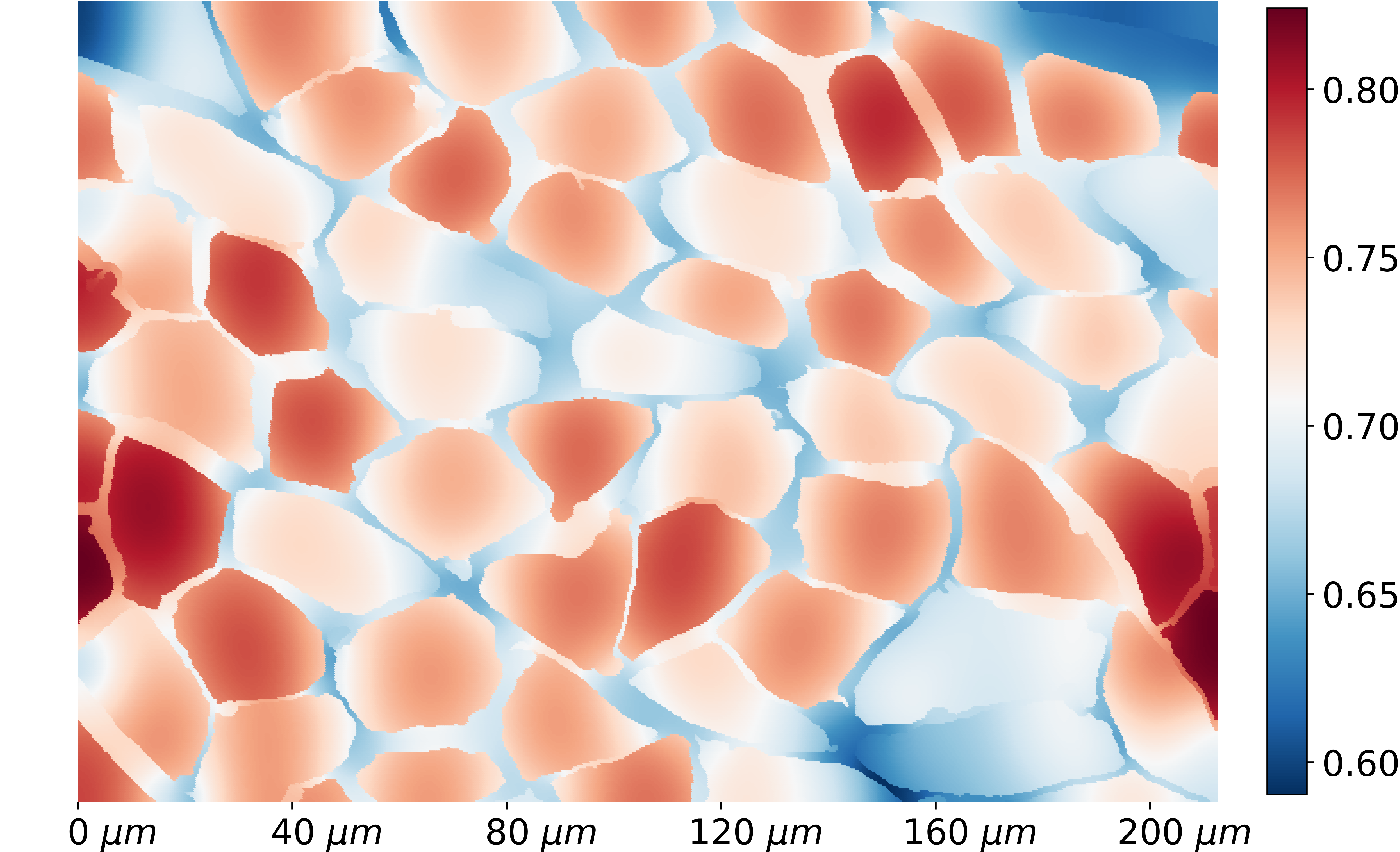}
		\caption{}
		\label{fig:biphasic_signal}
	\end{subfigure}
	\begin{subfigure}[t]{0.44\textwidth}
		\includegraphics[height=2.1in]{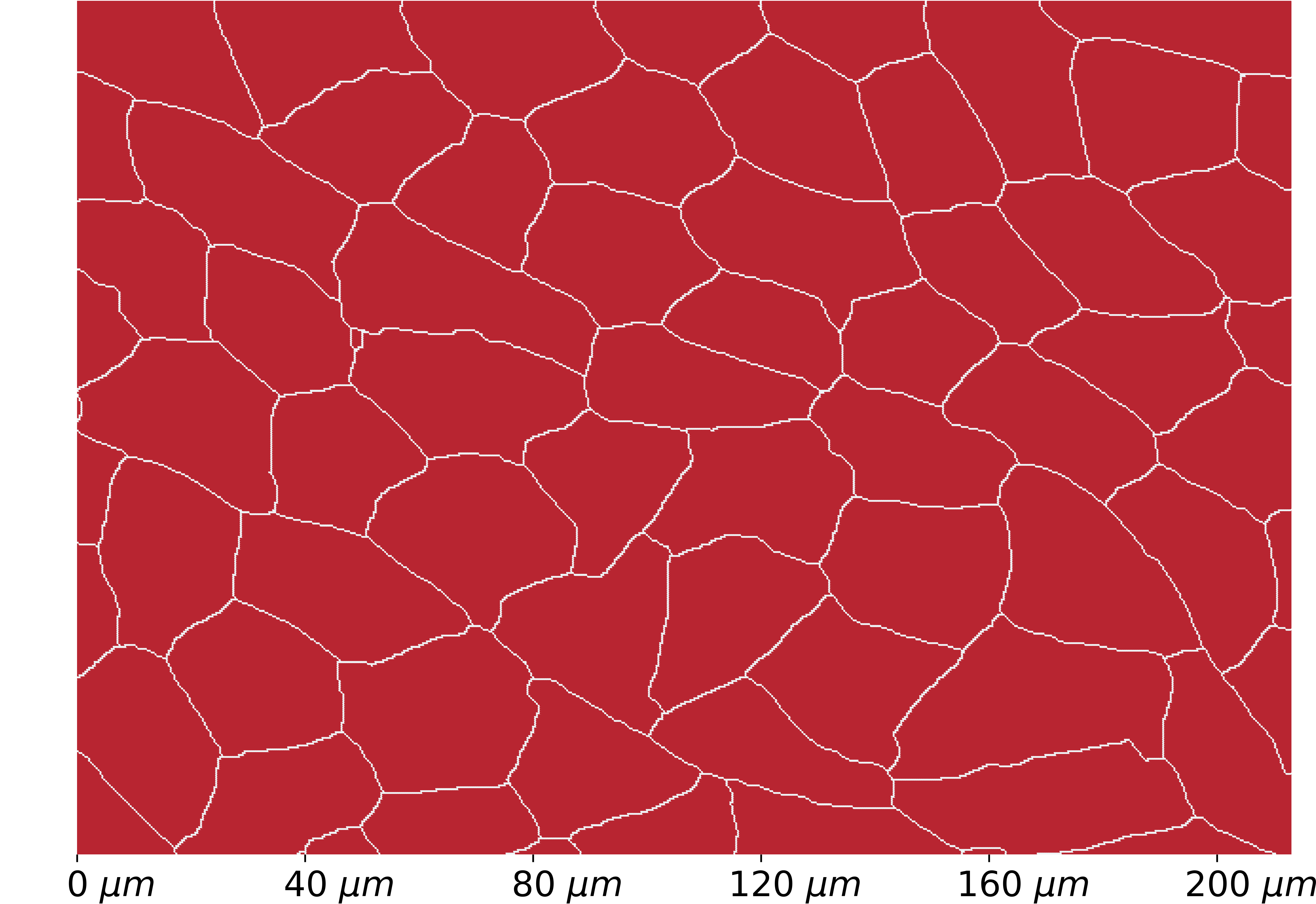}
		\caption{}
		\label{fig:skeleton_geo}
	\end{subfigure}\hfill
	\begin{subfigure}[t]{0.525\textwidth}
		\includegraphics[height=2.1in]{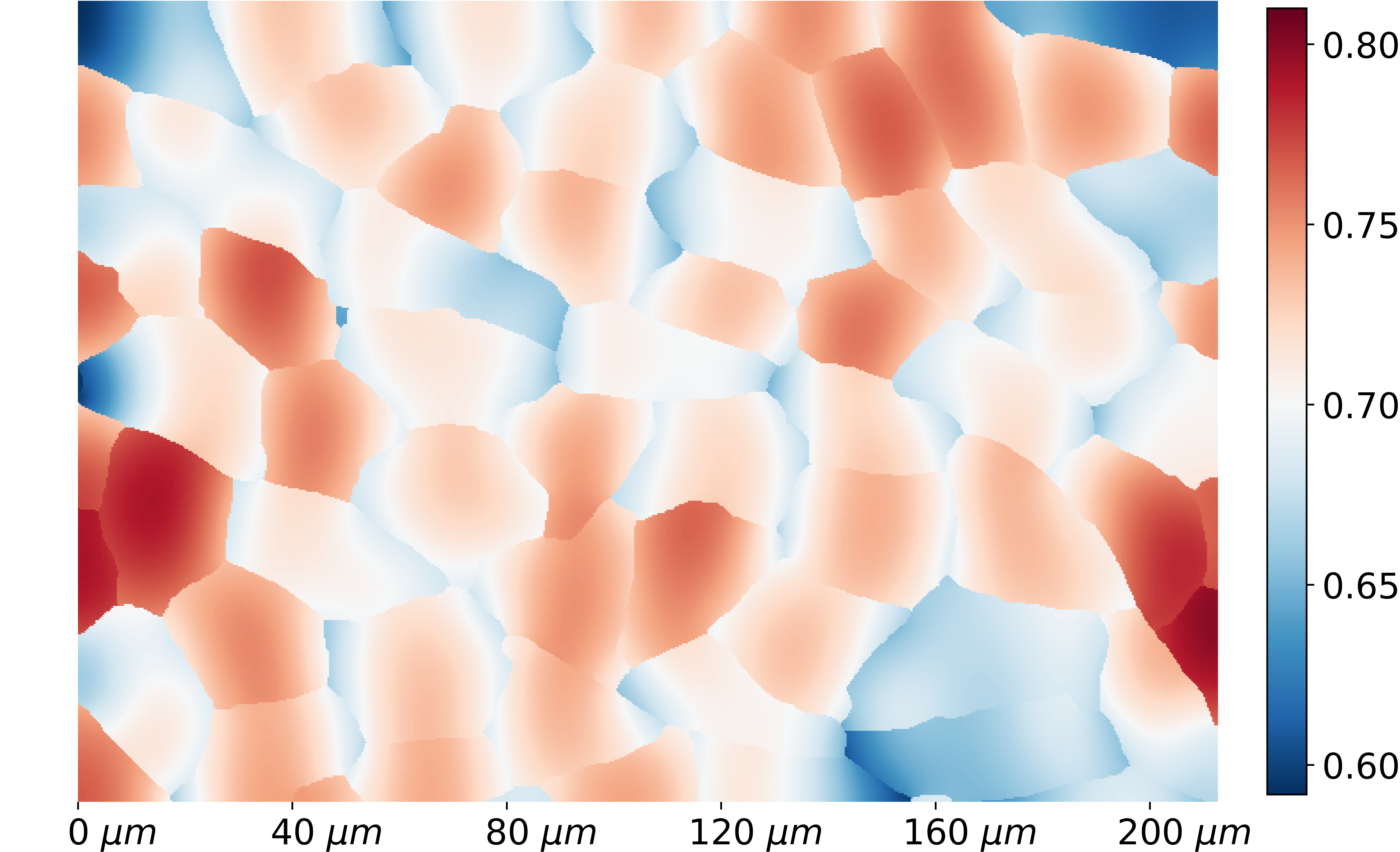}
		\caption{}
		\label{fig:skeleton_signal}
	\end{subfigure}
	\caption{a) Biphasic geometry domain of skeletal muscle cross-section. b) dMRI signal map (arbitrary units) of biphasic geometry at t=TE with $\Delta$=100 ms with the gradient applied in the horizontal direction. c) Skeletonized geometry domain of skeletal muscle cross-section. d) dMRI signal map (arbitrary units) of skeletonized geometry at t=TE with $\Delta$=100 ms with the gradient applied in the horizontal direction. For both geometries, $\delta x$ = 0.333 $\mu m$ and $\delta t$ = 8.33 $\mu s$.}
\end{figure*}

\begin{figure}[h]
	\includegraphics[height=2.3in]{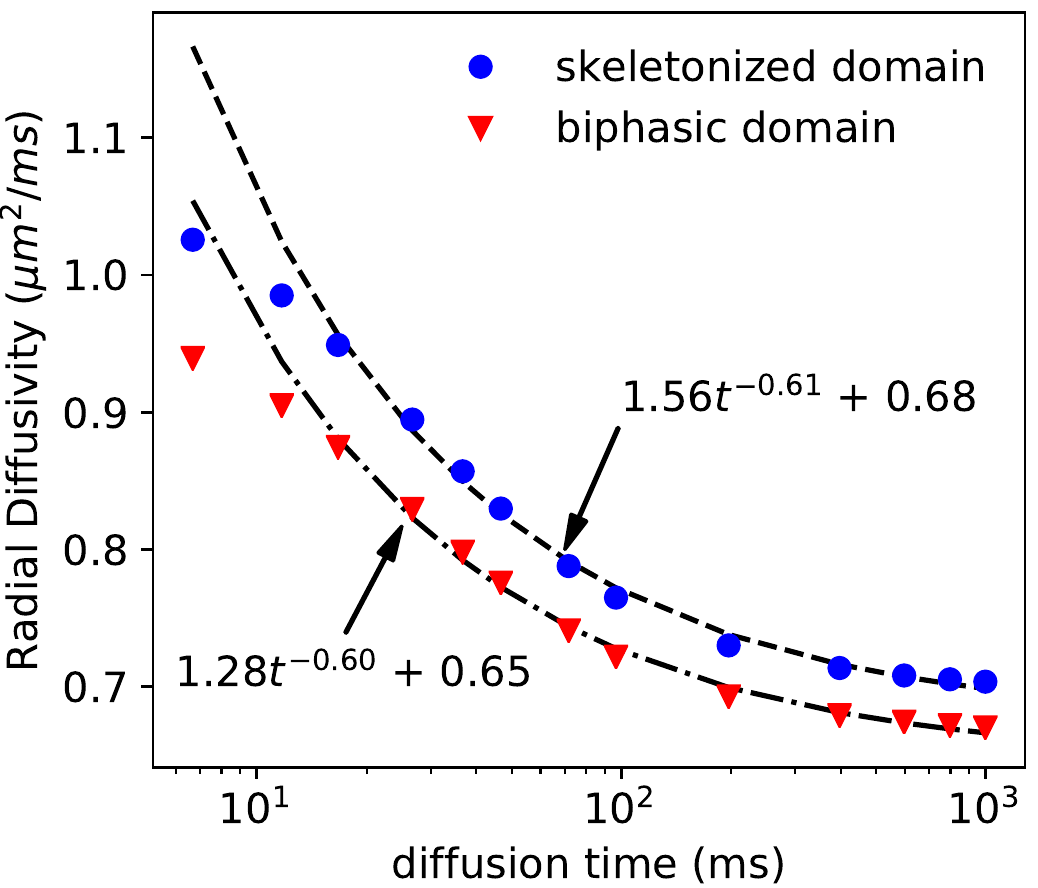}
	\caption{Time-dependent radial diffusivity of biphasic and skeletonized geometries with associated power law fits demonstrating agreement in the fitted exponents to experimentally observed measurements \cite{novikov2014revealing}. Fitting of the power law was performed using all data points except the two shortest diffusion times.}
	\label{fig:RD}
\end{figure}

One area where dMRI has found success in measuring tissue microstructure is in skeletal muscle \cite{fieremans2017vivo,karampinos2009myofiber}. Additionally, these measurements are often made with relatively low b-values, identifying it as an area where the simplified half-link membrane boundary condition can be applied to analyze complex, curved domains derived from tissue micrographs of skeletal muscle cross-sections. Here a micrograph was obtained from digital images available in the open literature \cite{muscle_image1}. To derive realistic REVs, morphological image processing of the micrograph was performed. This involved thresholding with ImageJ \cite{imageJ} and segmenting manually to produce a biphasic geometry consisting of extra- and intra-cellular domains (Figure \ref{fig:biphasic_Geo}). Skeletal muscle is tightly packed and often assumed to fill the entire domain. To analyze this case, another domain was created. Using a custom Matlab script, a watershed transform was performed on this image to dilate each cell so that the extracellular space was reduced to a skeletonized outline with the interface between the cells defined by a single permeable membrane (Figure \ref{fig:skeleton_geo}). 

LBM simulations of the dMRI signal in the REVs of Figure \ref{fig:biphasic_Geo} and Figure \ref{fig:skeleton_geo} were performed using the hybrid LBM scheme over a uniform, non-boundary conforming grid for diffusion times in the range 10-1000 ms. Membrane boundary conditions using the half-link membrane boundary condition were applied at the interfaces between cells while mirror boundary conditions were applied at the edges of the domains. Other dMRI sequence parameters were b = 400 s/$mm^2$, $\delta$ = 10 ms, and TE = $\Delta+\delta$. For long diffusion times, a stimulated echo (STEAM) sequence is often employed. To approximate this sequence, a generalized diffusion-weighted sequence, which emulates the STEAM sequence by ignoring the effects of $T_1$ and $T_2$ relaxation \cite{rose2019novel,Naughton2020MRM}, was employed. For the biphasic domain, the intracellular and extracellular diffusion coefficients were equal ($D_{in}=D_{ex}$ = 1.5 $\mu m^2$/ms) while the skeletonized domain only had one diffusion coefficient ($D = 1.5$ $\mu m^2/ms$). The membrane permeability was set to $\kappa$ = 50 $\mu m$/s. We note that if one desired to directly compare these two domains, one would need to adjust the membrane permeability in the skeletonized domain to account for two membranes sandwiched together.  Mirroring boundary condition's were applied on all sides. Figures \ref{fig:biphasic_signal} and \ref{fig:skeleton_signal} give the corresponding field maps for $\Delta$ = 100 ms.

Novikov et al. \cite{novikov2014revealing} showed that the radial diffusivity of skeletal muscle, that is, the average diffusion coefficient in the cross-sectional plane, demonstrates a diffusion time dependence that is $\sim$$t^{-1/2}$. Fitting a power law to the LBM scheme's results shows that both the biphasic and skeletonized domain exhibit time dependence that is consistent with this experimentally observed behavior (Figure \ref{fig:RD}). This match with experimentally observed results suggests that the hybrid LBM scheme may be a useful tool to examine how the dMRI signal evolves in biologically realistic domains and under different assumptions, For example, it can facilitate the quantification of the effect of the extracellular space on the signal.

\section{Discussion}

	Reporting on the first implementation of the lattice Boltzmann method to solve the Bloch-Torrey equation (\ref{LBM-B-Teq}), we proposed the hybrid LBM scheme summarized by Eq. (\ref{LBM_process_list_2}). The fundamental difference with the classical version of LBM, which is given by replacing Eq. (\ref{tau_def}) by Eq. (\ref{A12}), is the factorization of the operator, Eq. (\ref{hybrid_formulation}), which splits the reaction and diffusion temporal discretization. This splitting addresses the stiffness of the problem, which is characterized by the disparity between reaction and diffusion time scales, as discussed in the first paragraph of Appendix A. Based on comparisons of two cases with analytical solutions, we demonstrated that the hybrid scheme is more accurate than the classical LBM scheme. Both schemes are second-order accurate in space and first-order in time. When  $\tau$ and D are kept constant, Eq. (\ref{tau_def}) implies that $\delta t \sim \delta x^2$. In other words, we need to decrease $\delta t$ while maintaining $\delta t/\delta x^2$ constant in order to increase the approximation accuracy. As Eq. (\ref{A15}) indicates, the time step limitation for the classical LBM is more restrictive than that for the hybrid LBM, which is given by Eq. (\ref{A10}). This implies that the accurate integration of Eq. (\ref{A1}) with classical LBM requires an order of magnitude smaller time step than with the hybrid scheme for $L \sim 100\ \mu m$, with the concomitant requirement that the grid size has to decrease by two orders of magnitude ${(\delta x}^2 \sim \delta t)$. This result explains why the truncation error of the classical scheme is higher for the larger domain size, cf. Figure \ref{fig5}. 

	The hybrid LBM method shares the same clear advantage as LBM methods in terms of using uniform grids to discretize curved boundaries between various compartments in the REV while retaining second-order spatial accuracy and stability. Consistent with the kinetic nature of the LBM scheme, the membrane boundary conditions (Eqs. (\ref{B2a}) and (\ref{B2b})) connect the particle distribution functions on either side of the interface directly with the membrane permeability $\kappa$ and avoid the need to approximate transmembrane derivatives, as is the case with finite difference or finite element schemes. The full membrane boundary condition is capable of accurately maintaining the second-order spatial accuracy of the scheme when considering the effects of curvilinear boundaries that intersect the lattice at arbitrary angles. Additionally, the presented membrane boundary condition is valid in other heat and mass transfer conditions as well as when advection is considered. Simplifying the membrane boundary condition to assume that the membrane is perpendicular to the lattice ($\theta=0$) introduces error that increases for increasing gradient strength. While this assumption introduces error into the simulation, its simplicity of implementation as well as not requiring the angle of the membrane-lattice intersection make it appealing for use with more complicated tissue geometries such as those derived from histological images. 

	Like the classical LBM scheme, the hybrid version based on the time splitting scheme described by Eqs. (\ref{hybrid_formulation}) and (\ref{LBM_process_list}) is unconditionally stable for $\tau>0.6$ (because both time splits are stable), so the step sizes are determined by approximation accuracy considerations. Concerning the choice of $\delta t$ and $\delta t^\prime$, and in view of Eq. (\ref{A10}), let us consider the limitations placed on the diffusion time step $\delta t$ for realistic values of the diffusion coefficients in tissue. For typical values $\tau\sim0.6$ and $D\sim2$ $\mu m^2$/ms, Eq. (\ref{tau_def}) requires $\delta t/{\delta x}^2\sim0.02$\ $ms/\mu m^2$. For a spatial resolution of $\delta x\sim1$\ $\mu m$ in an REV with $L\sim100$\ $\mu m$, this requires $\delta t\sim0.02$\ ms. Given that this diffusion step also satisfies the requirement for the reaction step in Eq. (\ref{A10}), we have employed $\delta t$=$\delta t^\prime$ in the present study. Formal optimization of the hybrid LBM would involve a study of the variation of the truncation error as a function of $\tau$, like in \cite{holdych2004truncation,ayodele2011lattice}, and an investigation of varying the sequence or the step size of the diffusion and reaction splits $\left(\delta t\neq\delta t^\prime\right)$ \cite{alemani2005lbgk}, but both are outside the scope of the present study. 

	While improved explicit \cite{van2014finite,li2013numerical} and implicit \cite{beltrachini2015parametric,nguyen2018partition} temporal integration schemes have been proposed for the solution of the Bloch-Torrey equation with the finite element method, the disadvantage of LBM schemes relative to higher-order temporal schemes is offset by the amenability of the former to parallelization. Parallel computing is where LBM schemes have a performance advantage over finite difference or finite element schemes. Our LBM code for solving the Bloch-Torrey equation on a $1000 \times 1000 \times 1$ grid exhibits a parallel efficiency of 77\% at 168 cores, vs. an efficiency of $\sim$55\% for the finite element scheme with $\sim$3.5 million degrees of freedom at 256 cores \cite{nguyen2018partition}. Our speedup is optimal (linear) for the whole range, cf. Figure \ref{fig14}. Our parallelization algorithm relies on an uncomplicated domain decomposition scheme and one-to-one mapping of the MPI processes to CPU cores. Further gains in performance are anticipated by accounting for special computer architecture or by employing GPU cores \cite{zhao2008lattice,campos2016lattice,Naughton2019PEARC}.

	In terms of memory allocation, the LBM scheme requires a total allocation of  $44\times N^3$\ for the numerical integration of the Bloch-Torrey equation for $N\times N \times N$ lattices (in 3D). Based on these estimates we can describe the complexity of the hybrid LBM scheme as ${44\times N}_t\times N^3$, which is $\sim$${10}^{10}$, given discretization parameters discussed previously. Analysis of the computing performance of Monte Carlo methods to integrate the Bloch-Torrey equation in two-compartment tissue models morphologically similar to the ones used in this study indicates that a minimum complexity of $\sim$${10}^9$ is needed in order to avoid statistical error \cite{yeh2013diffusion,hall2009convergence}. Moreover, Yeh et al. \cite{yeh2013diffusion} repeat each simulation 10 times, thus raising this minimum to $\sim$${10}^{10}$. Consideration of the effect of thin cell membranes on the computing performance increases the complexity significantly with lattice-free Monte Carlo methods being required to adequately model the effect of curvilinear membranes. It is here that the advantages of a LBM scheme become evident in comparison to more widely adopted Monte Carlo methods. By being able to accurately resolve the effects of curvilinear membranes, LBM allows retention of the computational efficiently gained from using a structured grid. Further, the imposition of the external periodic conditions is straightforward for LBM schemes, in contrast with the finite element scheme \cite{nguyen2018partition} where they had to be approximated by introducing an artificial permeability to mimic diffusion at the external boundaries. A comprehensive comparison of the relative advantages of each of these three approaches will be the subject of future work.

	The hybrid LBM scheme is able to accurately match analytical solutions of the Bloch-Torrey equation in both the short and long-time limit as well as for increasing gradient strength. It is also able to match experimentally observed diffusion-time dependence when applied to a domain derived from histological images of skeletal muscle. The extension from 2D (D2Q5 stencil) to 3D (D3Q7 stencil) is straightforward owing to the simplicity of the spatial discretization and implementation of the boundary conditions. These results demonstrate the ability of the LBM scheme to be applied in a variety of cases. In particular, its ability to straightforwardly integrate a histologically-derived domain illustrates the flexibility of the LBM scheme to incorporate complex tissue domains. Further, the hybrid LBM scheme is not limited to PGSE sequences or linear gradients. It is capable of simulating arbitrary pulse sequences as well as non-linear magnetic gradients by modifying the term in brackets in Eq. \ref{reaction_LBM}, though the inclusion of non-linear gradients would introduce a domain size dependence in the error term of the hybrid LBM scheme (see Eq. (\ref{A9})). Future applications of the hybrid-LBM scheme in the field of dMRI include investigation and development of reduced analytical models \cite{Naughton2019PMB,Naughton2020MRM}, analysis of increasing complex tissue structures, particularly those derived from histological images, and even analysis of the inverse problem of estimating microstructure properties from dMRI measurements \cite{Naughton2019PEARC,Naughton2019THESIS}. 

	We conclude with several comments regarding possible extensions of the hybrid LBM scheme developed here. The presented hybrid LBM scheme is very general and can accommodate other transport phenomena, multiple tissue compartments, and other MRI sequences. The scheme can be readily extended to incorporate more complex physics and be applied in the study of a variety of biological tissues such as neural tissue, cardiac and skeletal muscle, liver, and cancer tumors. It can also accommodate more complex dMRI gradient waveforms and sequences involving other MRI contrast mechanisms (e.g. perfusion, magnetic susceptibility, elastography, etc.), or imaging gradients (slice selection, phase encoding, or readout). Starting with its first application for modeling unrestricted diffusion \cite{wolf1995lattice}, LBM has since accommodated anisotropic diffusion and advection \cite{ginzburg2005equilibrium,yoshida2010multiple}, coupled diffusion \cite{huber2010lattice}, coupled reaction-diffusion between multiple species \cite{alemani2005lbgk,ayodele2011lattice}, finite cell membrane permeability \cite{aho2016diffusion}, phase field models \cite{holdych2001migration}, and interstitial flow \cite{georgiadis1996questions,jurczuk2013computational,khirevich2015coarse}. Such processes are pertinent to biophysics problems involving transport and evolution of large biomolecules in blood-perfused cellular systems as well as physics other than diffusion. For example, LBM was applied to model protein diffusion inside mammalian cells \cite{kuhn2011protein} and to model cardiac electrophysiology \cite{campos2016lattice}. As an additional indication of its versatility, LBM has been recently employed to integrate fractional order diffusion-advection-reaction equations \cite{zhou2016lattice} and has also shown promise in incorporating fluid-structure interactions \cite{owen2011efficient}. This is not to say that other numerical methods could not have been employed for the phenomena mentioned above, but rather, owing to the local nature of the operations, the same LBM formulation can be easily adapted to accurately and efficiently simulate a vast range of physical phenomena. The combination of the numerical accuracy, efficiency, and ability to incorporate additional physical phenomena make lattice Boltzmann schemes an exciting alternative to currently used Monte Carlo and finite element based schemes in the simulation of diffusion-weighted MRI.

\section{Conclusions}

	Motivated by the need to interpret the dMRI signal from biological tissue, we have laid the foundation and performed the analysis of a hybrid implementation of the LBM to integrate the Bloch-Torrey equation in heterogeneous tissue models. In its current implementation, the hybrid LBM scheme accommodates finite membrane permeability, piece-wise uniform diffusion coefficients, a wide range of dMRI parameters, periodic and mirroring boundary conditions, and interphase conditions accounting for flux continuity. By splitting the reaction and diffusion time steps, the algorithm maintains the explicit nature of the (classical) LBM implementation. We have shown via truncation error analysis and numerical tests that this splitting obviates the requirement of small temporal steps introduced by the strong reaction term in the Bloch-Torrey equation. 

	Another attractive feature of the classical LBM scheme is also maintained here: the phase boundaries are discretized on uniform 2D and 3D lattices while still maintaining the ability to accurately solve for the effects of curvilinear permeable membranes located at arbitrary angles to the lattice. We have shown that the hybrid scheme retains second-order spatial accuracy and stability for a wide range of membrane orientations and typical dMRI parameter values. Further, the presented hybrid LBM scheme accurately replicates the behavior of analytical solutions in a variety of limiting cases, illustrating the robustness of the presented scheme. Our results indicate that the associated LBM code is very compact and can be easily parallelized and executed efficiently on a general multi-core computer with a excellent scaling for up to 168 CPU cores and a parallel efficiency above 77\%. Additionally, the LBM scheme is based on a uniform grid mesh, which, when combined with the efficient parallelization of the scheme, allows for straightforward application of the scheme to large, realistic tissue structures such as those derived from histological images. 

	Overall, the proposed lattice Boltzmann method present an exciting development in the numerical simulation of diffusion-weighted nuclear magnetic resonance physics. LBM schemes allow accurate treatments of general curvilinear membranes while retaining the computational efficiency and advantages of a structured grid-based scheme. Additionally, the ability of LBM schemes to incorporate a wide variety of additional physical phenomena such as advection, susceptibility, fluid-structure interaction illustrate the flexible and extensible nature of these schemes.

\section{Acknowledgments}
	We are grateful to Mr A. Z. Wang for histology image processing. We acknowledge the financial support by NSF (grants CBET-1236451 \& CMMI-1437113, and a Graduate Research Fellowship to NMN) and NIH (grants HL090455 and EB018107).  Partial support by the R.A. Pritzker chair fund is also acknowledged. Parallel computations were performed on the Extreme Science and Engineering Discovery Environment’s (XSEDE) Comet cluster located at the San Diego Supercomputer Center, which is supported by National Science Foundation grant number ACI-1548562. The contributions of two anonymous reviewers are also gratefully acknowledged.
	
\appendix
\lhead{APPENDICES} 
\setcounter{equation}{0}
\renewcommand{\theequation}{A.\arabic{equation}}
\section*{Appendix A}

The first part of this appendix presents a truncation error analysis of the time-splitting method implemented in the hybrid LBM scheme. There are two methods to derive the macroscopic equation from the evolution of the particle probability distribution function: multiple time scales (Chapman-Enskog expansion), and asymptotic analysis \cite{holdych2004truncation,yoshida2010multiple}; here we employ a combination of them. 

First, we present a scaling analysis of the Bloch-Torrey differential equation (\ref{LBM-B-Teq}), rewritten for a general gradient pulse $\bm{G}\left(t\right)=\bm{G}_0\ f(t)$ and for piece-wise uniform diffusion coefficients as follows
\begin{equation}
\label{A1}
\stacklines{
\frac{\partial\bm{M}}{\partial t} = \ -i\ \gamma\left[\bm{x}\cdot\bm{G}_0\ f \left(t\right)\right]\bm{M} 
\quad-\quad \frac{\bm{M}}{T_2} 
\quad+\quad \ D\nabla\cdot\nabla\bm{M}
}{
\quad\quad\quad\quad 
\gamma \delta x \left|\bm{G}_0\right|\sim\frac{\ 10^{-2}}{\text{ms}}\ 
\quad\quad\ \ \ 
\frac{1}{T_2}\sim\frac{ \ {10}^{-2}}{\text{ms}}
\quad\ \ \ 
\frac{D}{\delta x^2}\sim\frac{{1}}{\text{ms}}
} {.}
\end{equation}
The above equation is a homogeneous reaction-diffusion differential equation so the relative order of magnitude of the various terms does not depend on the magnitude of $\bm{M}$. Eq. (\ref{A1}) is defined in $t\in\left[0,TE\right]$ and $\bm{x}\in REV$. Using a $\delta x$ of $\sim$1 $\mu m$, typical whole-body MRI scanner parameters, and typical properties of biological tissue ($T_2$ = 100 ms and D = 1.0 $\mu m^2/ms$), an order of magnitude analysis of the terms in the right hand side of Eq. (\ref{A1}) reveals the disparity between the reaction (first and second term) and diffusion (third term) time scales. The reaction rates are more than two orders of magnitude slower than the diffusion rate. This is the motivation for the splitting scheme associated with the hybrid LBM, which is a concept that has been explored in prior studies of such models \cite{alemani2005lbgk}. We also note that as $\delta x$ decreases, the disparity between the diffusion and reaction rates increases, allowing the difference in scales to be maintained even for large gradient strengths provided a sufficiently small $\delta x$.

Second, we can rename the linear reaction operator in Eq. (\ref{A1}) as follows
\begin{equation}
\label{A2}
\frac{\partial\bm{M}}{\partial t}=\bm{R}\left(\bm{x},t\right)\ \bm{M}\ +\ D\nabla\cdot\nabla\bm{M}\ ,\quad \text{with} \quad \bm{R}\left(\bm{x},t\right)=-i\ \gamma\left[\bm{x}\cdot\bm{G}_0\ f\left(t\right)\right]-\frac{1}{T_2}) {,}
\end{equation}
and generalize Eq. (\ref{hybrid_formulation}) to express the evolution of magnetization starting past a reference time instant $t^k \in\left[0,t_E\right]$
\begin{equation}
\label{A3}
\bm{M}\left(\bm{x},t\right)=\exp{\left[\int_{t^k}^{t}\bm{R}\left(\bm{x},t^{\prime\prime}\right)\ dt^{\prime\prime}\right]}\bm{M}^\prime\left(\bm{x},t\right) \quad\text{for}\quad t\geq t^k {.}
\end{equation}
Note that $\bm{M}\left(\bm{x},t^k\right)=\bm{M}^\prime\left(\bm{x},t^k\right)$. Differentiate Eq. (\ref{A3}) with respect to time to obtain
\begin{equation}
\label{A4}
\frac{\partial}{\partial t}\bm{M}\left(\bm{x},t\right)=\bm{R}\left(\bm{x},t\right)\bm{M}\left(\bm{x},t\right)+\ \bm{E}\left(\bm{x},t^k;\delta t^\prime\right)\frac{\partial}{\partial t}\bm{M}^\prime\left(\bm{x},t\right)
\end{equation}
where
\begin{equation}
\label{A5}
\bm{E}\left(\bm{x},t^k;\delta t^\prime\right)=\exp{\left[\int_{t^k}^{t^k+\delta t^\prime}{\bm{R}\left(\bm{x},t^{\prime\prime}\right)\ dt^{\prime\prime}}\right]}, \quad\text{and}\quad \delta t^\prime\geq 0 {.}
\end{equation}
Let us suppress all dependent variables everywhere except in the expression $\bm{E}\left(\bm{x},t^k;\delta t^\prime\right)$, and require that $\bm{M}^\prime$ obeys the diffusion equation, within a certain truncation error ${TE}_D$
\begin{equation}
\label{A6}
\ \frac{\partial}{\partial t}\bm{M}^\prime=D\nabla\cdot\nabla\bm{M}^\prime+{TE}_D\ 
\end{equation}
so Eq. (\ref{A4}) becomes
\begin{equation}
\label{A7}
\frac{\partial}{\partial t}\bm{M}=\bm{R}\ \ \bm{M}+\ \ \ \bm{E}\left(\bm{x},t^k;\delta t^\prime\right)\ \ \left[\ D\nabla\cdot\nabla\bm{M}^\prime+{TE}_D\right] {.}
\end{equation}
Using Eq. (\ref{A3}), we can show that 
\begin{equation}
\label{A7.1}
\bm{E}\left(\bm{x},t^k;\delta t^\prime\right) \nabla\cdot\nabla\bm{M}^\prime=\ \nabla\cdot\nabla\bm{M}+2i\gamma F\left(t^k;\delta t^\prime\right) \bm{G}_0\cdot\nabla\bm{M} - {\gamma^2F}^2\left(t^k;\delta t^\prime\right)\ \left|\bm{G}_0\right|^2 \bm{M}      
\end{equation}
where 
\begin{equation}
\label{A8}
F\left(t^k;\delta t^\prime\right)=\int_{t^k}^{t^k+\delta t^\prime}{f(t")dt"\ } {.}
\end{equation}
This allows casting Eq. (\ref{A7}) in the form of Eq. (\ref{A2}) 
\begin{equation}
\begin{aligned}
\label{A9}
\frac{\partial}{\partial t}\bm{M}= \ \bm{R}\ \bm{M}+\ D\nabla\cdot\nabla\bm{M}+\left[2i\gamma D\ F\left(t^k;\delta t^\prime\right) \bm{G}_0\cdot\nabla\bm{M}-\ {\gamma^2D\ F}^2\left(t^k;\delta t^\prime\right) \left|\bm{G}_0\right|^2\bm{M}\right]& \\
+\ \Big[\bm{E}\left(\bm{x},t^k;\delta t^\prime\right)\ {TE}_D\Big] {.}
\end{aligned}
\end{equation}
The terms contained in the two square brackets constitute the truncation error of the hybrid LBM scheme proposed here. The terms in the first bracket containing Eq. (\ref{A8}) correspond to the error introduced in the treatment of the reaction part of Eq. (\ref{A2}) according to Eq. (\ref{A3}), and their magnitude can be estimated by assessing the magnitude of the integral in Eq. (\ref{A8}). By recognizing that $f(t)\sim O\left(1\right)$, we can easily see that $F\left(t^k;\delta t^\prime\right)\sim O\left(\delta t^\prime\right)$,  so that the formal order of magnitude of the two reaction truncation error terms is $\gamma D\left|\bm{G}_0\right| \delta t^\prime$ and  $\gamma^2D\left|\bm{G}_0\right|^2 {\delta t^\prime}^2$. In order for Eq. (\ref{A9}) to be consistent with Eq. (\ref{A2}), both these terms have to be much smaller than the smallest term in Eq. (\ref{A2}), which is the reaction term according to the order of magnitude analysis of Eq. (\ref{A1}) (for clarity, we will only consider the part of the reaction term related to the gradient, for ultra-short $T_2$, its effect can be straightforwardly incorporated). This requirement implies $D \ \delta t^\prime / \delta x^2\ll 1$ and $\gamma \left|\bm{G}_0\right| D \ \delta {t^\prime}^2 /\delta x \ll 1$, leading to constraints on the time step of 
\begin{equation}
\label{A10}
\begin{split}
&\delta t^\prime \ll \frac{\delta x^2}{D} \ \sim \ 1 \ ms\\
&\delta t^\prime \ll \sqrt{\frac{\delta x}{\gamma \left|\bm{G}_0\right| D}}\ \sim \ 10\  ms {.}
\end{split}
\end{equation}

Since the diffusion problem Eq. (\ref{A6}) is integrated with the classical LBM scheme, we can estimate ${TE}_D$ from a truncation error analysis of that scheme. This error can be obtained by modifying, according to our Eqs. (\ref{diff_eq}) and (\ref{tau_def}), the expression (A23) obtained by the Chapman-Enskog expansion in the Appendix of \cite{ayodele2011lattice}, and by separating the effect of the diffusion time step $\delta t$ from the lattice grid size $\delta x$
\begin{equation}
\label{A11}
{TE}_D=3\ \delta t\ \ \frac{\tau^2-\tau+\frac{1}{6}}{\tau-\frac{1}{2}}\  \frac{\partial^2\bm{M}^\prime}{\partial t^2} + h.o.t. =\ 3\epsilon_D \frac{{\delta x}^2}{D} \left(\tau^2-\tau+\frac{1}{6}\right) \frac{\partial^2\bm{M}^\prime}{\partial t^2} + h.o.t.
\end{equation}
The expressions in Eq. (\ref{A11}), which are equivalent via the use of Eq. (\ref{g_eq}), recover the known fact that the truncation error of the classical LBM is first-order in time and second-order in space. Returning to the last term in Eq. (\ref{A9}), we can see that this is also the contribution of ${TE}_D$ to the hybrid LBM error if we require that $\bm{E}\left(\bm{x},t^k;\delta t^\prime\right)\sim O\left(1\right)$. Referring to Eqs. (\ref{A2}) and (\ref{A5}), we note that the diffusion gradient term in $\bm{E}\left(\bm{x},t^k;\delta t^\prime\right)$ is periodic and so always $O\left(1\right)$, so the leading contribution to the overall truncation error of the diffusion term is $O\left(\delta t,{\delta x}^2\right)$, and of the reaction term is $O\left(\delta t^\prime\right)$.

The second part of this Appendix addresses the truncation error of the classical LBM scheme applied in the solution of the Bloch-Torrey (\ref{A2}), i.e. without the time splitting scheme Eq. (\ref{A3})). This involves a modification of the collision step of the LBM scheme as described in section \ref{hybrid-LBM-section}, which unlike Eq. (\ref{collision_LBM}), now reads
\begin{equation}
\label{A12}
{\hat{g}}_i\left(\bm{x},t\right)=g_i\left(\bm{x},t\right)-\frac{1}{\tau}\left[g_i\left(\bm{x},t\right)-g_i^{eq}\left(\bm{x},t\right)\right]+{\delta t\ \omega}_i\ \bm{R}\left(\bm{x},t\right)\ \bm{M}\left(\bm{x},t\right)\ 
\end{equation}
where $\bm{M}\left(\bm{x},t\right)$ is computed by summing over $g_i$,\ as shown in Eq. (\ref{M'summation}). Including the reaction term in Eq. (\ref{A12}), results in a different version of Eq. (\ref{A9}):
\begin{equation}
\label{A13}
\frac{\partial}{\partial t}\bm{M}=\bm{R}\ \ \bm{M}+\ D\nabla\cdot\nabla\bm{M}+{TE}_D+{TE}_R {.}
\end{equation}
The reaction truncation term, ${TE}_R$, can be evaluated by starting from the relevant truncation error expression (A23) in the Appendix of \cite{ayodele2011lattice} (after correcting an error; the reaction term is only $O\left(\delta t\right)$ and not $O\left({\delta t}^2\right))$.
\begin{equation}
\label{13.1}
{TE}_R=\tau\ \delta t\frac{\partial}{\partial t}\left[\bm{R}\ \ \bm{M}\right]=\ \tau\ \delta t\ \left[\frac{\partial\bm{R}}{\partial t}\bm{M}+\bm{R}\frac{\partial\bm{M}}{\partial t}\right]\  
\end{equation}
and using Eq. (\ref{A2}) 
\begin{equation}
\label{A14}
{TE}_R=\tau\ \delta t\ \left[\frac{\partial\bm{R}}{\partial t}\bm{M}+\bm{R}\left(\bm{R}\ \bm{M}\ +\ D\nabla\cdot\nabla\bm{M}\ \right)\right] {.}
\end{equation}
Referring to the order of magnitude analysis performed for Eq. (\ref{A1}), for the classical LBM scheme, the length scale of the domain (L) should be used in the reaction term instead of $\delta x$ because the strength of the reaction term $\bm{R}$ at each node in Eq. (\ref{A13}) is determined from the nodes location (Eq. (\ref{A2})), rather than using a periodic function as in the hybrid splitting scheme. In this case, the leading order term in ${TE}_R$ is the second term in Eq. (\ref{A14}), 
\begin{equation}
\label{14.1}
\tau\ \delta t\ \bm{R}\ \bm{R}\ \sim \ O\left\{\tau\ \delta t\ {(\gamma L\left|\bm{G}_0\right|)}^2\right\}
\end{equation}
Again, for consistency, this truncation error term has to be much smaller than the smallest term in Eq. (\ref{A2}), which now becomes the diffusion term. Since $\tau\sim O\left(1\right)$, this requirement implies that  
\begin{equation}
\label{A15}
\delta t\ {(\gamma L\left|\bm{G}_0\right|)}^2\ll\ \frac{D}{\delta x^2}\ \rightarrow\delta t\ll\frac{D}{(\gamma L\left|\bm{G}_0\right| \delta x)^2 }\ \sim \ {10}^{-1}\ \text{ms}
\end{equation}
for a typical domain length $L=100$ $\mu m$. This reaction error term scales with domain size as $L^2$, explaining the domain dependent results for the classical LBM scheme observed in Figure \ref{fig5}. \\

\setcounter{equation}{0}
\renewcommand{\theequation}{B.\arabic{equation}}
\section*{Appendix B}
In this appendix, the derivation of the membrane boundary condition is presented. For a general curved interface, Li et al. \cite{li2013boundary} proposed Dirichlet and Neumann boundary conditions describing the treatment of the nodes closest to the interface. The Dirchlet boundary conditions can be written as
\begin{equation}
\label{dirichlet_1}
g^{\prime}_{\bar{\alpha}}\left(\bm{x}_i,t\right) = c_{d1} \hat{g}_{\alpha}\left(\bm{x}_i,t\right) + c_{d2} \hat{g}_{\alpha}\left(\bm{x}_{ii},t\right) + c_{d3} \hat{g}_{\bar{\alpha}}\left(\bm{x}_i,t\right) + c_{d4} \epsilon_D \Phi_{d,in}
\end{equation}
\begin{equation}
\label{dirichlet_2}
g^{\prime}_{{\alpha}}\left(\bm{x}_e,t\right) = c_{d1}^\ast \hat{g}_{\alpha}\left(\bm{x}_e,t\right) + c_{d2}^\ast \hat{g}_{\alpha}\left(\bm{x}_{ee},t\right) + c_{d3}^\ast \hat{g}_{\bar{\alpha}}\left(\bm{x}_e,t\right) + c_{d4}^\ast \epsilon_D \Phi_{d,ex}
\end{equation}
while the Neumann boundary conditions are
\begin{equation}
\label{neumann_1}
g^{\prime}_{\bar{\alpha}}\left(\bm{x}_i,t\right) = c_{n1} \hat{g}_{\alpha}\left(\bm{x}_i,t\right) + c_{n2} \hat{g}_{\alpha}\left(\bm{x}_{ii},t\right) + c_{n3} \hat{g}_{\bar{\alpha}}\left(\bm{x}_i,t\right) + c_{n4} \frac{\delta t}{\delta x} \Phi_{n\bar{\alpha}}
\end{equation}
\begin{equation}
\label{neumann_2}
g^{\prime}_{{\alpha}}\left(\bm{x}_e,t\right) = c_{n1}^\ast \hat{g}_{\alpha}\left(\bm{x}_e,t\right) + c_{n2}^\ast \hat{g}_{\alpha}\left(\bm{x}_{ee},t\right) + c_{n3}^\ast \hat{g}_{\bar{\alpha}}\left(\bm{x}_e,t\right) + c_{n4}^\ast \frac{\delta t}{\delta x} \Phi_{n\alpha} {.}
\end{equation}
Here the coefficients $c_{d1}-c_{d4}$ and $c_{n1}-c_{n4}$ are coefficients related to the membrane lattice link distance $\Delta_m$ while $c_{d1}^\ast-c_{d4}^\ast$ and $c_{n1}^\ast-c_{n4}^\ast$ relate to $\Delta_m^\ast = 1-\Delta_m$. $\Phi_{d,in}$ and $\Phi_{d,ex}$ are the concentrations at the membrane on the intracellular and extracellular sides respectively while $\Phi_{n\alpha}$ and $\Phi_{n\alpha}$ are the fluxes in the lattice direction. To maintain second-order accuracy, the coefficients for the Neumann boundary condition must be 
\begin{equation}
\label{c_n}
c_{n1} = 1, \quad
c_{n2} = -\frac{2\Delta_m - 1}{2\Delta_m+1}, \quad
c_{n3} = \frac{2\Delta_m - 1}{2\Delta_m+1}, \quad \text{and} \quad
c_{n4} = \frac{2}{2\Delta_m+1} {.}
\end{equation}
For the Dirichlet case, the second-order accuracy is preserved under certain relationships between the coefficients. For definiteness, here we use 
\begin{equation}
\label{c_d}
c_{d1} = 2(\Delta_m-1), \quad
c_{d2} = -\frac{(2\Delta_m - 1)^2}{2\Delta_m+1}, \quad
c_{d3} = \frac{2(2\Delta_m - 1)}{2\Delta_m+1}, \quad \text{and} \quad
c_{d4} = \frac{3-2\Delta_m}{2\Delta_m+1} {.}
\end{equation}
In both cases, $c_{d1}^\ast$-$c_{d4}^\ast$ and $c_{n1}^\ast$-$c_{n4}^\ast$ are the same coefficients as those in Eq. (\ref{c_n}) and Eq. (\ref{c_d}) but with $\Delta_m^\ast$ substituted for $\Delta_m$. 

Eq. (\ref{neumann_1}) and Eq. (\ref{neumann_2}) describe the flux along the lattice direction, however, to implement the boundary condition, this flux must be related to the flux normal to the membrane. In the two-dimensional case, the following relationships exists
\begin{equation}
\label{phi_1}
\begin{split}
\Phi_{n\bar{\alpha}} = \Bigg\{&\frac{1}{c^\prime_{d4}}
	\big[
	(c^\prime_{n1}-c^\prime_{d1}) \hat{g}_{\beta}(\bm{x}_i^\prime,t) + (c^\prime_{n2}-c^\prime_{d2}) \hat{g}_{\beta}(\bm{x}_{ii}^\prime,t) +   (c^\prime_{n3}-c^\prime_{d3}) \hat{g}_{\bar{\beta}}(\bm{x}_i^\prime,t)  \big]\sin\theta \\ - &
	\frac{1}{c_{d4}} \big[
	(c_{n1}-c_{d1}) \hat{g}_{\alpha}(\bm{x}_i,t) + 
	(c_{n2}-c_{d2}) \hat{g}_{\alpha}(\bm{x}_{ii},t) + 
	(c_{n3}-c_{d3}) \hat{g}_{\bar{\alpha}}(\bm{x}_i,t) \big]\sin\theta \\ + &  
	\frac{c^\prime_{n4}}{c^\prime_{d4}} \frac{\delta t}{\delta x}\Phi_{n,in}
	\Bigg\} \Bigg/ 
	\left[
	\frac{c_{n4}}{c_{d4}} \frac{\delta t}{\delta x}\sin\theta + \frac{c^\prime_{n4}}{c^\prime_{d4}} \frac{\delta t}{\delta x}\cos\theta
	\right]
\end{split}
\end{equation}
and
\begin{equation}
\label{phi_2}
\begin{split}
\Phi_{n{\alpha}} = \Bigg\{&\frac{1}{c^\prime_{d4}}
\big[
(c^\prime_{n1}-c^\prime_{d1}) \hat{g}_{\bar{\beta}}(\bm{x}_e^\prime,t) + (c^\prime_{n2}-c^\prime_{d2}) \hat{g}_{\bar{\beta}}(\bm{x}_{ee}^\prime,t) +   (c^\prime_{n3}-c^\prime_{d3}) \hat{g}_{{\beta}}(\bm{x}_e^\prime,t)  \big]\sin\theta \\ - &
\frac{1}{c^\ast_{d4}} \big[
(c^\ast_{n1}-c^\ast_{d1}) \hat{g}_{\bar{\alpha}}(\bm{x}_e,t)  + 
(c^\ast_{n2}-c^\ast_{d2}) \hat{g}_{\bar{\alpha}}(\bm{x}_{ee},t) + 
(c^\ast_{n3}-c^\ast_{d3}) \hat{g}_{{\alpha}}(\bm{x}_e,t) \big]\sin\theta \\ + &  
\frac{c^\prime_{n4}}{c^\prime_{d4}} \frac{\delta t}{\delta x}\Phi_{n,ex}
\Bigg\} \Bigg/ 
\left[
\frac{c^\ast_{n4}}{c^\ast_{d4}} \frac{\delta t}{\delta x}\sin\theta + \frac{c^\prime_{n4}}{c^\prime_{d4}} \frac{\delta t}{\delta x}\cos\theta
\right] {,}
\end{split}
\end{equation}
where $c_{d1}^\prime$-$c_{d4}^\prime$ and $c_{n1}^\prime$-$c_{n4}^\prime$ are the coefficients from Eq. (\ref{c_n}) and Eq. (\ref{c_d}) for $\Delta_m=0$, and $\Phi_{n,in}$ and $\Phi_{n,ex}$ are the fluxes normal to the membrane \cite{li2013boundary}. The subscripts `e' and `ee' in $\bm{x}$ denote lattice nodes immediately adjacent to the membrane in the extracellular domain, while the subscripts `i' and `ii' denote corresponding adjacent nodes in the intracellular domain. The superscripted $\bm{x}^\prime$ refers to the extrapolated values from nodes within a respective domain for 
${\bm{x}^\prime}_e=\ {\bm{x}^\prime}_i=\ \bm{x}_{m}$, 
${\bm{x}^\prime}_{ii}=\ {\bm{x}^\prime}_i+\bm{e}_{\bar{\beta}}\ \delta t$, and 
${\bm{x}^\prime}_{ee}=\ {\bm{x}^\prime}_e+\bm{e}_{\beta}\delta t$,
where $\bm{e}_{\beta}$ and $\bm{e}_{\bar{\beta}}$ are in directions orthogonal to the lattice direction (cf. Figure \ref{figB1}).

To define the membrane boundary condition, we begin by considering the interfacial conditions, Eq. (\ref{permeability_flux}), which result in two relations that can be expressed in terms of Dirichlet ${(\Phi}_d =\bm{M})$\ and Neumann ${(\Phi}_n =D\ \bm{n}\cdot\nabla\bm{M})$\ boundary conditions at either side of the interface,
\begin{equation}
\label{B1}
\Phi_{n,in} =\ \kappa \left(\Phi_{d,ex} -\Phi_{d,in} \right) 
\end{equation}
and
\begin{equation}
\label{B1.5}
\Phi_{n,ex} = D_{ex} \frac{\partial\bm{M}_{ex}}{\partial n} = 
D_{in} \frac{\partial\bm{M}_{in}}{\partial n} = -\Phi_{n,in} {.}
\end{equation}

We consider the distribution functions representing particles towards the membrane in extra-and intra-cellular domains, denoted by $g_{\bar{\alpha}}^\prime{(\bm{x}}_i,\ t)$ and $g_{\alpha}^\prime{(\bm{x}}_e,\ t)$, respectively (the reaction initialization step of $g_i^\prime\left(\bm{x}_n,t\right)={\bar{g}}_i\ \left(\bm{x}_n,t+\delta t\right)$ is implied). Substituting Eq. (\ref{phi_1}) and Eq. (\ref{phi_2}) into Eqs. (\ref{dirichlet_1} -- \ref{neumann_2}), combining with Eq. (\ref{B1}) and Eq. (\ref{B1.5}), and rearranging yields
\begin{equation}
\label{B2a}
\begin{split}
{g^\prime}_{\bar{\alpha}}\left(\bm{x}_i,t\right)= \
& 
A_1^i{\hat{g}}_\alpha\left(\bm{x}_i,t\right)+
A_2^i{\hat{g}}_\alpha\left(\bm{x}_{ii},t\right)+
A_3^i{\hat{g}}_{\bar{\alpha}}\left(\bm{x}_i,t\right)+\\ 
&
B_1^i{\hat{g}}_{\bar{\alpha}}\left(\bm{x}_e,t\right)+\ 
B_2^i{\hat{g}}_{\bar{\alpha}}\left(\bm{x}_{ee},t\right)+\ 
B_3^i{\hat{g}}_\alpha\left(\bm{x}_e,t\right)+ \\
&
C_1^i{\hat{g}}_\beta\left({\bm{x}^\prime}_i,t\right)+\ 
C_2^i{\hat{g}}_\beta\left({\bm{x}^\prime}_{ii},t\right)+ 
C_3^i{\hat{g}}_{\bar{\beta}}\left({\bm{x}^\prime}_i,t\right)+\\ 
&
D_1^i{\hat{g}}_{\bar{\beta}}\left({\bm{x}^\prime}_e,t\right)+\ 
D_2^i{\hat{g}}_{\bar{\beta}}\left({\bm{x}^\prime}_{ee},t\right)+\ 
D_3^i{\hat{g}}_\beta\left({\bm{x}^\prime}_e,t\right)
\end{split}
\end{equation}
and
\begin{equation}
\label{B2b}
\begin{split}
{g^\prime}_\alpha\left(\bm{x}_e,t\right)= \
&
A_1^e{\hat{g}}_{\bar{\alpha}}\left(\bm{x}_e,t\right)+\
A_2^e{\hat{g}}_{\bar{\alpha}}\left(\bm{x}_{ee},t\right)+\
A_3^e{\hat{g}}_\alpha\left(\bm{x}_e,t\right)+\\
&
B_1^e{\hat{g}}_\alpha\left(\bm{x}_i,t\right)+\ 
B_2^e{\hat{g}}_\alpha\left(\bm{x}_{ii},t\right)+\
B_3^e{\hat{g}}_{\bar{\alpha}}\left(\bm{x}_i,t\right)+\\
&
C_1^e{\hat{g}}_{\bar{\beta}}\left({\bm{x}^\prime}_e,t\right)+\ 
C_2^e{\hat{g}}_{\bar{\beta}}\left({\bm{x}^\prime}_{ee},t\right)+\ 
C_3^e{\hat{g}}_\beta\left({\bm{x}^\prime}_e,t\right)+\\
&
D_1^e{\hat{g}}_\beta\left({\bm{x}^\prime}_i,t\right)+\ 
D_2^e{\hat{g}}_\beta\left({\bm{x}^\prime}_{ii},t\right)+\ 
D_3^e{\hat{g}}_{\bar{\beta}}\left({\bm{x}^\prime}_i,t\right)  {.}
\end{split}	
\end{equation}
The coefficients in Eq. (\ref{B2a}) and Eq. (\ref{B2b}) (which are the same as Eq. (\ref{B2a_text}) and Eq. (\ref{B2b_text}) in the main text) are  
\begin{equation}
\label{B3a}
\begin{split}
& \begin{split}
A_j^i =\ \Big(c_{d4}c_{n4}^\prime\cos{\theta}\left(c_{n4}^\prime c_{d4}^\ast c_{nj}\epsilon_D\delta x\cos{\theta}+c_{n4}^\prime{\ c}_{n4}^\ast\ c_{nj}\kappa\ \delta t+c_{d4}^\prime\ c_{n4}^\ast c_{nj}\ \epsilon_D\delta x\sin{\theta}\right)&\\
+ c_{n4}\Big(\cos{\theta}c_{n4}^\prime\ c_{d4}^\ast c_{dj}\ \left(c_{n4}^\prime\ \kappa\ \delta t
+\ c_{d4}^\prime\epsilon_D\delta x\sin{\theta}\right)&\\
+ c_{d4}^\prime c_{n4}^\ast c_{dj}\sin{\theta}\left(2c_{n4}^\prime\ \kappa\ \delta t+\ c_{d4}^\prime\epsilon_D\ \delta x\sin{\theta}\right)&\Big)\Big)\Big/F \end{split}
\\
&B_j^i = \ \Big(c_{d4}c_{n4}{c^\prime}_{n4}^2\ \kappa\ \delta t\cos{\theta}\ \Big({c_{nj}^\ast-c}_{dj}^\ast \Big)\Big)\Big/F \\
&C_j^i = \ \left(c_{d4}c_{n4}\sin{\theta}\left(c_{nj}^\prime-c_{dj}^\prime\right)
\left(c_{n4}^\prime c_{d4}^\ast\ \epsilon_D\ \delta x\cos{\theta}+\ c_{n4}^\prime c_{n4}^\ast\ \kappa\ \delta t+\ c_{d4}^\prime c_{n4}^\ast\ \epsilon_D\ \delta x\sin{\theta}\right)\right)\Big/F \\
&D_j^i = \ \Big(c_{d4}c_{n4}c_{n4}^\ast c_{n4}^\prime\ \kappa\ \delta t\sin{\theta}\ \Big({c_{nj}^\prime-c}_{dj}^\prime \Big)\Big)\Big/F
\end{split}
\end{equation}
and
\begin{equation}
\label{B3b}
\begin{split}
& \begin{split}
A_j^e=\ \Big(c_{d4}c_{n4}^\prime\cos{\theta}\left(c_{n4}^\prime c_{d4}^\ast c_{nj}^\ast\epsilon_D\delta x\cos{\theta}+c_{n4}^\prime{\ c}_{n4}^\ast\ c_{dj}^\ast \kappa\ \delta t+c_{d4}^\prime\ c_{n4}^\ast c_{dj}^\ast\ \epsilon_D\delta x\sin{\theta}\right)&\\
+\ c_{n4}\Big(\cos{\theta}c_{n4}^\prime\ c_{d4}^\ast c_{nj}^\ast\ \left(c_{n4}^\prime\ \kappa\ \delta t+\ c_{d4}^\prime\epsilon_D\delta x\sin{\theta}\right)&\\
+c_{d4}^\prime c_{n4}^\ast c_{nj}^\ast\sin{\theta}\left(2c_{n4}^\prime\ \kappa\ \delta t+\ c_{d4}^\prime\epsilon_D\ \delta x\sin{\theta}\right)&\Big)\Big)\Big/F 
\end{split}
\\
&B_j^e=\Big(c_{d4}^\ast c_{n4}^\ast{c^\prime}_{n4}^2\ \kappa\ \delta t\cos{\theta}\ \Big({c_{nj} -c}_{dj} \Big)\Big)\Big/F \\
&C_j^e= \Big(c_{d4}^\ast c_{n4}^\ast\sin{\theta}\left(c_{nj}^\prime-c_{dj}^\prime\right)\left(c_{n4}^\prime c_{d4} \ \epsilon_D\ \delta x\cos{\theta}+c_{n4}^\prime c_{n4} \ \kappa\ \delta t+\ c_{d4}^\prime c_{n4} \ \epsilon_D\ \delta x\sin{\theta}\right)\Big)\Big/F \\
&D_j^e= \Big(c_{d4}^\ast c_{n4}c_{n4}^\ast c_{n4}^\prime\ \kappa\ \delta t\sin{\theta}\ \Big({c_{nj}^\prime-c}_{dj}^\prime \Big)\Big)\Big/F
\end{split}
\end{equation}
with 
\begin{equation}
\label{B3c}
\begin{split}
F\ =\ \Big(c_{d4}\ c_{n4}^\prime\cos{\theta}\left(\cos{\theta}c_{n4}^\prime c_{d4}^\ast\ \epsilon_D dx+c_{n4}^\prime{\ c}_{n4}^\ast\ \kappa\ \delta t+c_{d4}^\prime\ c_{n4}^\ast\ \epsilon_D\delta x\sin{\theta}\right) & \\
\ \ + c_{n4}\Big(\cos{\theta}c_{n4}^\prime\ c_{d4}^\ast\ \left(c_{n4}^\prime\ \kappa\ \delta t+ \ c_{d4}^\prime\epsilon_D\delta x\sin{\theta}\right) & \\
\ + c_{d4}^\prime c_{n4}^\ast\sin{\theta}\left(2c_{n4}^\prime\ \kappa\ \delta t+\ c_{d4}^\prime\epsilon_D\ \delta x\sin{\theta}\right)& \Big)\Big)  {.}
\end{split}
\end{equation}
Here $\theta$ is the angle between the normal to the membrane and lattice direction $\bm{e}_{\bar{\alpha}}$ while the coefficients  $c_{dj}$  and $c_{nj}$  are the coefficients of the Dirichlet and Neumann boundary conditions for a lattice membrane distance of $\Delta_m$, $c_{dj}^\ast$ and $c_{nj}^\ast$ are the same coefficients for the  extracellular fraction $\Delta_m^\ast=(1-\Delta_m)$, and $c_{dj}^\prime$ and $c_{nj}^\prime$ are the coefficients for $\Delta_m=0$. 

\lhead{REFERENCES} 
{\small
\bibliography{PRE_ref}}

\begin{thebibliography}{100}%
\makeatletter
\providecommand \@ifxundefined [1]{%
 \@ifx{#1\undefined}
}%
\providecommand \@ifnum [1]{%
 \ifnum #1\expandafter \@firstoftwo
 \else \expandafter \@secondoftwo
 \fi
}%
\providecommand \@ifx [1]{%
 \ifx #1\expandafter \@firstoftwo
 \else \expandafter \@secondoftwo
 \fi
}%
\providecommand \natexlab [1]{#1}%
\providecommand \enquote  [1]{``#1''}%
\providecommand \bibnamefont  [1]{#1}%
\providecommand \bibfnamefont [1]{#1}%
\providecommand \citenamefont [1]{#1}%
\providecommand \href@noop [0]{\@secondoftwo}%
\providecommand \href [0]{\begingroup \@sanitize@url \@href}%
\providecommand \@href[1]{\@@startlink{#1}\@@href}%
\providecommand \@@href[1]{\endgroup#1\@@endlink}%
\providecommand \@sanitize@url [0]{\catcode `\\12\catcode `\$12\catcode
  `\&12\catcode `\#12\catcode `\^12\catcode `\_12\catcode `\%12\relax}%
\providecommand \@@startlink[1]{}%
\providecommand \@@endlink[0]{}%
\providecommand \url  [0]{\begingroup\@sanitize@url \@url }%
\providecommand \@url [1]{\endgroup\@href {#1}{\urlprefix }}%
\providecommand \urlprefix  [0]{URL }%
\providecommand \Eprint [0]{\href }%
\providecommand \doibase [0]{https://doi.org/}%
\providecommand \selectlanguage [0]{\@gobble}%
\providecommand \bibinfo  [0]{\@secondoftwo}%
\providecommand \bibfield  [0]{\@secondoftwo}%
\providecommand \translation [1]{[#1]}%
\providecommand \BibitemOpen [0]{}%
\providecommand \bibitemStop [0]{}%
\providecommand \bibitemNoStop [0]{.\EOS\space}%
\providecommand \EOS [0]{\spacefactor3000\relax}%
\providecommand \BibitemShut  [1]{\csname bibitem#1\endcsname}%
\let\auto@bib@innerbib\@empty
\bibitem [{\citenamefont {Callaghan}(1993)}]{paul1993principles}%
  \BibitemOpen
  \bibfield  {author} {\bibinfo {author} {\bibfnamefont {P.~T.}\ \bibnamefont
  {Callaghan}},\ }\href@noop {} {\emph {\bibinfo {title} {Principles of nuclear
  magnetic resonance microscopy}}}\ (\bibinfo  {publisher} {Oxford University
  Press on Demand},\ \bibinfo {year} {1993})\BibitemShut {NoStop}%
\bibitem [{\citenamefont {Behroozmand}\ \emph {et~al.}(2015)\citenamefont
  {Behroozmand}, \citenamefont {Keating},\ and\ \citenamefont
  {Auken}}]{behroozmand2015review}%
  \BibitemOpen
  \bibfield  {author} {\bibinfo {author} {\bibfnamefont {A.~A.}\ \bibnamefont
  {Behroozmand}}, \bibinfo {author} {\bibfnamefont {K.}~\bibnamefont
  {Keating}},\ and\ \bibinfo {author} {\bibfnamefont {E.}~\bibnamefont
  {Auken}},\ }\href@noop {} {\bibfield  {journal} {\bibinfo  {journal} {Surveys
  in Geophysics}\ }\textbf {\bibinfo {volume} {36}},\ \bibinfo {pages} {27}
  (\bibinfo {year} {2015})}\BibitemShut {NoStop}%
\bibitem [{\citenamefont {Vogt}\ \emph {et~al.}(2002)\citenamefont {Vogt},
  \citenamefont {Galvosas}, \citenamefont {Klitzsch},\ and\ \citenamefont
  {Stallmach}}]{vogt2002self}%
  \BibitemOpen
  \bibfield  {author} {\bibinfo {author} {\bibfnamefont {C.}~\bibnamefont
  {Vogt}}, \bibinfo {author} {\bibfnamefont {P.}~\bibnamefont {Galvosas}},
  \bibinfo {author} {\bibfnamefont {N.}~\bibnamefont {Klitzsch}},\ and\
  \bibinfo {author} {\bibfnamefont {F.}~\bibnamefont {Stallmach}},\ }\href@noop
  {} {\bibfield  {journal} {\bibinfo  {journal} {Journal of Applied
  Geophysics}\ }\textbf {\bibinfo {volume} {50}},\ \bibinfo {pages} {455}
  (\bibinfo {year} {2002})}\BibitemShut {NoStop}%
\bibitem [{\citenamefont {Brownstein}\ and\ \citenamefont
  {Tarr}(1979)}]{brownstein1979importance}%
  \BibitemOpen
  \bibfield  {author} {\bibinfo {author} {\bibfnamefont {K.~R.}\ \bibnamefont
  {Brownstein}}\ and\ \bibinfo {author} {\bibfnamefont {C.}~\bibnamefont
  {Tarr}},\ }\href@noop {} {\bibfield  {journal} {\bibinfo  {journal} {Physical
  Review A}\ }\textbf {\bibinfo {volume} {19}},\ \bibinfo {pages} {2446}
  (\bibinfo {year} {1979})}\BibitemShut {NoStop}%
\bibitem [{\citenamefont {Le~Bihan}\ and\ \citenamefont
  {Iima}(2015)}]{le2015diffusion}%
  \BibitemOpen
  \bibfield  {author} {\bibinfo {author} {\bibfnamefont {D.}~\bibnamefont
  {Le~Bihan}}\ and\ \bibinfo {author} {\bibfnamefont {M.}~\bibnamefont
  {Iima}},\ }\href@noop {} {\bibfield  {journal} {\bibinfo  {journal} {PLoS
  Biology}\ }\textbf {\bibinfo {volume} {13}} (\bibinfo {year}
  {2015})}\BibitemShut {NoStop}%
\bibitem [{\citenamefont {Le~Bihan}\ \emph {et~al.}(1986)\citenamefont
  {Le~Bihan}, \citenamefont {Breton}, \citenamefont {Lallemand}, \citenamefont
  {Grenier}, \citenamefont {Cabanis},\ and\ \citenamefont
  {Laval-Jeantet}}]{le1986mr}%
  \BibitemOpen
  \bibfield  {author} {\bibinfo {author} {\bibfnamefont {D.}~\bibnamefont
  {Le~Bihan}}, \bibinfo {author} {\bibfnamefont {E.}~\bibnamefont {Breton}},
  \bibinfo {author} {\bibfnamefont {D.}~\bibnamefont {Lallemand}}, \bibinfo
  {author} {\bibfnamefont {P.}~\bibnamefont {Grenier}}, \bibinfo {author}
  {\bibfnamefont {E.}~\bibnamefont {Cabanis}},\ and\ \bibinfo {author}
  {\bibfnamefont {M.}~\bibnamefont {Laval-Jeantet}},\ }\href@noop {} {\bibfield
   {journal} {\bibinfo  {journal} {Radiology}\ }\textbf {\bibinfo {volume}
  {161}},\ \bibinfo {pages} {401} (\bibinfo {year} {1986})}\BibitemShut
  {NoStop}%
\bibitem [{\citenamefont {Pierpaoli}\ \emph {et~al.}(1996)\citenamefont
  {Pierpaoli}, \citenamefont {Jezzard}, \citenamefont {Basser}, \citenamefont
  {Barnett},\ and\ \citenamefont {Di~Chiro}}]{pierpaoli1996diffusion}%
  \BibitemOpen
  \bibfield  {author} {\bibinfo {author} {\bibfnamefont {C.}~\bibnamefont
  {Pierpaoli}}, \bibinfo {author} {\bibfnamefont {P.}~\bibnamefont {Jezzard}},
  \bibinfo {author} {\bibfnamefont {P.~J.}\ \bibnamefont {Basser}}, \bibinfo
  {author} {\bibfnamefont {A.}~\bibnamefont {Barnett}},\ and\ \bibinfo {author}
  {\bibfnamefont {G.}~\bibnamefont {Di~Chiro}},\ }\href@noop {} {\bibfield
  {journal} {\bibinfo  {journal} {Radiology}\ }\textbf {\bibinfo {volume}
  {201}},\ \bibinfo {pages} {637} (\bibinfo {year} {1996})}\BibitemShut
  {NoStop}%
\bibitem [{\citenamefont {Moseley}\ \emph {et~al.}(1990)\citenamefont
  {Moseley}, \citenamefont {Cohen}, \citenamefont {Mintorovitch}, \citenamefont
  {Chileuitt}, \citenamefont {Shimizu}, \citenamefont {Kucharczyk},
  \citenamefont {Wendland},\ and\ \citenamefont
  {Weinstein}}]{moseley1990early}%
  \BibitemOpen
  \bibfield  {author} {\bibinfo {author} {\bibfnamefont {M.}~\bibnamefont
  {Moseley}}, \bibinfo {author} {\bibfnamefont {Y.}~\bibnamefont {Cohen}},
  \bibinfo {author} {\bibfnamefont {J.}~\bibnamefont {Mintorovitch}}, \bibinfo
  {author} {\bibfnamefont {L.}~\bibnamefont {Chileuitt}}, \bibinfo {author}
  {\bibfnamefont {H.}~\bibnamefont {Shimizu}}, \bibinfo {author} {\bibfnamefont
  {J.}~\bibnamefont {Kucharczyk}}, \bibinfo {author} {\bibfnamefont
  {M.}~\bibnamefont {Wendland}},\ and\ \bibinfo {author} {\bibfnamefont
  {P.}~\bibnamefont {Weinstein}},\ }\href@noop {} {\bibfield  {journal}
  {\bibinfo  {journal} {{Magnetic Resonance in Medicine}}\ }\textbf {\bibinfo
  {volume} {14}},\ \bibinfo {pages} {330} (\bibinfo {year} {1990})}\BibitemShut
  {NoStop}%
\bibitem [{\citenamefont {Cleveland}\ \emph {et~al.}(1976)\citenamefont
  {Cleveland}, \citenamefont {Chang}, \citenamefont {Hazlewood},\ and\
  \citenamefont {Rorschach}}]{cleveland1976nuclear}%
  \BibitemOpen
  \bibfield  {author} {\bibinfo {author} {\bibfnamefont {G.}~\bibnamefont
  {Cleveland}}, \bibinfo {author} {\bibfnamefont {D.}~\bibnamefont {Chang}},
  \bibinfo {author} {\bibfnamefont {C.}~\bibnamefont {Hazlewood}},\ and\
  \bibinfo {author} {\bibfnamefont {H.}~\bibnamefont {Rorschach}},\ }\href@noop
  {} {\bibfield  {journal} {\bibinfo  {journal} {Biophysical Journal}\ }\textbf
  {\bibinfo {volume} {16}},\ \bibinfo {pages} {1043} (\bibinfo {year}
  {1976})}\BibitemShut {NoStop}%
\bibitem [{\citenamefont {Tanner}(1979)}]{tanner1979self}%
  \BibitemOpen
  \bibfield  {author} {\bibinfo {author} {\bibfnamefont {J.}~\bibnamefont
  {Tanner}},\ }\href@noop {} {\bibfield  {journal} {\bibinfo  {journal}
  {Biophysical Journal}\ }\textbf {\bibinfo {volume} {28}},\ \bibinfo {pages}
  {107} (\bibinfo {year} {1979})}\BibitemShut {NoStop}%
\bibitem [{\citenamefont {Garrido}\ \emph {et~al.}(1994)\citenamefont
  {Garrido}, \citenamefont {Wedeen}, \citenamefont {Kwong}, \citenamefont
  {Spencer},\ and\ \citenamefont {Kantor}}]{garrido1994anisotropy}%
  \BibitemOpen
  \bibfield  {author} {\bibinfo {author} {\bibfnamefont {L.}~\bibnamefont
  {Garrido}}, \bibinfo {author} {\bibfnamefont {V.~J.}\ \bibnamefont {Wedeen}},
  \bibinfo {author} {\bibfnamefont {K.~K.}\ \bibnamefont {Kwong}}, \bibinfo
  {author} {\bibfnamefont {U.~M.}\ \bibnamefont {Spencer}},\ and\ \bibinfo
  {author} {\bibfnamefont {H.~L.}\ \bibnamefont {Kantor}},\ }\href@noop {}
  {\bibfield  {journal} {\bibinfo  {journal} {Circulation Research}\ }\textbf
  {\bibinfo {volume} {74}},\ \bibinfo {pages} {789} (\bibinfo {year}
  {1994})}\BibitemShut {NoStop}%
\bibitem [{\citenamefont {Englander}\ \emph {et~al.}(1997)\citenamefont
  {Englander}, \citenamefont {Ulug}, \citenamefont {Brem}, \citenamefont
  {Glickson},\ and\ \citenamefont {van Zilj}}]{englander1997diffusion}%
  \BibitemOpen
  \bibfield  {author} {\bibinfo {author} {\bibfnamefont {S.~A.}\ \bibnamefont
  {Englander}}, \bibinfo {author} {\bibfnamefont {A.~M.}\ \bibnamefont {Ulug}},
  \bibinfo {author} {\bibfnamefont {R.}~\bibnamefont {Brem}}, \bibinfo {author}
  {\bibfnamefont {J.~D.}\ \bibnamefont {Glickson}},\ and\ \bibinfo {author}
  {\bibfnamefont {P.~C.}\ \bibnamefont {van Zilj}},\ }\href@noop {} {\bibfield
  {journal} {\bibinfo  {journal} {NMR in Biomedicine}\ }\textbf {\bibinfo
  {volume} {10}},\ \bibinfo {pages} {348} (\bibinfo {year} {1997})}\BibitemShut
  {NoStop}%
\bibitem [{\citenamefont {Sinha}\ \emph {et~al.}(2002)\citenamefont {Sinha},
  \citenamefont {Lucas-Quesada}, \citenamefont {Sinha}, \citenamefont
  {DeBruhl},\ and\ \citenamefont {Bassett}}]{sinha2002vivo}%
  \BibitemOpen
  \bibfield  {author} {\bibinfo {author} {\bibfnamefont {S.}~\bibnamefont
  {Sinha}}, \bibinfo {author} {\bibfnamefont {F.~A.}\ \bibnamefont
  {Lucas-Quesada}}, \bibinfo {author} {\bibfnamefont {U.}~\bibnamefont
  {Sinha}}, \bibinfo {author} {\bibfnamefont {N.}~\bibnamefont {DeBruhl}},\
  and\ \bibinfo {author} {\bibfnamefont {L.~W.}\ \bibnamefont {Bassett}},\
  }\href@noop {} {\bibfield  {journal} {\bibinfo  {journal} {Journal of
  Magnetic Resonance Imaging}\ }\textbf {\bibinfo {volume} {15}},\ \bibinfo
  {pages} {693} (\bibinfo {year} {2002})}\BibitemShut {NoStop}%
\bibitem [{\citenamefont {Kim}\ \emph {et~al.}(1999)\citenamefont {Kim},
  \citenamefont {Murakami}, \citenamefont {Takahashi}, \citenamefont {Hori},
  \citenamefont {Tsuda},\ and\ \citenamefont {Nakamura}}]{kim1999diffusion}%
  \BibitemOpen
  \bibfield  {author} {\bibinfo {author} {\bibfnamefont {T.}~\bibnamefont
  {Kim}}, \bibinfo {author} {\bibfnamefont {T.}~\bibnamefont {Murakami}},
  \bibinfo {author} {\bibfnamefont {S.}~\bibnamefont {Takahashi}}, \bibinfo
  {author} {\bibfnamefont {M.}~\bibnamefont {Hori}}, \bibinfo {author}
  {\bibfnamefont {K.}~\bibnamefont {Tsuda}},\ and\ \bibinfo {author}
  {\bibfnamefont {H.}~\bibnamefont {Nakamura}},\ }\href@noop {} {\bibfield
  {journal} {\bibinfo  {journal} {American Journal of Roentgenology}\ }\textbf
  {\bibinfo {volume} {173}},\ \bibinfo {pages} {393} (\bibinfo {year}
  {1999})}\BibitemShut {NoStop}%
\bibitem [{\citenamefont {Taouli}\ and\ \citenamefont
  {Koh}(2010)}]{taouli2010diffusion}%
  \BibitemOpen
  \bibfield  {author} {\bibinfo {author} {\bibfnamefont {B.}~\bibnamefont
  {Taouli}}\ and\ \bibinfo {author} {\bibfnamefont {D.-M.}\ \bibnamefont
  {Koh}},\ }\href@noop {} {\bibfield  {journal} {\bibinfo  {journal}
  {Radiology}\ }\textbf {\bibinfo {volume} {254}},\ \bibinfo {pages} {47}
  (\bibinfo {year} {2010})}\BibitemShut {NoStop}%
\bibitem [{\citenamefont {Chenevert}\ \emph {et~al.}(2002)\citenamefont
  {Chenevert}, \citenamefont {Meyer}, \citenamefont {Moffat}, \citenamefont
  {Rehemtulla}, \citenamefont {Mukherji}, \citenamefont {Gebarski},
  \citenamefont {Quint}, \citenamefont {Robertson}, \citenamefont {Lawrence},
  \citenamefont {Junck}, \citenamefont {Taylor}, \citenamefont {Johnson},
  \citenamefont {Dong}, \citenamefont {Muraszko}, \citenamefont {Brunberg},\
  and\ \citenamefont {Ross}}]{chenevert2002diffusion}%
  \BibitemOpen
  \bibfield  {author} {\bibinfo {author} {\bibfnamefont {T.~L.}\ \bibnamefont
  {Chenevert}}, \bibinfo {author} {\bibfnamefont {C.~R.}\ \bibnamefont
  {Meyer}}, \bibinfo {author} {\bibfnamefont {B.~A.}\ \bibnamefont {Moffat}},
  \bibinfo {author} {\bibfnamefont {A.}~\bibnamefont {Rehemtulla}}, \bibinfo
  {author} {\bibfnamefont {S.~K.}\ \bibnamefont {Mukherji}}, \bibinfo {author}
  {\bibfnamefont {S.~S.}\ \bibnamefont {Gebarski}}, \bibinfo {author}
  {\bibfnamefont {D.~J.}\ \bibnamefont {Quint}}, \bibinfo {author}
  {\bibfnamefont {P.~L.}\ \bibnamefont {Robertson}}, \bibinfo {author}
  {\bibfnamefont {T.~S.}\ \bibnamefont {Lawrence}}, \bibinfo {author}
  {\bibfnamefont {L.}~\bibnamefont {Junck}}, \bibinfo {author} {\bibfnamefont
  {J.~M.~G.}\ \bibnamefont {Taylor}}, \bibinfo {author} {\bibfnamefont {T.~D.}\
  \bibnamefont {Johnson}}, \bibinfo {author} {\bibfnamefont {Q.}~\bibnamefont
  {Dong}}, \bibinfo {author} {\bibfnamefont {K.~M.}\ \bibnamefont {Muraszko}},
  \bibinfo {author} {\bibfnamefont {J.~A.}\ \bibnamefont {Brunberg}},\ and\
  \bibinfo {author} {\bibfnamefont {B.~D.}\ \bibnamefont {Ross}},\ }\href@noop
  {} {\bibfield  {journal} {\bibinfo  {journal} {Molecular Imaging}\ }\textbf
  {\bibinfo {volume} {1}},\ \bibinfo {pages} {15353500200221482} (\bibinfo
  {year} {2002})}\BibitemShut {NoStop}%
\bibitem [{\citenamefont {Lyng}\ \emph {et~al.}(2000)\citenamefont {Lyng},
  \citenamefont {Haraldseth},\ and\ \citenamefont
  {Rofstad}}]{lyng2000measurement}%
  \BibitemOpen
  \bibfield  {author} {\bibinfo {author} {\bibfnamefont {H.}~\bibnamefont
  {Lyng}}, \bibinfo {author} {\bibfnamefont {O.}~\bibnamefont {Haraldseth}},\
  and\ \bibinfo {author} {\bibfnamefont {E.~K.}\ \bibnamefont {Rofstad}},\
  }\href@noop {} {\bibfield  {journal} {\bibinfo  {journal} {Magnetic Resonance
  in Medicine}\ }\textbf {\bibinfo {volume} {43}},\ \bibinfo {pages} {828}
  (\bibinfo {year} {2000})}\BibitemShut {NoStop}%
\bibitem [{\citenamefont {Torrey}(1956)}]{torrey1956bloch}%
  \BibitemOpen
  \bibfield  {author} {\bibinfo {author} {\bibfnamefont {H.~C.}\ \bibnamefont
  {Torrey}},\ }\href@noop {} {\bibfield  {journal} {\bibinfo  {journal}
  {Physical Review}\ }\textbf {\bibinfo {volume} {104}},\ \bibinfo {pages}
  {563} (\bibinfo {year} {1956})}\BibitemShut {NoStop}%
\bibitem [{\citenamefont {Stallmach}\ and\ \citenamefont
  {Galvosas}(2007)}]{stallmach2007spin}%
  \BibitemOpen
  \bibfield  {author} {\bibinfo {author} {\bibfnamefont {F.}~\bibnamefont
  {Stallmach}}\ and\ \bibinfo {author} {\bibfnamefont {P.}~\bibnamefont
  {Galvosas}},\ }\href@noop {} {\bibfield  {journal} {\bibinfo  {journal}
  {Annual Reports on NMR Spectroscopy}\ }\textbf {\bibinfo {volume} {61}},\
  \bibinfo {pages} {51} (\bibinfo {year} {2007})}\BibitemShut {NoStop}%
\bibitem [{\citenamefont {Norris}(2001)}]{norris2001effects}%
  \BibitemOpen
  \bibfield  {author} {\bibinfo {author} {\bibfnamefont {D.~G.}\ \bibnamefont
  {Norris}},\ }\href@noop {} {\bibfield  {journal} {\bibinfo  {journal} {NMR in
  Biomedicine}\ }\textbf {\bibinfo {volume} {14}},\ \bibinfo {pages} {77}
  (\bibinfo {year} {2001})}\BibitemShut {NoStop}%
\bibitem [{\citenamefont {Grebenkov}(2007)}]{grebenkov2007nmr}%
  \BibitemOpen
  \bibfield  {author} {\bibinfo {author} {\bibfnamefont {D.~S.}\ \bibnamefont
  {Grebenkov}},\ }\href@noop {} {\bibfield  {journal} {\bibinfo  {journal}
  {Reviews of Modern Physics}\ }\textbf {\bibinfo {volume} {79}},\ \bibinfo
  {pages} {1077} (\bibinfo {year} {2007})}\BibitemShut {NoStop}%
\bibitem [{\citenamefont {Novikov}\ and\ \citenamefont
  {Kiselev}(2010)}]{novikov2010effective}%
  \BibitemOpen
  \bibfield  {author} {\bibinfo {author} {\bibfnamefont {D.~S.}\ \bibnamefont
  {Novikov}}\ and\ \bibinfo {author} {\bibfnamefont {V.~G.}\ \bibnamefont
  {Kiselev}},\ }\href@noop {} {\bibfield  {journal} {\bibinfo  {journal} {NMR
  in Biomedicine}\ }\textbf {\bibinfo {volume} {23}},\ \bibinfo {pages} {682}
  (\bibinfo {year} {2010})}\BibitemShut {NoStop}%
\bibitem [{\citenamefont {Novikov}\ \emph {et~al.}(2011)\citenamefont
  {Novikov}, \citenamefont {Fieremans}, \citenamefont {Jensen},\ and\
  \citenamefont {Helpern}}]{novikov2011random}%
  \BibitemOpen
  \bibfield  {author} {\bibinfo {author} {\bibfnamefont {D.~S.}\ \bibnamefont
  {Novikov}}, \bibinfo {author} {\bibfnamefont {E.}~\bibnamefont {Fieremans}},
  \bibinfo {author} {\bibfnamefont {J.~H.}\ \bibnamefont {Jensen}},\ and\
  \bibinfo {author} {\bibfnamefont {J.~A.}\ \bibnamefont {Helpern}},\
  }\href@noop {} {\bibfield  {journal} {\bibinfo  {journal} {Nature Physics}\
  }\textbf {\bibinfo {volume} {7}},\ \bibinfo {pages} {508} (\bibinfo {year}
  {2011})}\BibitemShut {NoStop}%
\bibitem [{\citenamefont {Sigmund}\ \emph {et~al.}(2014)\citenamefont
  {Sigmund}, \citenamefont {Novikov}, \citenamefont {Sui}, \citenamefont
  {Ukpebor}, \citenamefont {Baete}, \citenamefont {Babb}, \citenamefont {Liu},
  \citenamefont {Feiweier}, \citenamefont {Kwon}, \citenamefont {McGorty} \emph
  {et~al.}}]{sigmund2014time}%
  \BibitemOpen
  \bibfield  {author} {\bibinfo {author} {\bibfnamefont {E.~E.}\ \bibnamefont
  {Sigmund}}, \bibinfo {author} {\bibfnamefont {D.~S.}\ \bibnamefont
  {Novikov}}, \bibinfo {author} {\bibfnamefont {D.}~\bibnamefont {Sui}},
  \bibinfo {author} {\bibfnamefont {O.}~\bibnamefont {Ukpebor}}, \bibinfo
  {author} {\bibfnamefont {S.}~\bibnamefont {Baete}}, \bibinfo {author}
  {\bibfnamefont {J.~S.}\ \bibnamefont {Babb}}, \bibinfo {author}
  {\bibfnamefont {K.}~\bibnamefont {Liu}}, \bibinfo {author} {\bibfnamefont
  {T.}~\bibnamefont {Feiweier}}, \bibinfo {author} {\bibfnamefont
  {J.}~\bibnamefont {Kwon}}, \bibinfo {author} {\bibfnamefont {K.}~\bibnamefont
  {McGorty}}, \emph {et~al.},\ }\href@noop {} {\bibfield  {journal} {\bibinfo
  {journal} {NMR in Biomedicine}\ }\textbf {\bibinfo {volume} {27}},\ \bibinfo
  {pages} {519} (\bibinfo {year} {2014})}\BibitemShut {NoStop}%
\bibitem [{\citenamefont {Novikov}\ \emph {et~al.}(2014)\citenamefont
  {Novikov}, \citenamefont {Jensen}, \citenamefont {Helpern},\ and\
  \citenamefont {Fieremans}}]{novikov2014revealing}%
  \BibitemOpen
  \bibfield  {author} {\bibinfo {author} {\bibfnamefont {D.~S.}\ \bibnamefont
  {Novikov}}, \bibinfo {author} {\bibfnamefont {J.~H.}\ \bibnamefont {Jensen}},
  \bibinfo {author} {\bibfnamefont {J.~A.}\ \bibnamefont {Helpern}},\ and\
  \bibinfo {author} {\bibfnamefont {E.}~\bibnamefont {Fieremans}},\ }\href@noop
  {} {\bibfield  {journal} {\bibinfo  {journal} {Proceedings of the National
  Academy of Sciences}\ }\textbf {\bibinfo {volume} {111}},\ \bibinfo {pages}
  {5088} (\bibinfo {year} {2014})}\BibitemShut {NoStop}%
\bibitem [{\citenamefont {Kiselev}(2017)}]{kiselev2017fundamentals}%
  \BibitemOpen
  \bibfield  {author} {\bibinfo {author} {\bibfnamefont {V.~G.}\ \bibnamefont
  {Kiselev}},\ }\href@noop {} {\bibfield  {journal} {\bibinfo  {journal} {NMR
  in Biomedicine}\ }\textbf {\bibinfo {volume} {30}},\ \bibinfo {pages} {e3602}
  (\bibinfo {year} {2017})}\BibitemShut {NoStop}%
\bibitem [{\citenamefont {Novikov}\ \emph {et~al.}(2018)\citenamefont
  {Novikov}, \citenamefont {Kiselev},\ and\ \citenamefont
  {Jespersen}}]{novikov2018modeling}%
  \BibitemOpen
  \bibfield  {author} {\bibinfo {author} {\bibfnamefont {D.~S.}\ \bibnamefont
  {Novikov}}, \bibinfo {author} {\bibfnamefont {V.~G.}\ \bibnamefont
  {Kiselev}},\ and\ \bibinfo {author} {\bibfnamefont {S.~N.}\ \bibnamefont
  {Jespersen}},\ }\href@noop {} {\bibfield  {journal} {\bibinfo  {journal}
  {Magnetic Resonance in Medicine}\ }\textbf {\bibinfo {volume} {79}},\
  \bibinfo {pages} {3172} (\bibinfo {year} {2018})}\BibitemShut {NoStop}%
\bibitem [{\citenamefont {Novikov}\ \emph {et~al.}(2019)\citenamefont
  {Novikov}, \citenamefont {Fieremans}, \citenamefont {Jespersen},\ and\
  \citenamefont {Kiselev}}]{novikov2019quantifying}%
  \BibitemOpen
  \bibfield  {author} {\bibinfo {author} {\bibfnamefont {D.~S.}\ \bibnamefont
  {Novikov}}, \bibinfo {author} {\bibfnamefont {E.}~\bibnamefont {Fieremans}},
  \bibinfo {author} {\bibfnamefont {S.~N.}\ \bibnamefont {Jespersen}},\ and\
  \bibinfo {author} {\bibfnamefont {V.~G.}\ \bibnamefont {Kiselev}},\
  }\href@noop {} {\bibfield  {journal} {\bibinfo  {journal} {NMR in
  Biomedicine}\ }\textbf {\bibinfo {volume} {32}},\ \bibinfo {pages} {e3998}
  (\bibinfo {year} {2019})}\BibitemShut {NoStop}%
\bibitem [{\citenamefont {Alexander}\ \emph {et~al.}(2019)\citenamefont
  {Alexander}, \citenamefont {Dyrby}, \citenamefont {Nilsson},\ and\
  \citenamefont {Zhang}}]{alexander2019imaging}%
  \BibitemOpen
  \bibfield  {author} {\bibinfo {author} {\bibfnamefont {D.~C.}\ \bibnamefont
  {Alexander}}, \bibinfo {author} {\bibfnamefont {T.~B.}\ \bibnamefont
  {Dyrby}}, \bibinfo {author} {\bibfnamefont {M.}~\bibnamefont {Nilsson}},\
  and\ \bibinfo {author} {\bibfnamefont {H.}~\bibnamefont {Zhang}},\
  }\href@noop {} {\bibfield  {journal} {\bibinfo  {journal} {NMR in
  Biomedicine}\ }\textbf {\bibinfo {volume} {32}},\ \bibinfo {pages} {e3841}
  (\bibinfo {year} {2019})}\BibitemShut {NoStop}%
\bibitem [{\citenamefont {Jelescu}\ and\ \citenamefont
  {Budde}(2017)}]{jelescu2017design}%
  \BibitemOpen
  \bibfield  {author} {\bibinfo {author} {\bibfnamefont {I.~O.}\ \bibnamefont
  {Jelescu}}\ and\ \bibinfo {author} {\bibfnamefont {M.~D.}\ \bibnamefont
  {Budde}},\ }\href@noop {} {\bibfield  {journal} {\bibinfo  {journal}
  {Frontiers in Physics}\ }\textbf {\bibinfo {volume} {5}},\ \bibinfo {pages}
  {61} (\bibinfo {year} {2017})}\BibitemShut {NoStop}%
\bibitem [{\citenamefont {Berry}\ \emph {et~al.}(2018)\citenamefont {Berry},
  \citenamefont {Regner}, \citenamefont {Galinsky}, \citenamefont {Ward},\ and\
  \citenamefont {Frank}}]{berry2018relationships}%
  \BibitemOpen
  \bibfield  {author} {\bibinfo {author} {\bibfnamefont {D.~B.}\ \bibnamefont
  {Berry}}, \bibinfo {author} {\bibfnamefont {B.}~\bibnamefont {Regner}},
  \bibinfo {author} {\bibfnamefont {V.}~\bibnamefont {Galinsky}}, \bibinfo
  {author} {\bibfnamefont {S.~R.}\ \bibnamefont {Ward}},\ and\ \bibinfo
  {author} {\bibfnamefont {L.~R.}\ \bibnamefont {Frank}},\ }\href@noop {}
  {\bibfield  {journal} {\bibinfo  {journal} {Magnetic Resonance in Medicine}\
  }\textbf {\bibinfo {volume} {80}},\ \bibinfo {pages} {317} (\bibinfo {year}
  {2018})}\BibitemShut {NoStop}%
\bibitem [{\citenamefont {Rose}\ \emph {et~al.}(2019)\citenamefont {Rose},
  \citenamefont {Nielles-Vallespin}, \citenamefont {Ferreira}, \citenamefont
  {Firmin}, \citenamefont {Scott},\ and\ \citenamefont
  {Doorly}}]{rose2019novel}%
  \BibitemOpen
  \bibfield  {author} {\bibinfo {author} {\bibfnamefont {J.~N.}\ \bibnamefont
  {Rose}}, \bibinfo {author} {\bibfnamefont {S.}~\bibnamefont
  {Nielles-Vallespin}}, \bibinfo {author} {\bibfnamefont {P.~F.}\ \bibnamefont
  {Ferreira}}, \bibinfo {author} {\bibfnamefont {D.~N.}\ \bibnamefont
  {Firmin}}, \bibinfo {author} {\bibfnamefont {A.~D.}\ \bibnamefont {Scott}},\
  and\ \bibinfo {author} {\bibfnamefont {D.~J.}\ \bibnamefont {Doorly}},\
  }\href@noop {} {\bibfield  {journal} {\bibinfo  {journal} {Magnetic Resonance
  in Medicine}\ }\textbf {\bibinfo {volume} {81}},\ \bibinfo {pages} {2759}
  (\bibinfo {year} {2019})}\BibitemShut {NoStop}%
\bibitem [{\citenamefont {Szafer}\ \emph {et~al.}(1995)\citenamefont {Szafer},
  \citenamefont {Zhong},\ and\ \citenamefont {Gore}}]{szafer1995theoretical}%
  \BibitemOpen
  \bibfield  {author} {\bibinfo {author} {\bibfnamefont {A.}~\bibnamefont
  {Szafer}}, \bibinfo {author} {\bibfnamefont {J.}~\bibnamefont {Zhong}},\ and\
  \bibinfo {author} {\bibfnamefont {J.~C.}\ \bibnamefont {Gore}},\ }\href@noop
  {} {\bibfield  {journal} {\bibinfo  {journal} {Magnetic Resonance in
  Medicine}\ }\textbf {\bibinfo {volume} {33}},\ \bibinfo {pages} {697}
  (\bibinfo {year} {1995})}\BibitemShut {NoStop}%
\bibitem [{\citenamefont {Balls}\ and\ \citenamefont
  {Frank}(2009)}]{balls2009simulation}%
  \BibitemOpen
  \bibfield  {author} {\bibinfo {author} {\bibfnamefont {G.~T.}\ \bibnamefont
  {Balls}}\ and\ \bibinfo {author} {\bibfnamefont {L.~R.}\ \bibnamefont
  {Frank}},\ }\href@noop {} {\bibfield  {journal} {\bibinfo  {journal}
  {Magnetic Resonance in Medicine}\ }\textbf {\bibinfo {volume} {62}},\
  \bibinfo {pages} {771} (\bibinfo {year} {2009})}\BibitemShut {NoStop}%
\bibitem [{\citenamefont {Fieremans}\ \emph {et~al.}(2010)\citenamefont
  {Fieremans}, \citenamefont {Novikov}, \citenamefont {Jensen},\ and\
  \citenamefont {Helpern}}]{fieremans2010monte}%
  \BibitemOpen
  \bibfield  {author} {\bibinfo {author} {\bibfnamefont {E.}~\bibnamefont
  {Fieremans}}, \bibinfo {author} {\bibfnamefont {D.~S.}\ \bibnamefont
  {Novikov}}, \bibinfo {author} {\bibfnamefont {J.~H.}\ \bibnamefont
  {Jensen}},\ and\ \bibinfo {author} {\bibfnamefont {J.~A.}\ \bibnamefont
  {Helpern}},\ }\href@noop {} {\bibfield  {journal} {\bibinfo  {journal} {NMR
  in Biomedicine}\ }\textbf {\bibinfo {volume} {23}},\ \bibinfo {pages} {711}
  (\bibinfo {year} {2010})}\BibitemShut {NoStop}%
\bibitem [{\citenamefont {Baxter}\ and\ \citenamefont
  {Frank}(2013)}]{baxter2013computational}%
  \BibitemOpen
  \bibfield  {author} {\bibinfo {author} {\bibfnamefont {G.~T.}\ \bibnamefont
  {Baxter}}\ and\ \bibinfo {author} {\bibfnamefont {L.~R.}\ \bibnamefont
  {Frank}},\ }\href@noop {} {\bibfield  {journal} {\bibinfo  {journal}
  {Neuroimage}\ }\textbf {\bibinfo {volume} {75}},\ \bibinfo {pages} {204}
  (\bibinfo {year} {2013})}\BibitemShut {NoStop}%
\bibitem [{\citenamefont {Yeh}\ \emph {et~al.}(2013)\citenamefont {Yeh},
  \citenamefont {Schmitt}, \citenamefont {Le~Bihan}, \citenamefont
  {Li-Schlittgen}, \citenamefont {Lin},\ and\ \citenamefont
  {Poupon}}]{yeh2013diffusion}%
  \BibitemOpen
  \bibfield  {author} {\bibinfo {author} {\bibfnamefont {C.-H.}\ \bibnamefont
  {Yeh}}, \bibinfo {author} {\bibfnamefont {B.}~\bibnamefont {Schmitt}},
  \bibinfo {author} {\bibfnamefont {D.}~\bibnamefont {Le~Bihan}}, \bibinfo
  {author} {\bibfnamefont {J.-R.}\ \bibnamefont {Li-Schlittgen}}, \bibinfo
  {author} {\bibfnamefont {C.-P.}\ \bibnamefont {Lin}},\ and\ \bibinfo {author}
  {\bibfnamefont {C.}~\bibnamefont {Poupon}},\ }\href@noop {} {\bibfield
  {journal} {\bibinfo  {journal} {PloS One}\ }\textbf {\bibinfo {volume} {8}},\
  \bibinfo {pages} {e76626} (\bibinfo {year} {2013})}\BibitemShut {NoStop}%
\bibitem [{\citenamefont {Bates}\ \emph {et~al.}(2017)\citenamefont {Bates},
  \citenamefont {Teh}, \citenamefont {McClymont}, \citenamefont {Kohl},
  \citenamefont {Schneider},\ and\ \citenamefont {Grau}}]{bates2017monte}%
  \BibitemOpen
  \bibfield  {author} {\bibinfo {author} {\bibfnamefont {J.}~\bibnamefont
  {Bates}}, \bibinfo {author} {\bibfnamefont {I.}~\bibnamefont {Teh}}, \bibinfo
  {author} {\bibfnamefont {D.}~\bibnamefont {McClymont}}, \bibinfo {author}
  {\bibfnamefont {P.}~\bibnamefont {Kohl}}, \bibinfo {author} {\bibfnamefont
  {J.~E.}\ \bibnamefont {Schneider}},\ and\ \bibinfo {author} {\bibfnamefont
  {V.}~\bibnamefont {Grau}},\ }\href@noop {} {\bibfield  {journal} {\bibinfo
  {journal} {IEEE Transactions on Medical Imaging}\ }\textbf {\bibinfo {volume}
  {36}},\ \bibinfo {pages} {1316} (\bibinfo {year} {2017})}\BibitemShut
  {NoStop}%
\bibitem [{\citenamefont {Hall}\ and\ \citenamefont
  {Alexander}(2009)}]{hall2009convergence}%
  \BibitemOpen
  \bibfield  {author} {\bibinfo {author} {\bibfnamefont {M.~G.}\ \bibnamefont
  {Hall}}\ and\ \bibinfo {author} {\bibfnamefont {D.~C.}\ \bibnamefont
  {Alexander}},\ }\href@noop {} {\bibfield  {journal} {\bibinfo  {journal}
  {IEEE Transactions on Medical Imaging}\ }\textbf {\bibinfo {volume} {28}},\
  \bibinfo {pages} {1354} (\bibinfo {year} {2009})}\BibitemShut {NoStop}%
\bibitem [{\citenamefont {Chin}\ \emph {et~al.}(2002)\citenamefont {Chin},
  \citenamefont {Wehrli}, \citenamefont {Hwang}, \citenamefont {Takahashi},\
  and\ \citenamefont {Hackney}}]{chin2002biexponential}%
  \BibitemOpen
  \bibfield  {author} {\bibinfo {author} {\bibfnamefont {C.-L.}\ \bibnamefont
  {Chin}}, \bibinfo {author} {\bibfnamefont {F.~W.}\ \bibnamefont {Wehrli}},
  \bibinfo {author} {\bibfnamefont {S.~N.}\ \bibnamefont {Hwang}}, \bibinfo
  {author} {\bibfnamefont {M.}~\bibnamefont {Takahashi}},\ and\ \bibinfo
  {author} {\bibfnamefont {D.~B.}\ \bibnamefont {Hackney}},\ }\href@noop {}
  {\bibfield  {journal} {\bibinfo  {journal} {Magnetic Resonance in Medicine}\
  }\textbf {\bibinfo {volume} {47}},\ \bibinfo {pages} {455} (\bibinfo {year}
  {2002})}\BibitemShut {NoStop}%
\bibitem [{\citenamefont {Hwang}\ \emph {et~al.}(2003)\citenamefont {Hwang},
  \citenamefont {Chin}, \citenamefont {Wehrli},\ and\ \citenamefont
  {Hackney}}]{hwang2003image}%
  \BibitemOpen
  \bibfield  {author} {\bibinfo {author} {\bibfnamefont {S.~N.}\ \bibnamefont
  {Hwang}}, \bibinfo {author} {\bibfnamefont {C.-L.}\ \bibnamefont {Chin}},
  \bibinfo {author} {\bibfnamefont {F.~W.}\ \bibnamefont {Wehrli}},\ and\
  \bibinfo {author} {\bibfnamefont {D.~B.}\ \bibnamefont {Hackney}},\
  }\href@noop {} {\bibfield  {journal} {\bibinfo  {journal} {Magnetic Resonance
  in Medicine}\ }\textbf {\bibinfo {volume} {50}},\ \bibinfo {pages} {373}
  (\bibinfo {year} {2003})}\BibitemShut {NoStop}%
\bibitem [{\citenamefont {Xu}\ \emph {et~al.}(2007)\citenamefont {Xu},
  \citenamefont {Does},\ and\ \citenamefont {Gore}}]{xu2007numerical}%
  \BibitemOpen
  \bibfield  {author} {\bibinfo {author} {\bibfnamefont {J.}~\bibnamefont
  {Xu}}, \bibinfo {author} {\bibfnamefont {M.~D.}\ \bibnamefont {Does}},\ and\
  \bibinfo {author} {\bibfnamefont {J.~C.}\ \bibnamefont {Gore}},\ }\href@noop
  {} {\bibfield  {journal} {\bibinfo  {journal} {Physics in Medicine \&
  Biology}\ }\textbf {\bibinfo {volume} {52}},\ \bibinfo {pages} {N111}
  (\bibinfo {year} {2007})}\BibitemShut {NoStop}%
\bibitem [{\citenamefont {Russell}\ \emph {et~al.}(2012)\citenamefont
  {Russell}, \citenamefont {Harkins}, \citenamefont {Secomb}, \citenamefont
  {Galons},\ and\ \citenamefont {Trouard}}]{russell2012finite}%
  \BibitemOpen
  \bibfield  {author} {\bibinfo {author} {\bibfnamefont {G.}~\bibnamefont
  {Russell}}, \bibinfo {author} {\bibfnamefont {K.~D.}\ \bibnamefont
  {Harkins}}, \bibinfo {author} {\bibfnamefont {T.~W.}\ \bibnamefont {Secomb}},
  \bibinfo {author} {\bibfnamefont {J.-P.}\ \bibnamefont {Galons}},\ and\
  \bibinfo {author} {\bibfnamefont {T.~P.}\ \bibnamefont {Trouard}},\
  }\href@noop {} {\bibfield  {journal} {\bibinfo  {journal} {Physics in
  Medicine \& Biology}\ }\textbf {\bibinfo {volume} {57}},\ \bibinfo {pages}
  {N35} (\bibinfo {year} {2012})}\BibitemShut {NoStop}%
\bibitem [{\citenamefont {Van~Nguyen}\ \emph {et~al.}(2014)\citenamefont
  {Van~Nguyen}, \citenamefont {Li}, \citenamefont {Grebenkov},\ and\
  \citenamefont {Le~Bihan}}]{van2014finite}%
  \BibitemOpen
  \bibfield  {author} {\bibinfo {author} {\bibfnamefont {D.}~\bibnamefont
  {Van~Nguyen}}, \bibinfo {author} {\bibfnamefont {J.-R.}\ \bibnamefont {Li}},
  \bibinfo {author} {\bibfnamefont {D.}~\bibnamefont {Grebenkov}},\ and\
  \bibinfo {author} {\bibfnamefont {D.}~\bibnamefont {Le~Bihan}},\ }\href@noop
  {} {\bibfield  {journal} {\bibinfo  {journal} {Journal of Computational
  Physics}\ }\textbf {\bibinfo {volume} {263}},\ \bibinfo {pages} {283}
  (\bibinfo {year} {2014})}\BibitemShut {NoStop}%
\bibitem [{\citenamefont {Nguyen}\ \emph {et~al.}(2018)\citenamefont {Nguyen},
  \citenamefont {Jansson}, \citenamefont {Hoffman},\ and\ \citenamefont
  {Li}}]{nguyen2018partition}%
  \BibitemOpen
  \bibfield  {author} {\bibinfo {author} {\bibfnamefont {V.-D.}\ \bibnamefont
  {Nguyen}}, \bibinfo {author} {\bibfnamefont {J.}~\bibnamefont {Jansson}},
  \bibinfo {author} {\bibfnamefont {J.}~\bibnamefont {Hoffman}},\ and\ \bibinfo
  {author} {\bibfnamefont {J.-R.}\ \bibnamefont {Li}},\ }\href@noop {}
  {\bibfield  {journal} {\bibinfo  {journal} {Journal of Computational
  Physics}\ }\textbf {\bibinfo {volume} {375}},\ \bibinfo {pages} {271}
  (\bibinfo {year} {2018})}\BibitemShut {NoStop}%
\bibitem [{\citenamefont {Beltrachini}\ \emph {et~al.}(2015)\citenamefont
  {Beltrachini}, \citenamefont {Taylor},\ and\ \citenamefont
  {Frangi}}]{beltrachini2015parametric}%
  \BibitemOpen
  \bibfield  {author} {\bibinfo {author} {\bibfnamefont {L.}~\bibnamefont
  {Beltrachini}}, \bibinfo {author} {\bibfnamefont {Z.~A.}\ \bibnamefont
  {Taylor}},\ and\ \bibinfo {author} {\bibfnamefont {A.~F.}\ \bibnamefont
  {Frangi}},\ }\href@noop {} {\bibfield  {journal} {\bibinfo  {journal}
  {Journal of Magnetic Resonance}\ }\textbf {\bibinfo {volume} {259}},\
  \bibinfo {pages} {126} (\bibinfo {year} {2015})}\BibitemShut {NoStop}%
\bibitem [{\citenamefont {Li}\ \emph {et~al.}(2013{\natexlab{a}})\citenamefont
  {Li}, \citenamefont {Calhoun}, \citenamefont {Poupon},\ and\ \citenamefont
  {Le~Bihan}}]{li2013numerical}%
  \BibitemOpen
  \bibfield  {author} {\bibinfo {author} {\bibfnamefont {J.-R.}\ \bibnamefont
  {Li}}, \bibinfo {author} {\bibfnamefont {D.}~\bibnamefont {Calhoun}},
  \bibinfo {author} {\bibfnamefont {C.}~\bibnamefont {Poupon}},\ and\ \bibinfo
  {author} {\bibfnamefont {D.}~\bibnamefont {Le~Bihan}},\ }\href@noop {}
  {\bibfield  {journal} {\bibinfo  {journal} {Physics in Medicine \& Biology}\
  }\textbf {\bibinfo {volume} {59}},\ \bibinfo {pages} {441} (\bibinfo {year}
  {2013}{\natexlab{a}})}\BibitemShut {NoStop}%
\bibitem [{\citenamefont {Chubynsky}\ and\ \citenamefont
  {Slater}(2012)}]{chubynsky2012optimizing}%
  \BibitemOpen
  \bibfield  {author} {\bibinfo {author} {\bibfnamefont {M.~V.}\ \bibnamefont
  {Chubynsky}}\ and\ \bibinfo {author} {\bibfnamefont {G.~W.}\ \bibnamefont
  {Slater}},\ }\href@noop {} {\bibfield  {journal} {\bibinfo  {journal}
  {Physical Review E}\ }\textbf {\bibinfo {volume} {85}},\ \bibinfo {pages}
  {016709} (\bibinfo {year} {2012})}\BibitemShut {NoStop}%
\bibitem [{\citenamefont {Noble}\ \emph {et~al.}(1995)\citenamefont {Noble},
  \citenamefont {Chen}, \citenamefont {Georgiadis},\ and\ \citenamefont
  {Buckius}}]{noble1995consistent}%
  \BibitemOpen
  \bibfield  {author} {\bibinfo {author} {\bibfnamefont {D.~R.}\ \bibnamefont
  {Noble}}, \bibinfo {author} {\bibfnamefont {S.}~\bibnamefont {Chen}},
  \bibinfo {author} {\bibfnamefont {J.~G.}\ \bibnamefont {Georgiadis}},\ and\
  \bibinfo {author} {\bibfnamefont {R.~O.}\ \bibnamefont {Buckius}},\
  }\href@noop {} {\bibfield  {journal} {\bibinfo  {journal} {Physics of
  Fluids}\ }\textbf {\bibinfo {volume} {7}},\ \bibinfo {pages} {203} (\bibinfo
  {year} {1995})}\BibitemShut {NoStop}%
\bibitem [{\citenamefont {Gallivan}\ \emph {et~al.}(1997)\citenamefont
  {Gallivan}, \citenamefont {Noble}, \citenamefont {Georgiadis},\ and\
  \citenamefont {Buckius}}]{gallivan1997evaluation}%
  \BibitemOpen
  \bibfield  {author} {\bibinfo {author} {\bibfnamefont {M.~A.}\ \bibnamefont
  {Gallivan}}, \bibinfo {author} {\bibfnamefont {D.~R.}\ \bibnamefont {Noble}},
  \bibinfo {author} {\bibfnamefont {J.~G.}\ \bibnamefont {Georgiadis}},\ and\
  \bibinfo {author} {\bibfnamefont {R.~O.}\ \bibnamefont {Buckius}},\
  }\href@noop {} {\bibfield  {journal} {\bibinfo  {journal} {International
  Journal for Numerical Methods in Fluids}\ }\textbf {\bibinfo {volume} {25}},\
  \bibinfo {pages} {249} (\bibinfo {year} {1997})}\BibitemShut {NoStop}%
\bibitem [{\citenamefont {Li}\ \emph {et~al.}(2013{\natexlab{b}})\citenamefont
  {Li}, \citenamefont {Mei},\ and\ \citenamefont {Klausner}}]{li2013boundary}%
  \BibitemOpen
  \bibfield  {author} {\bibinfo {author} {\bibfnamefont {L.}~\bibnamefont
  {Li}}, \bibinfo {author} {\bibfnamefont {R.}~\bibnamefont {Mei}},\ and\
  \bibinfo {author} {\bibfnamefont {J.~F.}\ \bibnamefont {Klausner}},\
  }\href@noop {} {\bibfield  {journal} {\bibinfo  {journal} {Journal of
  Computational Physics}\ }\textbf {\bibinfo {volume} {237}},\ \bibinfo {pages}
  {366} (\bibinfo {year} {2013}{\natexlab{b}})}\BibitemShut {NoStop}%
\bibitem [{\citenamefont {Zhang}\ \emph {et~al.}(2018)\citenamefont {Zhang},
  \citenamefont {Yang}, \citenamefont {Zeng},\ and\ \citenamefont
  {Chew}}]{zhang2018consistent}%
  \BibitemOpen
  \bibfield  {author} {\bibinfo {author} {\bibfnamefont {L.}~\bibnamefont
  {Zhang}}, \bibinfo {author} {\bibfnamefont {S.}~\bibnamefont {Yang}},
  \bibinfo {author} {\bibfnamefont {Z.}~\bibnamefont {Zeng}},\ and\ \bibinfo
  {author} {\bibfnamefont {J.~W.}\ \bibnamefont {Chew}},\ }\href@noop {}
  {\bibfield  {journal} {\bibinfo  {journal} {Physical Review E}\ }\textbf
  {\bibinfo {volume} {97}},\ \bibinfo {pages} {023302} (\bibinfo {year}
  {2018})}\BibitemShut {NoStop}%
\bibitem [{\citenamefont {Zhao}(2008)}]{zhao2008lattice}%
  \BibitemOpen
  \bibfield  {author} {\bibinfo {author} {\bibfnamefont {Y.}~\bibnamefont
  {Zhao}},\ }\href@noop {} {\bibfield  {journal} {\bibinfo  {journal} {The
  Visual Computer}\ }\textbf {\bibinfo {volume} {24}},\ \bibinfo {pages} {323}
  (\bibinfo {year} {2008})}\BibitemShut {NoStop}%
\bibitem [{\citenamefont {Clausen}\ \emph {et~al.}(2010)\citenamefont
  {Clausen}, \citenamefont {Reasor~Jr},\ and\ \citenamefont
  {Aidun}}]{clausen2010parallel}%
  \BibitemOpen
  \bibfield  {author} {\bibinfo {author} {\bibfnamefont {J.~R.}\ \bibnamefont
  {Clausen}}, \bibinfo {author} {\bibfnamefont {D.~A.}\ \bibnamefont
  {Reasor~Jr}},\ and\ \bibinfo {author} {\bibfnamefont {C.~K.}\ \bibnamefont
  {Aidun}},\ }\href@noop {} {\bibfield  {journal} {\bibinfo  {journal}
  {Computer Physics Communications}\ }\textbf {\bibinfo {volume} {181}},\
  \bibinfo {pages} {1013} (\bibinfo {year} {2010})}\BibitemShut {NoStop}%
\bibitem [{\citenamefont {Campos}\ \emph {et~al.}(2016)\citenamefont {Campos},
  \citenamefont {Oliveira}, \citenamefont {dos Santos},\ and\ \citenamefont
  {Rocha}}]{campos2016lattice}%
  \BibitemOpen
  \bibfield  {author} {\bibinfo {author} {\bibfnamefont {J.}~\bibnamefont
  {Campos}}, \bibinfo {author} {\bibfnamefont {R.~S.}\ \bibnamefont
  {Oliveira}}, \bibinfo {author} {\bibfnamefont {R.~W.}\ \bibnamefont {dos
  Santos}},\ and\ \bibinfo {author} {\bibfnamefont {B.~M.}\ \bibnamefont
  {Rocha}},\ }\href@noop {} {\bibfield  {journal} {\bibinfo  {journal} {Journal
  of Computational and Applied Mathematics}\ }\textbf {\bibinfo {volume}
  {295}},\ \bibinfo {pages} {70} (\bibinfo {year} {2016})}\BibitemShut
  {NoStop}%
\bibitem [{\citenamefont {Hiorth}\ \emph {et~al.}(2009)\citenamefont {Hiorth},
  \citenamefont {a~Lad}, \citenamefont {Evje},\ and\ \citenamefont
  {Skjaeveland}}]{hiorth2009lattice}%
  \BibitemOpen
  \bibfield  {author} {\bibinfo {author} {\bibfnamefont {A.}~\bibnamefont
  {Hiorth}}, \bibinfo {author} {\bibfnamefont {U.}~\bibnamefont {a~Lad}},
  \bibinfo {author} {\bibfnamefont {S.}~\bibnamefont {Evje}},\ and\ \bibinfo
  {author} {\bibfnamefont {S.}~\bibnamefont {Skjaeveland}},\ }\href@noop {}
  {\bibfield  {journal} {\bibinfo  {journal} {International Journal for
  Numerical Methods in Fluids}\ }\textbf {\bibinfo {volume} {59}},\ \bibinfo
  {pages} {405} (\bibinfo {year} {2009})}\BibitemShut {NoStop}%
\bibitem [{\citenamefont {Stejskal}\ and\ \citenamefont
  {Tanner}(1965)}]{stejskal1965spin}%
  \BibitemOpen
  \bibfield  {author} {\bibinfo {author} {\bibfnamefont {E.~O.}\ \bibnamefont
  {Stejskal}}\ and\ \bibinfo {author} {\bibfnamefont {J.~E.}\ \bibnamefont
  {Tanner}},\ }\href@noop {} {\bibfield  {journal} {\bibinfo  {journal} {The
  Journal of Chemical Physics}\ }\textbf {\bibinfo {volume} {42}},\ \bibinfo
  {pages} {288} (\bibinfo {year} {1965})}\BibitemShut {NoStop}%
\bibitem [{\citenamefont {Kr{\"u}ger}\ \emph {et~al.}(2017)\citenamefont
  {Kr{\"u}ger}, \citenamefont {Kusumaatmaja}, \citenamefont {Kuzmin},
  \citenamefont {Shardt}, \citenamefont {Silva},\ and\ \citenamefont
  {Viggen}}]{kruger2017lattice}%
  \BibitemOpen
  \bibfield  {author} {\bibinfo {author} {\bibfnamefont {T.}~\bibnamefont
  {Kr{\"u}ger}}, \bibinfo {author} {\bibfnamefont {H.}~\bibnamefont
  {Kusumaatmaja}}, \bibinfo {author} {\bibfnamefont {A.}~\bibnamefont
  {Kuzmin}}, \bibinfo {author} {\bibfnamefont {O.}~\bibnamefont {Shardt}},
  \bibinfo {author} {\bibfnamefont {G.}~\bibnamefont {Silva}},\ and\ \bibinfo
  {author} {\bibfnamefont {E.~M.}\ \bibnamefont {Viggen}},\ }\href@noop {}
  {\emph {\bibinfo {title} {The lattice Boltzmann method}}}\ (\bibinfo
  {publisher} {Springer International Publishing},\ \bibinfo {year}
  {2017})\BibitemShut {NoStop}%
\bibitem [{\citenamefont {Bhatnagar}\ \emph {et~al.}(1954)\citenamefont
  {Bhatnagar}, \citenamefont {Gross},\ and\ \citenamefont
  {Krook}}]{bhatnagar1954model}%
  \BibitemOpen
  \bibfield  {author} {\bibinfo {author} {\bibfnamefont {P.~L.}\ \bibnamefont
  {Bhatnagar}}, \bibinfo {author} {\bibfnamefont {E.~P.}\ \bibnamefont
  {Gross}},\ and\ \bibinfo {author} {\bibfnamefont {M.}~\bibnamefont {Krook}},\
  }\href@noop {} {\bibfield  {journal} {\bibinfo  {journal} {Physical Review}\
  }\textbf {\bibinfo {volume} {94}},\ \bibinfo {pages} {511} (\bibinfo {year}
  {1954})}\BibitemShut {NoStop}%
\bibitem [{\citenamefont {Qian}\ \emph {et~al.}(1992)\citenamefont {Qian},
  \citenamefont {d'Humi{\`e}res},\ and\ \citenamefont
  {Lallemand}}]{qian1992lattice}%
  \BibitemOpen
  \bibfield  {author} {\bibinfo {author} {\bibfnamefont {Y.-H.}\ \bibnamefont
  {Qian}}, \bibinfo {author} {\bibfnamefont {D.}~\bibnamefont
  {d'Humi{\`e}res}},\ and\ \bibinfo {author} {\bibfnamefont {P.}~\bibnamefont
  {Lallemand}},\ }\href@noop {} {\bibfield  {journal} {\bibinfo  {journal}
  {Europhysics Letters ({EPL})}\ }\textbf {\bibinfo {volume} {17}},\ \bibinfo
  {pages} {479} (\bibinfo {year} {1992})}\BibitemShut {NoStop}%
\bibitem [{\citenamefont {Chen}\ and\ \citenamefont
  {Doolen}(1998)}]{chen1998lattice}%
  \BibitemOpen
  \bibfield  {author} {\bibinfo {author} {\bibfnamefont {S.}~\bibnamefont
  {Chen}}\ and\ \bibinfo {author} {\bibfnamefont {G.~D.}\ \bibnamefont
  {Doolen}},\ }\href@noop {} {\bibfield  {journal} {\bibinfo  {journal} {Annual
  Review of Fluid Mechanics}\ }\textbf {\bibinfo {volume} {30}},\ \bibinfo
  {pages} {329} (\bibinfo {year} {1998})}\BibitemShut {NoStop}%
\bibitem [{\citenamefont {Ayodele}\ \emph {et~al.}(2011)\citenamefont
  {Ayodele}, \citenamefont {Varnik},\ and\ \citenamefont
  {Raabe}}]{ayodele2011lattice}%
  \BibitemOpen
  \bibfield  {author} {\bibinfo {author} {\bibfnamefont {S.}~\bibnamefont
  {Ayodele}}, \bibinfo {author} {\bibfnamefont {F.}~\bibnamefont {Varnik}},\
  and\ \bibinfo {author} {\bibfnamefont {D.}~\bibnamefont {Raabe}},\
  }\href@noop {} {\bibfield  {journal} {\bibinfo  {journal} {Physical Review
  E}\ }\textbf {\bibinfo {volume} {83}},\ \bibinfo {pages} {016702} (\bibinfo
  {year} {2011})}\BibitemShut {NoStop}%
\bibitem [{\citenamefont {Aidun}\ and\ \citenamefont
  {Clausen}(2010)}]{aidun2010lattice}%
  \BibitemOpen
  \bibfield  {author} {\bibinfo {author} {\bibfnamefont {C.~K.}\ \bibnamefont
  {Aidun}}\ and\ \bibinfo {author} {\bibfnamefont {J.~R.}\ \bibnamefont
  {Clausen}},\ }\href@noop {} {\bibfield  {journal} {\bibinfo  {journal}
  {Annual Review of Fluid Mechanics}\ }\textbf {\bibinfo {volume} {42}},\
  \bibinfo {pages} {439} (\bibinfo {year} {2010})}\BibitemShut {NoStop}%
\bibitem [{\citenamefont {Perumal}\ and\ \citenamefont
  {Dass}(2015)}]{perumal2015review}%
  \BibitemOpen
  \bibfield  {author} {\bibinfo {author} {\bibfnamefont {D.~A.}\ \bibnamefont
  {Perumal}}\ and\ \bibinfo {author} {\bibfnamefont {A.~K.}\ \bibnamefont
  {Dass}},\ }\href@noop {} {\bibfield  {journal} {\bibinfo  {journal}
  {Alexandria Engineering Journal}\ }\textbf {\bibinfo {volume} {54}},\
  \bibinfo {pages} {955} (\bibinfo {year} {2015})}\BibitemShut {NoStop}%
\bibitem [{\citenamefont {He}\ \emph {et~al.}(2019)\citenamefont {He},
  \citenamefont {Liu}, \citenamefont {Li},\ and\ \citenamefont
  {Tao}}]{he2019lattice}%
  \BibitemOpen
  \bibfield  {author} {\bibinfo {author} {\bibfnamefont {Y.-L.}\ \bibnamefont
  {He}}, \bibinfo {author} {\bibfnamefont {Q.}~\bibnamefont {Liu}}, \bibinfo
  {author} {\bibfnamefont {Q.}~\bibnamefont {Li}},\ and\ \bibinfo {author}
  {\bibfnamefont {W.-Q.}\ \bibnamefont {Tao}},\ }\href@noop {} {\bibfield
  {journal} {\bibinfo  {journal} {International Journal of Heat and Mass
  Transfer}\ }\textbf {\bibinfo {volume} {129}},\ \bibinfo {pages} {160}
  (\bibinfo {year} {2019})}\BibitemShut {NoStop}%
\bibitem [{\citenamefont {Alemani}\ \emph {et~al.}(2005)\citenamefont
  {Alemani}, \citenamefont {Chopard}, \citenamefont {Galceran},\ and\
  \citenamefont {Buffle}}]{alemani2005lbgk}%
  \BibitemOpen
  \bibfield  {author} {\bibinfo {author} {\bibfnamefont {D.}~\bibnamefont
  {Alemani}}, \bibinfo {author} {\bibfnamefont {B.}~\bibnamefont {Chopard}},
  \bibinfo {author} {\bibfnamefont {J.}~\bibnamefont {Galceran}},\ and\
  \bibinfo {author} {\bibfnamefont {J.}~\bibnamefont {Buffle}},\ }\href@noop {}
  {\bibfield  {journal} {\bibinfo  {journal} {Physical Chemistry Chemical
  Physics}\ }\textbf {\bibinfo {volume} {7}},\ \bibinfo {pages} {3331}
  (\bibinfo {year} {2005})}\BibitemShut {NoStop}%
\bibitem [{\citenamefont {Yoshida}\ and\ \citenamefont
  {Nagaoka}(2010)}]{yoshida2010multiple}%
  \BibitemOpen
  \bibfield  {author} {\bibinfo {author} {\bibfnamefont {H.}~\bibnamefont
  {Yoshida}}\ and\ \bibinfo {author} {\bibfnamefont {M.}~\bibnamefont
  {Nagaoka}},\ }\href@noop {} {\bibfield  {journal} {\bibinfo  {journal}
  {Journal of Computational Physics}\ }\textbf {\bibinfo {volume} {229}},\
  \bibinfo {pages} {7774} (\bibinfo {year} {2010})}\BibitemShut {NoStop}%
\bibitem [{\citenamefont {Huber}\ \emph {et~al.}(2010)\citenamefont {Huber},
  \citenamefont {Chopard},\ and\ \citenamefont {Manga}}]{huber2010lattice}%
  \BibitemOpen
  \bibfield  {author} {\bibinfo {author} {\bibfnamefont {C.}~\bibnamefont
  {Huber}}, \bibinfo {author} {\bibfnamefont {B.}~\bibnamefont {Chopard}},\
  and\ \bibinfo {author} {\bibfnamefont {M.}~\bibnamefont {Manga}},\
  }\href@noop {} {\bibfield  {journal} {\bibinfo  {journal} {Journal of
  Computational Physics}\ }\textbf {\bibinfo {volume} {229}},\ \bibinfo {pages}
  {7956} (\bibinfo {year} {2010})}\BibitemShut {NoStop}%
\bibitem [{\citenamefont {Li}\ \emph {et~al.}(2017)\citenamefont {Li},
  \citenamefont {Mei},\ and\ \citenamefont {Klausner}}]{li2017lattice}%
  \BibitemOpen
  \bibfield  {author} {\bibinfo {author} {\bibfnamefont {L.}~\bibnamefont
  {Li}}, \bibinfo {author} {\bibfnamefont {R.}~\bibnamefont {Mei}},\ and\
  \bibinfo {author} {\bibfnamefont {J.~F.}\ \bibnamefont {Klausner}},\
  }\href@noop {} {\bibfield  {journal} {\bibinfo  {journal} {International
  Journal of Heat and Mass Transfer}\ }\textbf {\bibinfo {volume} {108}},\
  \bibinfo {pages} {41} (\bibinfo {year} {2017})}\BibitemShut {NoStop}%
\bibitem [{\citenamefont {Li}\ \emph {et~al.}(2014)\citenamefont {Li},
  \citenamefont {Chen}, \citenamefont {Mei},\ and\ \citenamefont
  {Klausner}}]{li2014conjugate}%
  \BibitemOpen
  \bibfield  {author} {\bibinfo {author} {\bibfnamefont {L.}~\bibnamefont
  {Li}}, \bibinfo {author} {\bibfnamefont {C.}~\bibnamefont {Chen}}, \bibinfo
  {author} {\bibfnamefont {R.}~\bibnamefont {Mei}},\ and\ \bibinfo {author}
  {\bibfnamefont {J.~F.}\ \bibnamefont {Klausner}},\ }\href@noop {} {\bibfield
  {journal} {\bibinfo  {journal} {Physical Review E}\ }\textbf {\bibinfo
  {volume} {89}},\ \bibinfo {pages} {043308} (\bibinfo {year}
  {2014})}\BibitemShut {NoStop}%
\bibitem [{\citenamefont {Guo}\ \emph {et~al.}(2015)\citenamefont {Guo},
  \citenamefont {Li}, \citenamefont {Xiao}, \citenamefont {AuYeung},\ and\
  \citenamefont {Mei}}]{guo2015lattice}%
  \BibitemOpen
  \bibfield  {author} {\bibinfo {author} {\bibfnamefont {K.}~\bibnamefont
  {Guo}}, \bibinfo {author} {\bibfnamefont {L.}~\bibnamefont {Li}}, \bibinfo
  {author} {\bibfnamefont {G.}~\bibnamefont {Xiao}}, \bibinfo {author}
  {\bibfnamefont {N.}~\bibnamefont {AuYeung}},\ and\ \bibinfo {author}
  {\bibfnamefont {R.}~\bibnamefont {Mei}},\ }\href@noop {} {\bibfield
  {journal} {\bibinfo  {journal} {International Journal of Heat and Mass
  Transfer}\ }\textbf {\bibinfo {volume} {88}},\ \bibinfo {pages} {306}
  (\bibinfo {year} {2015})}\BibitemShut {NoStop}%
\bibitem [{\citenamefont {Fieremans}\ and\ \citenamefont
  {Lee}(2018)}]{fieremans2018physical}%
  \BibitemOpen
  \bibfield  {author} {\bibinfo {author} {\bibfnamefont {E.}~\bibnamefont
  {Fieremans}}\ and\ \bibinfo {author} {\bibfnamefont {H.-H.}\ \bibnamefont
  {Lee}},\ }\href@noop {} {\bibfield  {journal} {\bibinfo  {journal}
  {Neuroimage}\ }\textbf {\bibinfo {volume} {182}},\ \bibinfo {pages} {39}
  (\bibinfo {year} {2018})}\BibitemShut {NoStop}%
\bibitem [{\citenamefont {Sharafi}\ and\ \citenamefont
  {Blemker}(2010)}]{sharafi2010micromechanical}%
  \BibitemOpen
  \bibfield  {author} {\bibinfo {author} {\bibfnamefont {B.}~\bibnamefont
  {Sharafi}}\ and\ \bibinfo {author} {\bibfnamefont {S.~S.}\ \bibnamefont
  {Blemker}},\ }\href@noop {} {\bibfield  {journal} {\bibinfo  {journal}
  {Journal of Biomechanics}\ }\textbf {\bibinfo {volume} {43}},\ \bibinfo
  {pages} {3207} (\bibinfo {year} {2010})}\BibitemShut {NoStop}%
\bibitem [{\citenamefont {Wang}\ and\ \citenamefont
  {Georgiadis}(1991)}]{wang1991parallel}%
  \BibitemOpen
  \bibfield  {author} {\bibinfo {author} {\bibfnamefont {M.}~\bibnamefont
  {Wang}}\ and\ \bibinfo {author} {\bibfnamefont {J.~G.}\ \bibnamefont
  {Georgiadis}},\ }\href@noop {} {\bibfield  {journal} {\bibinfo  {journal}
  {Numerical Heat Transfer, Part B Fundamentals}\ }\textbf {\bibinfo {volume}
  {20}},\ \bibinfo {pages} {41} (\bibinfo {year} {1991})}\BibitemShut {NoStop}%
\bibitem [{\citenamefont {Kenkre}\ \emph {et~al.}(1997)\citenamefont {Kenkre},
  \citenamefont {Fukushima},\ and\ \citenamefont
  {Sheltraw}}]{kenkre1997simple}%
  \BibitemOpen
  \bibfield  {author} {\bibinfo {author} {\bibfnamefont {V.}~\bibnamefont
  {Kenkre}}, \bibinfo {author} {\bibfnamefont {E.}~\bibnamefont {Fukushima}},\
  and\ \bibinfo {author} {\bibfnamefont {D.}~\bibnamefont {Sheltraw}},\
  }\href@noop {} {\bibfield  {journal} {\bibinfo  {journal} {Journal of
  Magnetic Resonance}\ }\textbf {\bibinfo {volume} {128}},\ \bibinfo {pages}
  {62} (\bibinfo {year} {1997})}\BibitemShut {NoStop}%
\bibitem [{\citenamefont {Sukstanskii}\ \emph {et~al.}(2004)\citenamefont
  {Sukstanskii}, \citenamefont {Yablonskiy},\ and\ \citenamefont
  {Ackerman}}]{sukstanskii2004effects}%
  \BibitemOpen
  \bibfield  {author} {\bibinfo {author} {\bibfnamefont {A.}~\bibnamefont
  {Sukstanskii}}, \bibinfo {author} {\bibfnamefont {D.}~\bibnamefont
  {Yablonskiy}},\ and\ \bibinfo {author} {\bibfnamefont {J.}~\bibnamefont
  {Ackerman}},\ }\href@noop {} {\bibfield  {journal} {\bibinfo  {journal}
  {Journal of Magnetic Resonance}\ }\textbf {\bibinfo {volume} {170}},\
  \bibinfo {pages} {56} (\bibinfo {year} {2004})}\BibitemShut {NoStop}%
\bibitem [{\citenamefont {Tanner}(1978)}]{tanner1978transient}%
  \BibitemOpen
  \bibfield  {author} {\bibinfo {author} {\bibfnamefont {J.~E.}\ \bibnamefont
  {Tanner}},\ }\href@noop {} {\bibfield  {journal} {\bibinfo  {journal} {The
  Journal of Chemical Physics}\ }\textbf {\bibinfo {volume} {69}},\ \bibinfo
  {pages} {1748} (\bibinfo {year} {1978})}\BibitemShut {NoStop}%
\bibitem [{\citenamefont {S{\"o}derman}\ and\ \citenamefont
  {J{\"o}nsson}(1995)}]{soderman1995restricted}%
  \BibitemOpen
  \bibfield  {author} {\bibinfo {author} {\bibfnamefont {O.}~\bibnamefont
  {S{\"o}derman}}\ and\ \bibinfo {author} {\bibfnamefont {B.}~\bibnamefont
  {J{\"o}nsson}},\ }\href@noop {} {\bibfield  {journal} {\bibinfo  {journal}
  {Journal of Magnetic Resonance, Series A}\ }\textbf {\bibinfo {volume}
  {117}},\ \bibinfo {pages} {94} (\bibinfo {year} {1995})}\BibitemShut
  {NoStop}%
\bibitem [{\citenamefont {Stoller}\ \emph {et~al.}(1991)\citenamefont
  {Stoller}, \citenamefont {Happer},\ and\ \citenamefont
  {Dyson}}]{stoller1991transverse}%
  \BibitemOpen
  \bibfield  {author} {\bibinfo {author} {\bibfnamefont {S.}~\bibnamefont
  {Stoller}}, \bibinfo {author} {\bibfnamefont {W.}~\bibnamefont {Happer}},\
  and\ \bibinfo {author} {\bibfnamefont {F.~J.}\ \bibnamefont {Dyson}},\
  }\href@noop {} {\bibfield  {journal} {\bibinfo  {journal} {Physical Review
  A}\ }\textbf {\bibinfo {volume} {44}},\ \bibinfo {pages} {7459} (\bibinfo
  {year} {1991})}\BibitemShut {NoStop}%
\bibitem [{\citenamefont {Mitra}\ \emph {et~al.}(1993)\citenamefont {Mitra},
  \citenamefont {Sen},\ and\ \citenamefont {Schwartz}}]{mitra1993short}%
  \BibitemOpen
  \bibfield  {author} {\bibinfo {author} {\bibfnamefont {P.~P.}\ \bibnamefont
  {Mitra}}, \bibinfo {author} {\bibfnamefont {P.~N.}\ \bibnamefont {Sen}},\
  and\ \bibinfo {author} {\bibfnamefont {L.~M.}\ \bibnamefont {Schwartz}},\
  }\href@noop {} {\bibfield  {journal} {\bibinfo  {journal} {Physical Review
  B}\ }\textbf {\bibinfo {volume} {47}},\ \bibinfo {pages} {8565} (\bibinfo
  {year} {1993})}\BibitemShut {NoStop}%
\bibitem [{\citenamefont {Towns}\ \emph {et~al.}(2014)\citenamefont {Towns},
  \citenamefont {Cockerill}, \citenamefont {Dahan}, \citenamefont {Foster},
  \citenamefont {Gaither}, \citenamefont {Grimshaw}, \citenamefont {Hazlewood},
  \citenamefont {Lathrop}, \citenamefont {Lifka}, \citenamefont {Peterson}
  \emph {et~al.}}]{towns2014xsede}%
  \BibitemOpen
  \bibfield  {author} {\bibinfo {author} {\bibfnamefont {J.}~\bibnamefont
  {Towns}}, \bibinfo {author} {\bibfnamefont {T.}~\bibnamefont {Cockerill}},
  \bibinfo {author} {\bibfnamefont {M.}~\bibnamefont {Dahan}}, \bibinfo
  {author} {\bibfnamefont {I.}~\bibnamefont {Foster}}, \bibinfo {author}
  {\bibfnamefont {K.}~\bibnamefont {Gaither}}, \bibinfo {author} {\bibfnamefont
  {A.}~\bibnamefont {Grimshaw}}, \bibinfo {author} {\bibfnamefont
  {V.}~\bibnamefont {Hazlewood}}, \bibinfo {author} {\bibfnamefont
  {S.}~\bibnamefont {Lathrop}}, \bibinfo {author} {\bibfnamefont
  {D.}~\bibnamefont {Lifka}}, \bibinfo {author} {\bibfnamefont {G.~D.}\
  \bibnamefont {Peterson}}, \emph {et~al.},\ }\href@noop {} {\bibfield
  {journal} {\bibinfo  {journal} {Computing in Science \& Engineering}\
  }\textbf {\bibinfo {volume} {16}},\ \bibinfo {pages} {62} (\bibinfo {year}
  {2014})}\BibitemShut {NoStop}%
\bibitem [{\citenamefont {Fieremans}\ \emph {et~al.}(2017)\citenamefont
  {Fieremans}, \citenamefont {Lemberskiy}, \citenamefont {Veraart},
  \citenamefont {Sigmund}, \citenamefont {Gyftopoulos},\ and\ \citenamefont
  {Novikov}}]{fieremans2017vivo}%
  \BibitemOpen
  \bibfield  {author} {\bibinfo {author} {\bibfnamefont {E.}~\bibnamefont
  {Fieremans}}, \bibinfo {author} {\bibfnamefont {G.}~\bibnamefont
  {Lemberskiy}}, \bibinfo {author} {\bibfnamefont {J.}~\bibnamefont {Veraart}},
  \bibinfo {author} {\bibfnamefont {E.~E.}\ \bibnamefont {Sigmund}}, \bibinfo
  {author} {\bibfnamefont {S.}~\bibnamefont {Gyftopoulos}},\ and\ \bibinfo
  {author} {\bibfnamefont {D.~S.}\ \bibnamefont {Novikov}},\ }\href@noop {}
  {\bibfield  {journal} {\bibinfo  {journal} {NMR in Biomedicine}\ }\textbf
  {\bibinfo {volume} {30}},\ \bibinfo {pages} {e3612} (\bibinfo {year}
  {2017})}\BibitemShut {NoStop}%
\bibitem [{\citenamefont {Karampinos}\ \emph {et~al.}(2009)\citenamefont
  {Karampinos}, \citenamefont {King}, \citenamefont {Sutton},\ and\
  \citenamefont {Georgiadis}}]{karampinos2009myofiber}%
  \BibitemOpen
  \bibfield  {author} {\bibinfo {author} {\bibfnamefont {D.~C.}\ \bibnamefont
  {Karampinos}}, \bibinfo {author} {\bibfnamefont {K.~F.}\ \bibnamefont
  {King}}, \bibinfo {author} {\bibfnamefont {B.~P.}\ \bibnamefont {Sutton}},\
  and\ \bibinfo {author} {\bibfnamefont {J.~G.}\ \bibnamefont {Georgiadis}},\
  }\href@noop {} {\bibfield  {journal} {\bibinfo  {journal} {Annals of
  biomedical engineering}\ }\textbf {\bibinfo {volume} {37}},\ \bibinfo {pages}
  {2532} (\bibinfo {year} {2009})}\BibitemShut {NoStop}%
\bibitem [{\citenamefont {Hill}(2017)}]{muscle_image1}%
  \BibitemOpen
  \bibfield  {author} {\bibinfo {author} {\bibfnamefont {M.}~\bibnamefont
  {Hill}},\ }\href@noop {} {\bibinfo {title} {Embryology skeletal muscle
  histology}},\ \bibinfo {howpublished}
  {\url{https://embryology.med.unsw.edu.au/embryology/index.php/File:Skeletal_muscle_histology_003.jpg}}
  (\bibinfo {year} {2017}),\ \bibinfo {note} {accessed: 2019-04-05}\BibitemShut
  {NoStop}%
\bibitem [{\citenamefont {Schneider}\ \emph {et~al.}(2012)\citenamefont
  {Schneider}, \citenamefont {Rasband},\ and\ \citenamefont
  {Eliceiri}}]{imageJ}%
  \BibitemOpen
  \bibfield  {author} {\bibinfo {author} {\bibfnamefont {C.~A.}\ \bibnamefont
  {Schneider}}, \bibinfo {author} {\bibfnamefont {W.~S.}\ \bibnamefont
  {Rasband}},\ and\ \bibinfo {author} {\bibfnamefont {K.~W.}\ \bibnamefont
  {Eliceiri}},\ }\href@noop {} {\bibfield  {journal} {\bibinfo  {journal}
  {Nature Methods}\ }\textbf {\bibinfo {volume} {9}},\ \bibinfo {pages} {671}
  (\bibinfo {year} {2012})}\BibitemShut {NoStop}%
\bibitem [{\citenamefont {Naughton}\ and\ \citenamefont
  {Georgiadis}(2020)}]{Naughton2020MRM}%
  \BibitemOpen
  \bibfield  {author} {\bibinfo {author} {\bibfnamefont {N.~M.}\ \bibnamefont
  {Naughton}}\ and\ \bibinfo {author} {\bibfnamefont {J.~G.}\ \bibnamefont
  {Georgiadis}},\ }\href@noop {} {\bibfield  {journal} {\bibinfo  {journal}
  {Magnetic Resonance in Medicine}\ }\textbf {\bibinfo {volume} {83}},\
  \bibinfo {pages} {1458} (\bibinfo {year} {2020})}\BibitemShut {NoStop}%
\bibitem [{\citenamefont {Holdych}\ \emph {et~al.}(2004)\citenamefont
  {Holdych}, \citenamefont {Noble}, \citenamefont {Georgiadis},\ and\
  \citenamefont {Buckius}}]{holdych2004truncation}%
  \BibitemOpen
  \bibfield  {author} {\bibinfo {author} {\bibfnamefont {D.~J.}\ \bibnamefont
  {Holdych}}, \bibinfo {author} {\bibfnamefont {D.~R.}\ \bibnamefont {Noble}},
  \bibinfo {author} {\bibfnamefont {J.~G.}\ \bibnamefont {Georgiadis}},\ and\
  \bibinfo {author} {\bibfnamefont {R.~O.}\ \bibnamefont {Buckius}},\
  }\href@noop {} {\bibfield  {journal} {\bibinfo  {journal} {Journal of
  Computational Physics}\ }\textbf {\bibinfo {volume} {193}},\ \bibinfo {pages}
  {595} (\bibinfo {year} {2004})}\BibitemShut {NoStop}%
\bibitem [{\citenamefont {Naughton}\ and\ \citenamefont
  {Georgiadis}(2019{\natexlab{a}})}]{Naughton2019PEARC}%
  \BibitemOpen
  \bibfield  {author} {\bibinfo {author} {\bibfnamefont {N.~M.}\ \bibnamefont
  {Naughton}}\ and\ \bibinfo {author} {\bibfnamefont {J.~G.}\ \bibnamefont
  {Georgiadis}},\ }in\ \href@noop {} {\emph {\bibinfo {booktitle} {Proceedings
  of the Practice and Experience in Advanced Research Computing on Rise of the
  Machines (learning)}}}\ (\bibinfo {year} {2019})\ pp.\ \bibinfo {pages}
  {1--7}\BibitemShut {NoStop}%
\bibitem [{\citenamefont {Naughton}\ and\ \citenamefont
  {Georgiadis}(2019{\natexlab{b}})}]{Naughton2019PMB}%
  \BibitemOpen
  \bibfield  {author} {\bibinfo {author} {\bibfnamefont {N.~M.}\ \bibnamefont
  {Naughton}}\ and\ \bibinfo {author} {\bibfnamefont {J.~G.}\ \bibnamefont
  {Georgiadis}},\ }\href@noop {} {\bibfield  {journal} {\bibinfo  {journal}
  {Physics in Medicine \& Biology}\ }\textbf {\bibinfo {volume} {64}},\
  \bibinfo {pages} {155004} (\bibinfo {year} {2019}{\natexlab{b}})}\BibitemShut
  {NoStop}%
\bibitem [{\citenamefont {Naughton}(2019)}]{Naughton2019THESIS}%
  \BibitemOpen
  \bibfield  {author} {\bibinfo {author} {\bibfnamefont {N.~M.}\ \bibnamefont
  {Naughton}},\ }\emph {\bibinfo {title} {Diffusion-weighted MRI of skeletal
  muscle: Estimation of microstructural parameters}},\ \href@noop {} {Ph.D.
  thesis},\ \bibinfo  {school} {University of Illinois at Urbana-Champaign}
  (\bibinfo {year} {2019})\BibitemShut {NoStop}%
\bibitem [{\citenamefont {Wolf-Gladrow}(1995)}]{wolf1995lattice}%
  \BibitemOpen
  \bibfield  {author} {\bibinfo {author} {\bibfnamefont {D.}~\bibnamefont
  {Wolf-Gladrow}},\ }\href@noop {} {\bibfield  {journal} {\bibinfo  {journal}
  {Journal of Statistical Physics}\ }\textbf {\bibinfo {volume} {79}},\
  \bibinfo {pages} {1023} (\bibinfo {year} {1995})}\BibitemShut {NoStop}%
\bibitem [{\citenamefont {Ginzburg}(2005)}]{ginzburg2005equilibrium}%
  \BibitemOpen
  \bibfield  {author} {\bibinfo {author} {\bibfnamefont {I.}~\bibnamefont
  {Ginzburg}},\ }\href@noop {} {\bibfield  {journal} {\bibinfo  {journal}
  {Advances in Water Resources}\ }\textbf {\bibinfo {volume} {28}},\ \bibinfo
  {pages} {1171} (\bibinfo {year} {2005})}\BibitemShut {NoStop}%
\bibitem [{\citenamefont {Aho}\ \emph {et~al.}(2016)\citenamefont {Aho},
  \citenamefont {Mattila}, \citenamefont {K{\"u}hn}, \citenamefont
  {Kek{\"a}l{\"a}inen}, \citenamefont {Pulkkinen}, \citenamefont {Minussi},
  \citenamefont {Vihinen-Ranta},\ and\ \citenamefont
  {Timonen}}]{aho2016diffusion}%
  \BibitemOpen
  \bibfield  {author} {\bibinfo {author} {\bibfnamefont {V.}~\bibnamefont
  {Aho}}, \bibinfo {author} {\bibfnamefont {K.}~\bibnamefont {Mattila}},
  \bibinfo {author} {\bibfnamefont {T.}~\bibnamefont {K{\"u}hn}}, \bibinfo
  {author} {\bibfnamefont {P.}~\bibnamefont {Kek{\"a}l{\"a}inen}}, \bibinfo
  {author} {\bibfnamefont {O.}~\bibnamefont {Pulkkinen}}, \bibinfo {author}
  {\bibfnamefont {R.~B.}\ \bibnamefont {Minussi}}, \bibinfo {author}
  {\bibfnamefont {M.}~\bibnamefont {Vihinen-Ranta}},\ and\ \bibinfo {author}
  {\bibfnamefont {J.}~\bibnamefont {Timonen}},\ }\href@noop {} {\bibfield
  {journal} {\bibinfo  {journal} {Physical Review E}\ }\textbf {\bibinfo
  {volume} {93}},\ \bibinfo {pages} {043309} (\bibinfo {year}
  {2016})}\BibitemShut {NoStop}%
\bibitem [{\citenamefont {Holdych}\ \emph {et~al.}(2001)\citenamefont
  {Holdych}, \citenamefont {Georgiadis},\ and\ \citenamefont
  {Buckius}}]{holdych2001migration}%
  \BibitemOpen
  \bibfield  {author} {\bibinfo {author} {\bibfnamefont {D.}~\bibnamefont
  {Holdych}}, \bibinfo {author} {\bibfnamefont {J.}~\bibnamefont
  {Georgiadis}},\ and\ \bibinfo {author} {\bibfnamefont {R.}~\bibnamefont
  {Buckius}},\ }\href@noop {} {\bibfield  {journal} {\bibinfo  {journal}
  {Physics of Fluids}\ }\textbf {\bibinfo {volume} {13}},\ \bibinfo {pages}
  {817} (\bibinfo {year} {2001})}\BibitemShut {NoStop}%
\bibitem [{\citenamefont {Georgiadis}\ \emph {et~al.}(1996)\citenamefont
  {Georgiadis}, \citenamefont {Noble}, \citenamefont {Uchanski},\ and\
  \citenamefont {Buckius}}]{georgiadis1996questions}%
  \BibitemOpen
  \bibfield  {author} {\bibinfo {author} {\bibfnamefont {J.~G.}\ \bibnamefont
  {Georgiadis}}, \bibinfo {author} {\bibfnamefont {D.~R.}\ \bibnamefont
  {Noble}}, \bibinfo {author} {\bibfnamefont {M.~R.}\ \bibnamefont
  {Uchanski}},\ and\ \bibinfo {author} {\bibfnamefont {R.~O.}\ \bibnamefont
  {Buckius}},\ }\href@noop {} {\bibfield  {journal} {\bibinfo  {journal}
  {Journal of Fluids Engineering}\ }\textbf {\bibinfo {volume} {118}},\
  \bibinfo {pages} {434} (\bibinfo {year} {1996})}\BibitemShut {NoStop}%
\bibitem [{\citenamefont {Jurczuk}\ \emph {et~al.}(2013)\citenamefont
  {Jurczuk}, \citenamefont {Kretowski}, \citenamefont {Bellanger},
  \citenamefont {Eliat}, \citenamefont {Saint-Jalmes},\ and\ \citenamefont
  {B{\'e}zy-Wendling}}]{jurczuk2013computational}%
  \BibitemOpen
  \bibfield  {author} {\bibinfo {author} {\bibfnamefont {K.}~\bibnamefont
  {Jurczuk}}, \bibinfo {author} {\bibfnamefont {M.}~\bibnamefont {Kretowski}},
  \bibinfo {author} {\bibfnamefont {J.-J.}\ \bibnamefont {Bellanger}}, \bibinfo
  {author} {\bibfnamefont {P.-A.}\ \bibnamefont {Eliat}}, \bibinfo {author}
  {\bibfnamefont {H.}~\bibnamefont {Saint-Jalmes}},\ and\ \bibinfo {author}
  {\bibfnamefont {J.}~\bibnamefont {B{\'e}zy-Wendling}},\ }\href@noop {}
  {\bibfield  {journal} {\bibinfo  {journal} {Magnetic Resonance Imaging}\
  }\textbf {\bibinfo {volume} {31}},\ \bibinfo {pages} {1163} (\bibinfo {year}
  {2013})}\BibitemShut {NoStop}%
\bibitem [{\citenamefont {Khirevich}\ \emph {et~al.}(2015)\citenamefont
  {Khirevich}, \citenamefont {Ginzburg},\ and\ \citenamefont
  {Tallarek}}]{khirevich2015coarse}%
  \BibitemOpen
  \bibfield  {author} {\bibinfo {author} {\bibfnamefont {S.}~\bibnamefont
  {Khirevich}}, \bibinfo {author} {\bibfnamefont {I.}~\bibnamefont
  {Ginzburg}},\ and\ \bibinfo {author} {\bibfnamefont {U.}~\bibnamefont
  {Tallarek}},\ }\href@noop {} {\bibfield  {journal} {\bibinfo  {journal}
  {Journal of Computational Physics}\ }\textbf {\bibinfo {volume} {281}},\
  \bibinfo {pages} {708} (\bibinfo {year} {2015})}\BibitemShut {NoStop}%
\bibitem [{\citenamefont {K{\"u}hn}\ \emph {et~al.}(2011)\citenamefont
  {K{\"u}hn}, \citenamefont {Ihalainen}, \citenamefont {Hyv{\"a}luoma},
  \citenamefont {Dross}, \citenamefont {Willman}, \citenamefont {Langowski},
  \citenamefont {Vihinen-Ranta},\ and\ \citenamefont
  {Timonen}}]{kuhn2011protein}%
  \BibitemOpen
  \bibfield  {author} {\bibinfo {author} {\bibfnamefont {T.}~\bibnamefont
  {K{\"u}hn}}, \bibinfo {author} {\bibfnamefont {T.~O.}\ \bibnamefont
  {Ihalainen}}, \bibinfo {author} {\bibfnamefont {J.}~\bibnamefont
  {Hyv{\"a}luoma}}, \bibinfo {author} {\bibfnamefont {N.}~\bibnamefont
  {Dross}}, \bibinfo {author} {\bibfnamefont {S.~F.}\ \bibnamefont {Willman}},
  \bibinfo {author} {\bibfnamefont {J.}~\bibnamefont {Langowski}}, \bibinfo
  {author} {\bibfnamefont {M.}~\bibnamefont {Vihinen-Ranta}},\ and\ \bibinfo
  {author} {\bibfnamefont {J.}~\bibnamefont {Timonen}},\ }\href@noop {}
  {\bibfield  {journal} {\bibinfo  {journal} {PloS One}\ }\textbf {\bibinfo
  {volume} {6}},\ \bibinfo {pages} {e22962} (\bibinfo {year}
  {2011})}\BibitemShut {NoStop}%
\bibitem [{\citenamefont {Zhou}\ \emph {et~al.}(2016)\citenamefont {Zhou},
  \citenamefont {Haygarth}, \citenamefont {Withers}, \citenamefont {Macleod},
  \citenamefont {Falloon}, \citenamefont {Beven}, \citenamefont {Ockenden},
  \citenamefont {Forber}, \citenamefont {Hollaway}, \citenamefont {Evans} \emph
  {et~al.}}]{zhou2016lattice}%
  \BibitemOpen
  \bibfield  {author} {\bibinfo {author} {\bibfnamefont {J.}~\bibnamefont
  {Zhou}}, \bibinfo {author} {\bibfnamefont {P.~M.}\ \bibnamefont {Haygarth}},
  \bibinfo {author} {\bibfnamefont {P.}~\bibnamefont {Withers}}, \bibinfo
  {author} {\bibfnamefont {C.}~\bibnamefont {Macleod}}, \bibinfo {author}
  {\bibfnamefont {P.~D.}\ \bibnamefont {Falloon}}, \bibinfo {author}
  {\bibfnamefont {K.~J.}\ \bibnamefont {Beven}}, \bibinfo {author}
  {\bibfnamefont {M.~C.}\ \bibnamefont {Ockenden}}, \bibinfo {author}
  {\bibfnamefont {K.~J.}\ \bibnamefont {Forber}}, \bibinfo {author}
  {\bibfnamefont {M.~J.}\ \bibnamefont {Hollaway}}, \bibinfo {author}
  {\bibfnamefont {R.}~\bibnamefont {Evans}}, \emph {et~al.},\ }\href@noop {}
  {\bibfield  {journal} {\bibinfo  {journal} {Physical Review E}\ }\textbf
  {\bibinfo {volume} {93}},\ \bibinfo {pages} {043310} (\bibinfo {year}
  {2016})}\BibitemShut {NoStop}%
\bibitem [{\citenamefont {Owen}\ \emph {et~al.}(2011)\citenamefont {Owen},
  \citenamefont {Leonardi},\ and\ \citenamefont {Feng}}]{owen2011efficient}%
  \BibitemOpen
  \bibfield  {author} {\bibinfo {author} {\bibfnamefont {D.}~\bibnamefont
  {Owen}}, \bibinfo {author} {\bibfnamefont {C.}~\bibnamefont {Leonardi}},\
  and\ \bibinfo {author} {\bibfnamefont {Y.}~\bibnamefont {Feng}},\ }\href@noop
  {} {\bibfield  {journal} {\bibinfo  {journal} {International Journal for
  Numerical Methods in Engineering}\ }\textbf {\bibinfo {volume} {87}},\
  \bibinfo {pages} {66} (\bibinfo {year} {2011})}\BibitemShut {NoStop}%
\end{thebibliography}%

\end{document}